\documentclass[12pt]{iopart}
\usepackage{graphicx}
\usepackage{amsfonts}
\usepackage{amstext}
\usepackage{amssymb}
\usepackage{cite}

\def\R{{\mathbb R}}
\def\L{{\mathcal L}}
\def\ve{\varepsilon}

\def\y{{\bf y}}
\def\ctan{{\rm ctan}}
\def\u{\tilde{u}}

\begin{document}

\title{Time-averaged MSD of Brownian motion}

\author{Alexei Andreanov$^1$ and Denis S. Grebenkov$^2$}

\address{$^1$ Abdus Salam ICTP - Strada Costiera 11, 34151, Trieste, Italy}
\ead{aandrean@ictp.it}

\address{$^2$ Laboratoire de Physique de la Mati\`{e}re Condens\'{e}e (UMR 7643), CNRS -- Ecole Polytechnique, F-91128 Palaiseau, France}
\ead{denis.grebenkov@polytechnique.edu}

\begin{abstract}
We study the statistical properties of the time-averaged mean-square
displacements (TAMSD).  This is a standard non-local quadratic
functional for inferring the diffusion coefficient from an individual
random trajectory of a diffusing tracer in single-particle tracking
experiments.  For Brownian motion, we derive an exact formula for the
Laplace transform of the probability density of the TAMSD by mapping
the original problem onto chains of coupled harmonic oscillators.
From this formula, we deduce the first four cumulant moments of the
TAMSD, the asymptotic behavior of the probability density and its
accurate approximation by a generalized Gamma distribution.
\end{abstract}
% 02.50.-r  (Probability theory, stochastic processes, and statistics)
% 05.60.-k  (Transport processes)
% 05.10.-a  (Computational methods in statistical physics and nonlinear dynamics) 
% 02.70.Rr  (General statistical methods)

%\begin{keyword}
%Brownian motion
%\PACS{76.60.-k, 82.56.Lz, 61.43.Gt, 76.60.Lz}
%  76.60.-k     Nuclear magnetic resonance and relaxation
%  51.20.+d 	Viscosity, diffusion, and thermal conductivity 
%- 61.18.Fs 	Magnetic resonance techniques; M¡ssbauer spectroscopy
%  61.43.Gt 	Powders, porous materials
%  76.60.Lz 	Spin echoes
%  76.60.Pc 	NMR imaging (for medical NMR imaging, see 87.61.-c)
%  33.25.+k 	Nuclear resonance and relaxation (see also 
%                 76.60.-k Nuclear magnetic resonance and relaxation in condensed matter and 
%                 82.56.-b in physical chemistry and chemical physics)
%  87.61.-c 	Magnetic resonance imaging
%  82.56.Lz 	Diffusion (by NMR)
%  82.56.Ub 	Structure determination with NMR
%\end{keyword}

% 83.10.Rs 	Computer simulation of molecular and particle dynamics 

% 87.16.-b 	Subcellular structure and processes
%   87.16.A- 	Theory, modeling, and simulations 

%\keywords{ Subdiffusion, MSD, RMS, generalized Langevin equation, Gaussian process, memory kernel}


%\submitto{\JSTAT}

\date{\today}

\maketitle

\section{Introduction}

The statistical properties of Brownian motion play a crucial role in
disciplines as various as mathematics, physics, chemistry, biology,
engineering, economics and ecology
\cite{Weiss,Ben-Avraham,Saxton97b,Majumdar05,Grebenkov07}.  In many
cases, an observation or an experiment can be repeated a large number
of times to get a representative statistics of outcomes.  In other
situations, one measures an observable which reflects a cumulative
effect of a large number of molecules (e.g., the macroscopic nuclear
magnetic resonance signal formed by an extremely large number of
nuclei \cite{Grebenkov07}).  In both cases, a system can be
characterized by ensemble-averaged quantities.  For instance, the
ensemble average (or expectation) of the mean-square displacement
(MSD) of a freely diffusing particle, $\langle X^2(t)\rangle$, yields
a measure of the diffusion coefficient $D$ through the Einstein's
relation: $\langle X^2(t)\rangle = 2Dt$.  However, there are other
situations when few or even single stochastic trajectory $X(t)$ is
available that makes the ensemble averaging inaccurate or impossible.
The common examples are stock prices in finance \cite{Bouchaud} or
trajectories of macromolecules inside living cells acquired by single
particle tracking (SPT) techniques
\cite{Saxton97b,Qian91,Saxton93,Saxton97,Goulian00,Tolic04,Golding06,Arcizet08,Wilhelm08,Wirtz09,Metzler09}.
In the former case, the observation (i.e., a price time series) is
evidently unique for each asset.  In the latter case, although an
experiment can be repeated, the ensemble average may still be
problematic, especially for tracers that move in spatially
heterogeneous and time-evaluating media such as living cells.  In
order to infer the dynamical information from such individual
trajectories, one needs therefore to replace ensemble averages by time
averages.  For instance, the diffusion coefficient is often inferred
by considering the time-averaged MSD (TAMSD) at a lag time $t$ over a
sample trajectory of duration $T$:
\begin{equation}
\label{eq:MSD0}
\chi_{t,T} = \frac{1}{T-t} \int\limits_0^{T-t} dt_0 (X(t_0+t) - X(t_0))^2 .
\end{equation}
Although this is still a random variable, the time average reduces its
fluctuations around the mean which, for Brownian motion, is $\langle
\chi_{t,T} \rangle = 2Dt$.

The statistical properties of quadratic functionals of a Gaussian
process have been intensively studied.  In mathematical statistics,
several series representations of the probability density of a general
quadratic form of a Gaussian process have been proposed, e.g. a
mixture of chi-squared distributions
\cite{Ruben62,Ruben63,Robbins48,Robbins49}, a power series expansion
\cite{Pachares55,Shah61}, and a series expansion over Laguerre
polynomials
\cite{Gurland53,Gurland55,Gurland56} (see also
\cite{Kotz67a,Kotz67b} for a review).  For biophysical applications,
Qian and co-workers analyzed the TAMSD of a discrete random walk which
is known to have a Gamma (or chi-squared) distribution only at the
smallest lag time \cite{Qian91}.  The distribution of diffusion
coefficients was also studied via Monte Carlo simulations by Saxton
\cite{Saxton93,Saxton97} (see also \cite{Goulian00,Duits09} for
experimental data).  More recently, Boyer and Dean investigated the
distribution of several estimators of diffusion coefficients for
Brownian motion \cite{Boyer11}.  They reduced the analysis of
least-squares and maximum likelihood estimates to the distribution of
a local quadratic functional of Brownian motion.  Mapping this problem
onto an imaginary time Schr\"odinger equation, they derived explicit
analytical results for the distribution of the diffusion coefficient
estimator.  In a separate paper, Boyer and co-workers discussed the
optimal way to extract diffusion coefficients from single trajectories
of Brownian motion and showed the superior efficiency of the maximum
likelihood estimator over least-squares estimates \cite{Boyer12}.  An
optimal Bayesian method for quantifying biomolecule diffusivity was
presented by Voisinne {\it et al.} \cite{Voisinne10}.

A different approach to the analysis of inferring schemes with
quadratic functionals was developed by Grebenkov
\cite{Grebenkov11a,Grebenkov11b}.  First, the explicit formulas for
the mean and variance of the TAMSD (and another quadratic functional)
were derived for a large class of Gaussian processes governed by a
generalized Langevin dynamics \cite{Grebenkov11a}.  The formula for
the variance allows one to estimate the spread of measurements and to
choose an appropriate sample duration $T$.  Second, a matrix
representation for the characteristic function of $\chi_{t,T}$ was
used in order to analyze the probability density $p(z)$ of the TAMSD
of a general discrete Gaussian process \cite{Grebenkov11b}.  Moreover,
an empirical approximation for $p(z)$ by a generalized Gamma
distribution (GGD) was proposed:
\begin{equation}
\label{eq:GGD}
p(z) = \frac{z^{\nu-1}}{2(ab)^{\nu/2} K_\nu(2\sqrt{a/b})} \exp\left(-\frac{a}{z} - \frac{z}{b}\right),
\end{equation}
with three parameters $a \geq 0$, $b > 0$ and $\nu \in \R$, and
$K_\nu(z)$ is the modified Bessel function of the second kind (in the
limit $a\to 0$, one retrieves the Gamma distribution $\frac{z^{\nu-1}
e^{-z/b}}{\Gamma(\nu) b^\nu}$).  Although the high accuracy of the GGD
approximation \eref{eq:GGD} was confirmed numerically for a discrete
Brownian motion, the theoretical status of this approximation remained
unclear.  In addition, the dependence of the parameters $a$, $b$ and
$\nu$ on the lag time $t$ was not analyzed.

In this paper, we rigorously characterize the TAMSD of Brownian
motion.  In Sect. \ref{sec:general}, we introduce a functional
integral representation for a Laplace transform $\varphi(s)$ of the
probability density $p(z)$ which is a cornerstone for all consecutive
derivations.  For $t > T/2$, it leads to a simple closed formula for
$\varphi(s)$.  For $t < T/2$, an exact representation of $\varphi(s)$
involves a determinant of an explicit matrix of size $[T/t] \times
[T/t]$, $[T/t]$ being an integer part of $T/t$.  In
Sect. \ref{sec:discussion}, we discuss some practical consequences of
these theoretical results.  In particular, we derive the explicit
formulas for the first four cumulant moments of the TAMSD, as well as
the asymptotic behavior of the probability density $p(z)$ at small and
large $z$.  For instance, we show that $p(z)$ decays as $e^{-a/z}$ at
small $z$ that justifies the use of a GGD as a convenient
approximation for $p(z)$.  It also explains a remarkable accuracy of
this approximation which correctly reproduces the asymptotic behavior
of $p(z)$ at both small and large $z$.  Finally, we derive the
asymptotic relations for the parameters $a$, $b$ and $\nu$ of the GGD
as functions of the lag time $t$ by matching the first moments of the
true and approximate distributions.  These relations allow one to get
an accurate approximation of the probability density $p(z)$ for a
given lag time $t$ that opens a number of practical applications for
single-particle tracking experiments.  These findings and future
perspectives are summarized in Conclusion.

\section{Theoretical results}
\label{sec:general}

Throughout this section, we consider $X(u)$ to be Brownian motion with
mean zero and covariance $\langle X(u),X(v)\rangle = \min\{u,v\}$,
with diffusion coefficient $D$ set to $1/2$.  The sample duration $T$
is fixed to $1$, while the lag time $t \in [0,1]$ is dimensionless.
We are interested in the probability density $p_t(z) =
\langle\delta(z-\chi_t)\rangle$ of the random variable $\chi_t$ which
denotes the TAMSD of $X(u)$:
\begin{equation}
\label{eq:MSD}
\chi_t = \frac{1}{1-t} \int\limits_0^{1-t} du (X(u+t) - X(u))^2 .
\end{equation}
We aim at calculating the Laplace transform $\varphi_t(s)$ of the
probability density $p_t(z)$
\begin{equation}
\label{eq:varphi_def}
\varphi_t(s) = \langle \exp(-s\chi_t)\rangle = \int\limits_0^\infty dz\,e^{-s z} p_t(z) ,
\end{equation}
Once $\varphi_t(s)$ is obtained, one can retrieve
the probability density $p_t(z)$ either through the inverse Laplace
transform of $\varphi_t(s)$, or through the inverse Fourier transform
of the characteristic function $\phi_t(k) \equiv \langle
e^{ik\chi_t}\rangle = \varphi_t(-ik)$ (although both ways are formally
equivalent, Fourier transforms are more convenient for numerical
implementation).  From a theoretical point of view, the knowledge of
$\varphi_t(s)$ is therefore fully equivalent to that of $p_t(z)$ (see
Sect. \ref{sec:discussion} for practical consequences).

We start by expressing the Laplace transform in
Eq. \eref{eq:varphi_def} as a functional integral over trajectories
$X(u)$ of a Brownian motion~\cite{Majumdar05,Feynman}
\begin{equation}
\label{eq:varphi_path}
\fl
\varphi_t(s)=\int\limits_\R dx_1 \hspace*{-3mm} \int\limits_{X(0)=0}^{X(1)=x_1}  \hspace*{-3mm} \mathcal{D}X(u) \exp\left(-\frac{1}{2}
\int\limits_0^1 (\partial_u X(u))^2 - \frac{s}{1-t}\int\limits_0^{1-t}(X(u+t)-X(u))^2\right),
\end{equation}
where we assume that Brownian motion starts from the origin: $X(0) =
0$ (anyway, the TAMSD is translation invariant and thus the above
functional integral is independent of the starting point).  The next
step would be to use the Feynman-Kac formula
\cite{Majumdar05,Freidlin} and map the functional integral to
a propagator of a certain quantum mechanical problem, to compute the
propagator and to evaluate $\varphi_t(s)$.  This method was
successfully used by Boyer and Dean for local quadratic functionals of
Brownian motion \cite{Boyer11}.  However, in our problem different
parts of the trajectory, at times $u$ and $u+t$, are coupled, and the
action is \emph{non-local}, forbiding a direct use of the Feynman-Kac
formalism.  To overcome this difficulty, we partition the trajectory
into pieces and relabel them as new Brownian walkers in order to
reduce it to an effectively many-body but \emph{local} problem.  The
latter is not surprising since a system with memory is often
equivalent to some many-body system, for example dynamics of a
classical spin-glass or quantum mean-field spin-glass are typically
mapped onto single particle problems with
memory~\cite{Bray80,Kirkpatrick87}.  The transformation appearing in
our problem is similar in spirit, but uses the inverse mapping to turn
a \emph{non-local} single particle problem into a \emph{local}
many-body problem.  We illustrate this in detail for the case
$t>\frac{1}{2}$ where the mapping and the computation are relatively
simple.  The case $t<\frac{1}{2}$ is a straightforward generalization
which is treated in the next Subsection.

\begin{figure}
\begin{center}
\includegraphics[width=100mm]{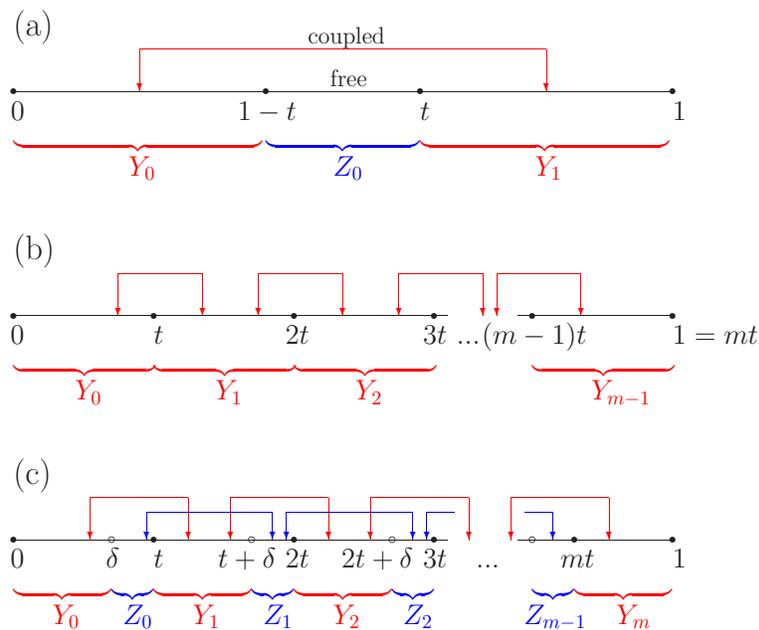}
\end{center}
\caption{
(Color online) Splitting of a trajectory into blocks. {\bf (a)} For
$t>\frac{1}{2}$, there are only three blocks: two coupled parts $Y_0$
and $Y_1$ and one free part $Z_0$; {\bf (b)} When $t = \frac{1}{m}$
($m = 2,3,4,...$), a trajectory is split into $m$ equal blocks
yielding a chain of $m$ successively coupled oscillators $Y_0$, ...,
$Y_{m-1}$; {\bf (c)} When $\frac{1}{1+m} < t < \frac{1}{m}$ (for some
$m = 2,3,4,...$), the time interval $[0,1]$ is covered by $m$
subintervals of duration $t$, plus a smaller interval $[1-mt,1]$ of
duration $\delta=1-mt$.  In order to split the interval $[0,1]$
regularly, each subinterval is divided into two parts, of durations
$\delta$ and $t-\delta$, respectively.  As a consequence, the
trajectory is mapped onto two chains of coupled oscillators $\{Y_k\}$
and $\{Z_k\}$. The two chains $\{Y_k\}$ and $\{Z_k\}$ are coupled
through boundary conditions for successive parts of the trajectory:
$Y_k(\delta) = Z_k(0)$, $k=0,...,m-1$ which express continuity of the
original trajectory $X(u)$.}
\label{fig:tsplit}
\end{figure}

\subsection{Case $t > \frac{1}{2}$}
\label{sec:m1}

The partitioning of the trajectory in the case $t < \frac12$ is
illustrated on Fig. \ref{fig:tsplit}a.  We split the whole trajectory
$X(u)$ into three parts:
\begin{eqnarray*}
Y_0(u) &=& X(u)  \hskip 22.5mm    u\in [0,1-t], \\
Z_0(u) &=& X(u + 1 - t)\qquad   u\in [0,2t-1],  \\
Y_1(u) &=& X(u + t)\hskip 16mm  u\in [0,1-t]. 
\end{eqnarray*}
The original trajectory is then replaced by a collection of two
harmonically coupled particles and a free walker.  Boundary conditions
express continuity of the original trajectory $X(u)$:
\begin{eqnarray*}
\fl
&& \varphi_t(s) = \int\limits_{\R^3} dx_{1-t} dx_t dx_1 \int\limits_{Y_0(0)=0}^{Y_0(1-t)=x_{1-t}}
\mathcal{D}Y_0\int\limits_{Z_0(0)=x_{1-t}}^{Z_0(2t-1)=x_t}\mathcal{D}Z_0\int\limits_{Y_1(0)=x_t}^{Y_1(1-t)=x_1}\mathcal{D}Y_1  \\
\fl
&& \times \exp\left(- \frac{1}{2}\int\limits_0^{2t-1}du\,(\partial_u Z_0)^2 
- \frac{1}{2}\int\limits_0^{1-t}du\,\left[(\partial_u Y_0)^2 + (\partial_u Y_1)^2\right] - \frac{g^2}{2}\int\limits_0^{1-t}du\,(Y_1-Y_0)^2\right) .
\end{eqnarray*}
Note that the subscripts of $x_{1-t}$, $x_t$ and $x_1$ are just
notations for these integration variables.  We have defined for
convenience
\begin{equation*}
g^2 = \frac{2s}{1-t}.
\end{equation*}
The original non-locality is translated into harmonic coupling of
particles $Y_0$, $Y_1$, the overall action is now local.  Since there
are just two particles involved, this case is easy to resolve: $Y_0$,
$Y_1$ are decoupled in a standard way by defining their normal modes
$A_0$, $A_1$
\begin{equation*}
A_0(u) = \frac{1}{\sqrt{2}}(Y_0(u)+Y_1(u)),\qquad A_1(u) = \frac{1}{\sqrt{2}}(Y_0(u)-Y_1(u)),
\end{equation*}
where $A_0$ represents the center of mass motion and $A_1$ describes
harmonic oscillations around the center of mass.  Boundary conditions
for $A_0$ and $A_1$ are:
\begin{eqnarray*}
a_{0,0} &=& A_0(0) = \frac{1}{\sqrt{2}}(Y_0(0)+Y_1(0)) = \frac{1}{\sqrt{2}}(0+x_{t}), \\ 
a_{0,1-t} &=& A_0(1-t) = \frac{1}{\sqrt{2}}(Y_0(1-t)+Y_1(1-t)) = \frac{1}{\sqrt{2}}(x_{1-t}+x_{1}) ,\\
a_{1,0} &=& A_1(0) = \frac{1}{\sqrt{2}}(Y_0(0)-Y_1(0)) = \frac{1}{\sqrt{2}}(0-x_{t}), \\
a_{1,1-t} &=& A_1(1-t) = \frac{1}{\sqrt{2}}(Y_0(1-t)-Y_1(1-t)) = \frac{1}{\sqrt{2}}(x_{1-t}-x_{1}) .
\end{eqnarray*}
Therefore we get a system of two free particles $Z_0$, $A_0$ and one
harmonic oscillator $A_1$:
\begin{eqnarray*}
\fl
&& \varphi_{t}(s)=\int\limits_{\R^3} dx_{1-t}\, dx_{t}\, d x_{1}\int_{A_1(0)=a_{1,0}}^{A_1(1-t)=a_{1,1-t}} \hspace*{-2mm} \mathcal{D}A_1\, 
\exp\left(-\frac{1}{2}\int_{0}^{1-t} \hspace*{-2mm} du (\partial_{u} A_1)^{2}- \frac{g^2}{2} \int_{0}^{1-t}\hspace*{-2mm} du A_1^{2} \right) \times \\
\fl
&& \int_{Z(0)=x_{1-t}}^{Z(2t-1)=x_t} \hspace*{-2mm} \mathcal{D}Z_0  \exp\left(-\frac{1}{2}\int_0^{2t-1} \hspace*{-3mm} du (\partial_{u} Z_0)^{2}\right)
\int_{A_0(0)=a_{0,0}}^{A_0(1-t)=a_{0,1-t}} \hspace*{-2mm} \mathcal{D}A_0  
\exp\left(-\frac{1}{2}\int_{0}^{1-t} \hspace*{-3mm} du (\partial_{u} A_0)^{2}\right) .
\end{eqnarray*}
The Feynman-Kac formula can now be applied to each particle separately
expressing $\varphi_t(s)$ as a product of propagators of respective
quantum problems:
\begin{equation*}
\fl 
\varphi_t(s) = \int\limits_{\R^3} dx_{1-t} dx_t dx_1  ~G_g(a_{1,1-t},1-t ~|~ a_{1,0},0) ~G_0(a_{0,1-t},1-t~|~ a_{0,0},0)~ G_0(x_t,t~|~x_{1-t},1-t),
\end{equation*}
i.e. two free propagators $G_0$ and one propagator $G_g$ for a
harmonic oscillator:
\begin{equation}
\label{eq:propagator}
G_g(x,t|y,u) = \frac{\sqrt{g}~ e^{-(t-u)g/2}}{\sqrt{\pi(1 - q^2)}} \exp\left(- \frac{\alpha g}{4} (x+y)^2 - 
\frac{g}{4\alpha} (x-y)^2\right) ,
\end{equation}
where $q = e^{-(t-u)g}$ and $\alpha = (1-q)/(1+q)$.  In the limit
$g\to 0$, one retrieves the diffusive propagator of a free walker:
\begin{equation}
G_0(x,t|y,u) = \frac{1}{\sqrt{2\pi (t-u)}} \exp\left(- \frac{(x-y)^2}{2(t-u)}\right) .
\end{equation}
Pluging expressions for propagators, one gets
\begin{eqnarray*}
\fl
&& \varphi_t(s) = \int\limits_{\R^3} dx_{1-t} dx_t dx_1 \frac{\sqrt{g}~ e^{-(1-t)g/2}}{\sqrt{\pi(1 - q^2)}} 
\exp\left(- \frac{\alpha g}{4} (a_{1,0} + a_{1,1-t})^2 - \frac{g}{4\alpha} (a_{1,0} - a_{1,1-t})^2\right)\times \\
\fl
&& \times \frac{\exp\left(- \frac{(a_{0,1-t} - a_{0,0})^2}{2(1-t)}\right)}{\sqrt{2\pi(1-t)}}
\times \frac{\exp\left( - \frac{(x_t - x_{1-t})^2}{2(2t - 1)}\right)}{\sqrt{2\pi (2t-1)}} ,
\end{eqnarray*}
where $q = e^{-(1-t)g}$, $\alpha = (1-q)/(1+q)$.  Substituting the
expressions for $a_{0,0}$, $a_{0,1-t}$, $a_{1,0}$, $a_{1,1-t}$ and
evaluating directly the Gaussian integrals, we get the following
closed formula
\begin{equation}
\label{eq:varphi_m1}
\varphi_t(s) = \frac{2 e^{-\sqrt{s(1-t)}}}{1 + e^{-2\sqrt{s(1-t)}}} ~ \left(1 + \sqrt{s(1-t)} ~\frac{3t-1}{1-t} ~
\frac{1- e^{-2\sqrt{s(1-t)}}}{1+ e^{-2\sqrt{s(1-t)}}} \right)^{-1/2} . 
\end{equation}
This formula is one of the main results of the paper.

\subsection{ Case $t < \frac{1}{2}$ }

Although the analysis for the case $t<\frac{1}{2}$ relies on the same
ideas and it is a straightforward generalization of the above
calculus, the computation is much more involved because the number of
particles is growing linearly to infinity as $t\to 0$.  Technically
one has to distinguish subcases of $\frac{1}{1+m}\leq
t\leq\frac{1}{m}$ for all integer $m>1$.  This is a reflection of the
fact that the smaller $t$ is, the more images $t+k\,u$ with integer
$k$ can be placed inside the interval $[0,1]$, yielding $X(u)$ coupled
to $X(u+t)$, which is coupled to $X(u+2t)$ and so forth.  As
previously, the original problem with self-interaction maps onto an
exactly solvable model of coupled harmonic oscillators.

For discrete lag times $t = \frac{1}{m}$ with $m = 2,3,4,...$, the
whole time interval $[0,1]$ is covered by $m$ equal subintervals so
that the trajectory can be mapped onto a single chain of $m$ coupled
harmonic oscillators as shown on Fig.~\ref{fig:tsplit}b.  Using this
mapping, we derive in \ref{sec:details1} the exact formula for the
Laplace transform $\varphi_t(s)$:
\begin{equation}
\label{eq:varphi_m}
\varphi_t(s) = \left(\prod\limits_{k=1}^{m-1} \frac{tg\sqrt{\lambda_k}}{\sinh(tg\sqrt{\lambda_k})}\right)^{1/2} \frac{1}{\sqrt{t^m \det(\mathbf{A})}} .
\end{equation}
where $\lambda_k = 2(1- \cos(\frac{\pi k}{m}))$, with $k = 1,...,m-1$.
The matrix $\mathbf{A}$ of size $m\times m$, which represents the
coupling of harmonic oscillators, is given by Eqs. \eref{eq:A} or
\eref{eq:A2}.  Although Eq. \eref{eq:varphi_m} provides an exact solution
for the original problem, it has a rather formal nature, as evaluation
of $\det(\mathbf{A})$ for large matrix sizes $m$ (i.e. small times
$t$) is quite complicated since $\mathbf{A}$ is a generic matrix with
no particular simplifying structure (see \ref{sec:case_m2}).  This is
the main difficulty encountered in the computation of $\varphi_t(s)$
for small values of $t$.

In the generic case $\frac{1}{1+m} < t < \frac{1}{m}$ (for some $m =
2,3,4,...$), the interval $[0,1]$ is covered by $m$ subintervals of
duration $t$, plus a smaller interval $[mt,1]$ of duration $\delta = 1
- mt < t$.  In order to split the interval $[0,1]$ regularly, each
subinterval is divided into two parts, of durations $\delta$ and
$t-\delta$.  This covering maps the whole trajectory onto two
interacting chains of $m+1$ and $m$ harmonic oscillators as
illustrated on Fig.~\ref{fig:tsplit}c.  In
\ref{sec:details2}, we derive the following exact formula
\begin{equation}
\label{eq:varphi_m2}
\fl
\varphi_t(s) = \left(\prod\limits_{k=1}^{m} 
\frac{\delta g\sqrt{\tilde{\lambda}_k}}{\sinh(\delta g\sqrt{\tilde{\lambda}_k})}\right)^{1/2}
\left(\prod\limits_{k=1}^{m-1} \frac{(t-\delta)g\sqrt{\lambda_k}}{\sinh((t-\delta)g\sqrt{\lambda_k})}\right)^{1/2}
\hspace*{-3mm} \frac{1}{\sqrt{\delta^{m+1}(t-\delta)^m \det(\mathbf{A})}} ,
\end{equation}
where $\tilde{\lambda}_k = 2(1- \cos(\frac{\pi k}{m+1}))$, with $k =
1,...,m$.  The matrix $\mathbf{A}$ of size $(2m+1)\times (2m+1)$,
which represents the coupling of harmonic oscillators, is given by
Eq. \eref{eq:A_gen}.  Once again, this solution is exact though still
quite formal, as the dependence of $\varphi_t(s)$ on $s$ and $t$ is
partly ``hidden'' in the determinant of the matrix $\mathbf{A}$.

\section{Discussion}
\label{sec:discussion}

Quite surprisingly, a theoretical analysis of the statistical
properties of the TAMSD turned out to be a challenging problem, even
in the simplest case of Brownian motion.  This reflects the
non-locality of the involved quadratic functional and the many-body
character of the underlying quantum system.  It is also worth
stressing that a ``resolution'' of this problem would mean different
things for theoreticians and experimentalists.  In the previous
Section, we have fully resolved the problem from a theoretical point
of view, by providing the exact formulas \eref{eq:varphi_m1},
\eref{eq:varphi_m}, \eref{eq:varphi_m2} for the Laplace transform
$\varphi_t(s)$ of the probability density.  However, this solution
remains somewhat formal since an explicit ``shape'' of the probability
density is still unknown, even for the simplest case $t > \frac12$.
In this Section, we discuss some practical consequences of these
results.

\subsection{Moments}

\subsubsection{Case $t > \frac12$.}

Using the exact solution \eref{eq:varphi_m1}, one can compute the
moments of the TAMSD:
\begin{equation}
\mu_k = (-1)^k \left(\frac{\partial^k}{\partial s^k} \varphi_t(s)\right)_{s=0} ,
\end{equation}
from which the cumulant moments are
\begin{eqnarray}
\label{eq:kappa1_1}
\kappa_1 &=& t ,\\
\label{eq:kappa2_1}
\kappa_2 &=& \frac{1}{3} (11t^2-6t+1), \\
\label{eq:kappa3_1}
\kappa_3 &=& \frac{1}{15} (286t^3-216t^2+54t-4), \\
\label{eq:kappa4_1}
\kappa_4 &=& \frac{2}{105} (8065t^4 - 8220t^3 + 3206t^2 - 572t + 41) .
\end{eqnarray}
Here, $\kappa_1$ is the mean value, $\kappa_2$ the variance, while the
skewness and kurtosis are given as $\kappa_3/\kappa_2^{3/2}$ and
$\kappa_4/\kappa_2^2$, respectively.  We note that the expression for
the variance $\kappa_2$ reproduces the result by Qian {\it et al.}
which was obtained for a discrete random walk \cite{Qian91}.  In turn,
the formulas for the third and fourth moments are derived for the
first time.

\begin{figure}
\begin{center}
\includegraphics[width=75mm]{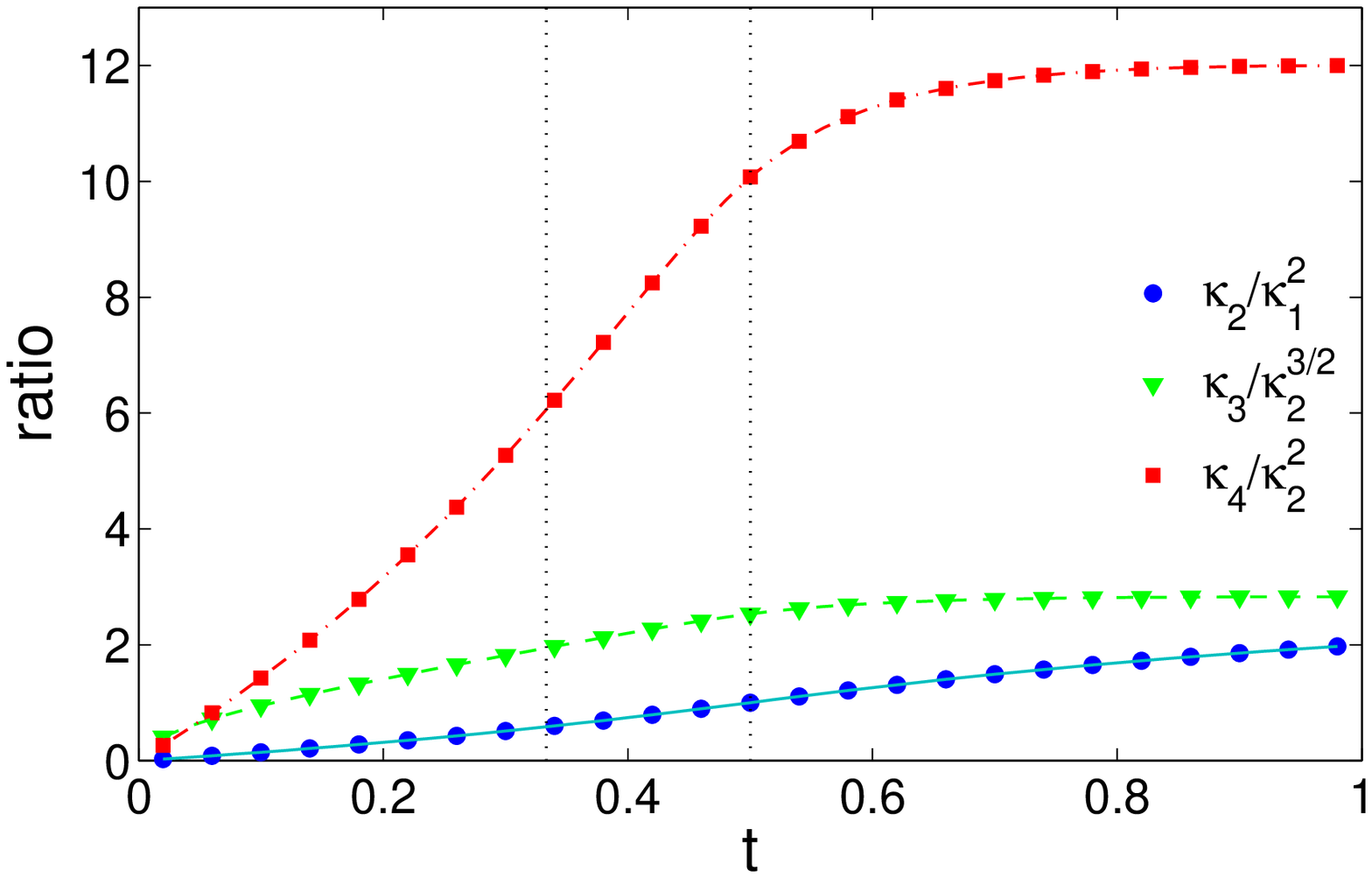}
\includegraphics[width=75mm]{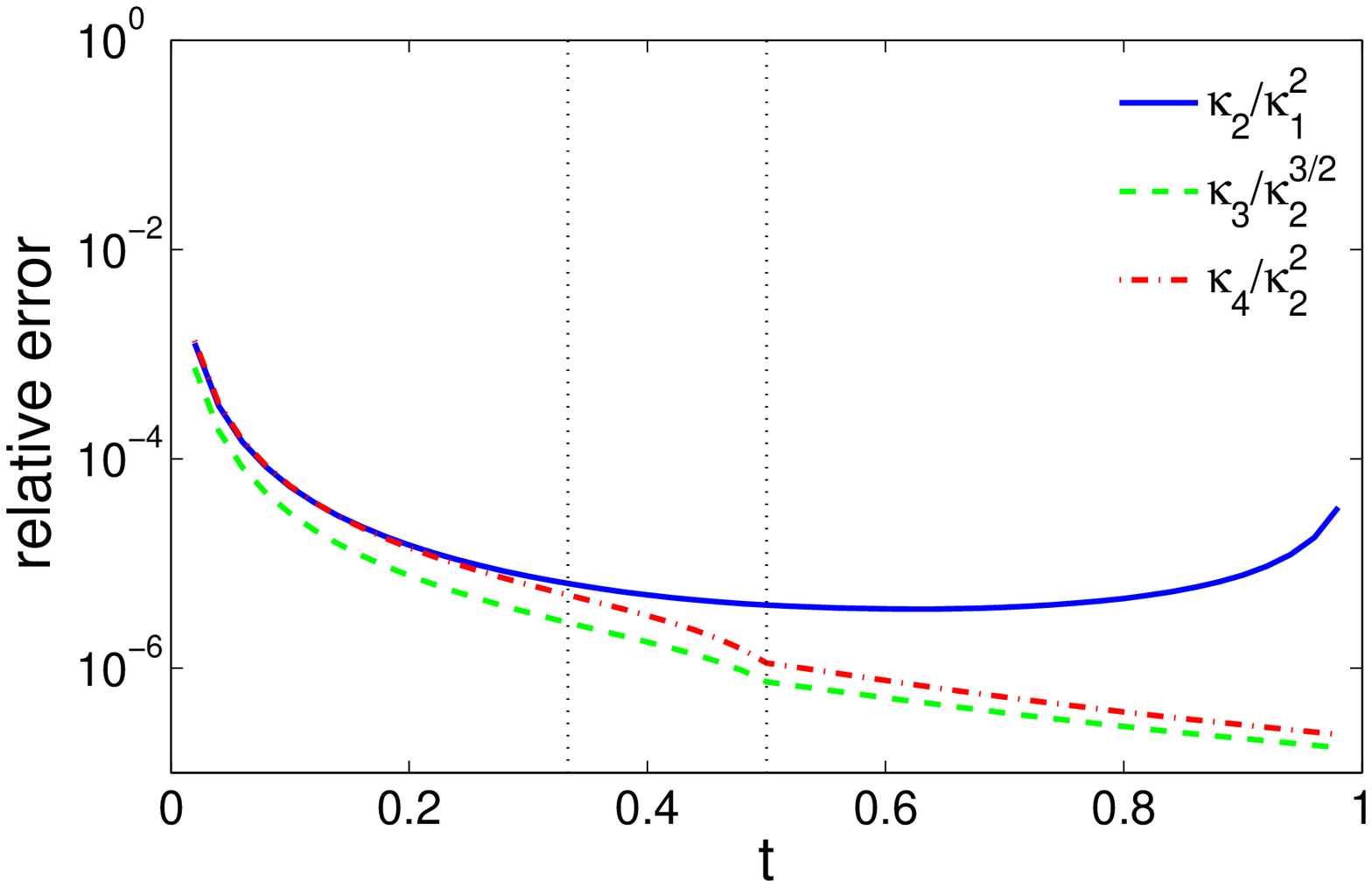}
\end{center}
\caption{
(Color online) {\bf (Left)} The reduced variance
$\kappa_2/\kappa_1^2$, skewness $\kappa_3/\kappa_2^{3/2}$ and kurtosis
$\kappa_4/\kappa_2^2$ of the TAMSD $\chi_t$ as functions of the lag
time $t$.  Symbols present the values obtained from an exact matrix
formula for a discrete random walk (see Eq. (6) from
\cite{Grebenkov11b}, the sample length $N = 1000$), while lines refer
to the analytical formulas
\eref{eq:kappa2_1}, \eref{eq:kappa3_1}, \eref{eq:kappa4_1} for $t >
1/2$ and \eref{eq:kappa2_2}, \eref{eq:kappa3_2}, \eref{eq:kappa4_2}
(or \eref{eq:kappa4_0}) for $t < 1/2$.  {\bf (Right)} The relative
error between the analytical formulas and their counterparts for a
discrete random walk.  The vertical dotted lines stand at $t = 1/3$
and $t = 1/2$.  All the formulas are accurate for the whole region (an
increase of the relative error at small $t$ is related to intrinsic
differences between discrete and continuous processes). }
\label{fig:kappa}
\end{figure}
%A_Andreanov_fit_kappa2.m;

\subsubsection{Case $t < 1/2$.}

For the case $1/3 < t < 1/2$ ($m = 2$), the explicit formula for
$\varphi_t(s)$ is provided in \ref{sec:case_m2}.  Taking the
derivatives with respect to $s$, we obtain the first cumulant moments:
\begin{eqnarray}
\kappa_1 &=& t , \\
\label{eq:kappa2_2}
\kappa_2 &=& \frac{1}{3}~ \frac{t^3(4-5t)}{(1-t)^2} ,   \\
\label{eq:kappa3_2}
\kappa_3 &=& \frac{2}{15}~ \frac{t^5 (33 - 47t)}{(1-t)^3} , \\
\label{eq:kappa4_2}
\kappa_4 &=& \frac{2}{105} (1-t)^{-4} \biggl[30913t^8 - 86272t^7 + 102060t^6 \\
\nonumber
&& - 68040t^5 + 28350t^4 - 7560t^3 + 1260t^2 - 120t + 5 \biggr] .
\end{eqnarray}
As the explicit formula (\ref{eq:varphi_m2}) for $\varphi_t(s)$ is too
cumbersome for $t < 1/3$, another representation \eref{eq:kappan_G}
for the cumulant moments was deduced in \ref{sec:cumulant}.  In
particular, we checked that Eqs. \eref{eq:kappa2_2} and
\eref{eq:kappa3_2} for $\kappa_2$ and $\kappa_3$ are also valid for $t
< 1/3$.  In turn, the formula for $\kappa_4$ for $t < 1/3$ becomes
\begin{equation}
\label{eq:kappa4_0}
\kappa_4 = \frac{8t^7}{105(1-t)^4} (302 - 473 t)   \qquad (t < 1/3) .
\end{equation}

Figure \ref{fig:kappa} shows the reduced variance
$\kappa_2/\kappa_1^2$, skewness $\kappa_3/\kappa_2^{3/2}$ and kurtosis
$\kappa_4/\kappa_2^2$ of the TAMSD as functions of the lag time $t$.
The values obtained from an exact matrix formula for a discrete random
walk (see Eq. (6) from \cite{Grebenkov11b}, the sample length $N =
1000$), are compared to the analytical formulas \eref{eq:kappa2_1},
\eref{eq:kappa3_1}, \eref{eq:kappa4_1} for $t > 1/2$ and
\eref{eq:kappa2_2}, \eref{eq:kappa3_2}, \eref{eq:kappa4_2} (or
\eref{eq:kappa4_0}) for $t < 1/2$.  Notice an excellent
agreement between the curves.

\subsection{Small $z$ asymptotic behavior}
\label{sec:pz_small}

\subsubsection{Case $t > \frac12$}

In the limit of large $s$, Eq. \eref{eq:varphi_m1} can be approximated
as
\begin{equation*}
\varphi_t(s) \approx \frac{2 e^{-\sqrt{s(1-t)}}}{\sqrt{\sqrt{s(1-t)} ~\frac{3t-1}{1-t}}}  \qquad (s \gg 1).
\end{equation*}
Using the asymptotic relation for the inverse Laplace transform
$\L^{-1}$,
\begin{equation}
\label{eq:Laplace}
\L^{-1}[s^{\kappa} e^{- 2\sqrt{as}}] \approx \frac{a^{\kappa+1/2}}{\sqrt{\pi}} z^{-2\kappa-3/2} e^{-a/z} ,
\end{equation}
with $\kappa = -1/4$ and $a = (1-t)/4$, one derives the asymptotic
behavior of the probability density $p_t(z)$ at small $z$:
\begin{equation}
\label{eq:pz_approx1}
p_t(z) \approx \frac{2}{\sqrt{2\pi \frac{3t-1}{1-t}}} ~  \frac{e^{-(1-t)/(4z)}}{z} \qquad (z\to 0) .
\end{equation}
In other words, the asymptotic decay of $\varphi_t(s)$ as
$e^{-2\sqrt{as}}$ at large $s$ implies the sharp decay of $p_t(z)$ as
$e^{-a/z}$ at small $z$.  This is the first rigorous justification for
using a GGD in Eq. \eref{eq:GGD} to approximate the probability
density $p_t(z)$.

\subsubsection{Case $t < \frac{1}{2}$}

In this case, we have to analyze the asymptotic behavior of
Eqs. \eref{eq:varphi_m} or \eref{eq:varphi_m2}.  For large $s$, one
can replace $\sinh(z)$ by $e^{z}/2$ (with exponentially small
corrections) to get in both cases
\begin{equation}
\label{eq:varphi_m2_ginf}
\varphi_t(s) \approx  \frac{\exp\left(-\sqrt{s} \frac{\eta(t)}{\sqrt{2(1-t)}}\right)}{\sqrt{I_t(g)}}  \qquad (s\gg 1).
\end{equation}
where the functions $\eta(t)$ and $I_t(g)$ are explicitly given in
\ref{sec:details}.  The asymptotic behavior of $\varphi_t(s)$ at large
$s$ is mainly determined by the exponential term, while $I_t(g)$ turns
out to be an algebraic correction.  In \ref{sec:Alarge}, we derive a
representation of $I_t(g)$ in the form $I_t(g) = c_0 + c_1 g$, where
$c_0$ and $c_1$ become independent of $g$ for large $g$ (or $s$).  As
a consequence, the asymptotic behavior of $\varphi_t(s)$ reads as
\begin{equation*}
\varphi_t(s) \simeq \frac{\exp\left(-\sqrt{s} \frac{\eta(t)}{\sqrt{2(1-t)}}\right)}
{\sqrt{c_0 + \frac{c_1\sqrt{2}}{\sqrt{1-t}} \sqrt{s}}} \qquad (s\gg 1).
\end{equation*}
Neglecting $c_0$ and using again Eq. \eref{eq:Laplace}, one deduces
the asymptotic behavior of $p_t(z)$:
\begin{equation}
\label{eq:pz_approx2}
p_t(z) \approx \frac{\sqrt{\eta(t)}}{\sqrt{2\pi  c_1(t)}}~ z^{-1} \exp\left(-\frac{\eta^2(t)}{8z(1-t)}\right)  \qquad (z\to 0),
\end{equation}
in which $c_1(t)$ is given by Eq. \eref{eq:c1t}.  Similar to the case
$t > \frac12$, the probability density sharply decays (as $e^{-a/z}$)
when $z\to 0$.

\subsection{Large $z$ asymptotic behavior}

Since the inverse Laplace transform can be written as the Bromwich
integral in the complex plane, the singularities of $\varphi_t(s)$ for
$s\in {\mathbb C}$ play an important role.  In particular, the
singularity with the largest (negative) real part, which is called the
abscissa of convergence of the Laplace transform, determines the
asymptotic behavior of the probability density $p_t(z)$ at large $z$
\cite{Nakagawa07}.  

For $t > \frac12$, one can easily check that all singularities of
$\varphi_t(s)$ are located on the negative real axis.  In fact, taking
$s = - \alpha^2/(1-t)$, one can rewrite Eq. (\ref{eq:varphi_m1}) as
\begin{equation}
\varphi_t(s) = \frac{2}{\cos \alpha} \left(1 - \beta^{-1} ~\alpha \tan \alpha\right)^{-1/2} ,
\end{equation}
where $\beta = (1-t)/(3t-1)$ is a positive parameter between $0$ and
$1$ (given that $\frac12 \leq t \leq 1$).  This function has two sets
of singularities at points $\tilde{\alpha}_k$ and $\alpha_k$ which are
solutions of the equations
\begin{equation*}
\cos \tilde{\alpha}_k = 0 , \qquad   \alpha_k \tan \alpha_k = \beta  \qquad (k = 1,2,3,...).
\end{equation*}
The smallest solution $\alpha_1$ of the second equation can be
determined perturbatively in powers of $\beta$:
\begin{equation*}
\alpha_1^2 = \beta - \frac13 \beta^2 + \frac{4}{45} \beta^3 + O(\beta^4).
\end{equation*}
The related singularity is thus located at
\begin{equation*}
s_1 = - \frac{\alpha_1^2}{1-t} = - \left(\frac{1}{3t-1} - \frac{1-t}{3(3t-1)^2} + \frac{4(1-t)^2}{45(3t-1)^3} + O((1-t)^3) \right) .
\end{equation*}
Note that $\alpha_1$ varies from $0$ at $t = 1$ to $0.8603...$ at $t =
0.5$ so that $|s_1|$ varies from $0.5$ at $t = 1$ to $1.4803$ at $t =
0.5$.  At the same time, the smallest solution $\tilde{\alpha}_1 =
\pi/2$ is independent of $t$ and is always larger than $\alpha_1$.

In close vicinity of $s_1$, one has
\begin{equation*}
\varphi_t(s) \simeq \sqrt{\frac{2(\beta^2 + \alpha_1^2)}{(\beta + \beta^2 + \alpha_1^2)(3t-1)}}~ (s - s_1)^{-1/2} 
\qquad (s \approx s_1) ,
\end{equation*}
from which the inverse Laplace transform yields an asymptotic
approximation
\begin{equation}
\label{eq:pz_short}
p_t(z) \simeq \sqrt{\frac{2(\beta^2 + \alpha_1^2)}{(\beta + \beta^2 + \alpha_1^2)(3t-1)}} ~ \frac{\exp(-z/b)}{\sqrt{\pi z}}   \qquad (z\gg 1) ,
\end{equation}
with $b = 1/|s_1| = (1-t)/\alpha_1^2$.

The analysis can be extended to the case $t < \frac12$.  For instance,
when $t = 1/m$, one can substitute $s = -(1-t)\alpha^2/(2t)$ into
Eq. \eref{eq:varphi_m} in order to locate the singularities of
$\varphi_t(s)$ either as solutions $\tilde{\alpha}_k$ of
$\sin(\tilde{\alpha}_k \sqrt{\lambda_k}) = 0$ (with $k = 1,...,m-1$),
or as solutions $\alpha_k$ of $\det(\mathbf{A}) = 0$.  For a general
situation $\frac{1}{m+1} < t < \frac{1}{m}$ for some $m = 2,3,4,...$,
the singularties of $\varphi_t(s)$ can be determined in a similar way.
In both cases, the large $z$ asymptotic behavior is determined by the
largest negative root $s_1$ (i.e., the root with the smallest absolute
value $|s_1|$) of the equation $\det(\mathbf{A}) = 0$.  Given the
square root type of singularity in Eq. \eref{eq:varphi_m2}, the
exponential asymptotic decay of $p_t(z)$ at large $z$ in the form
similar to Eq. \eref{eq:pz_short} is expected to be valid for any lag
time $t$.

\begin{figure}
\begin{center}
\includegraphics[width=70mm]{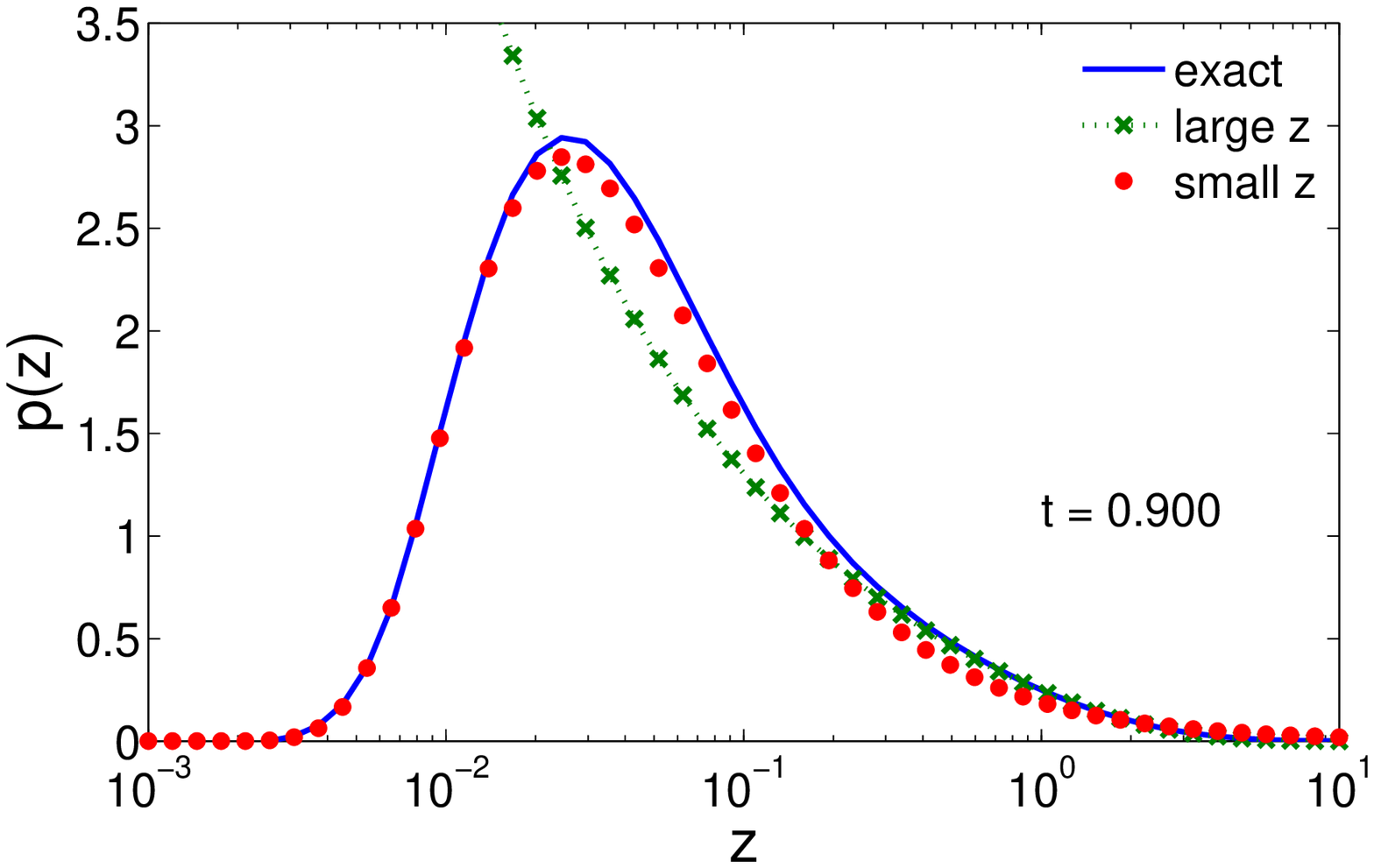}   \includegraphics[width=70mm]{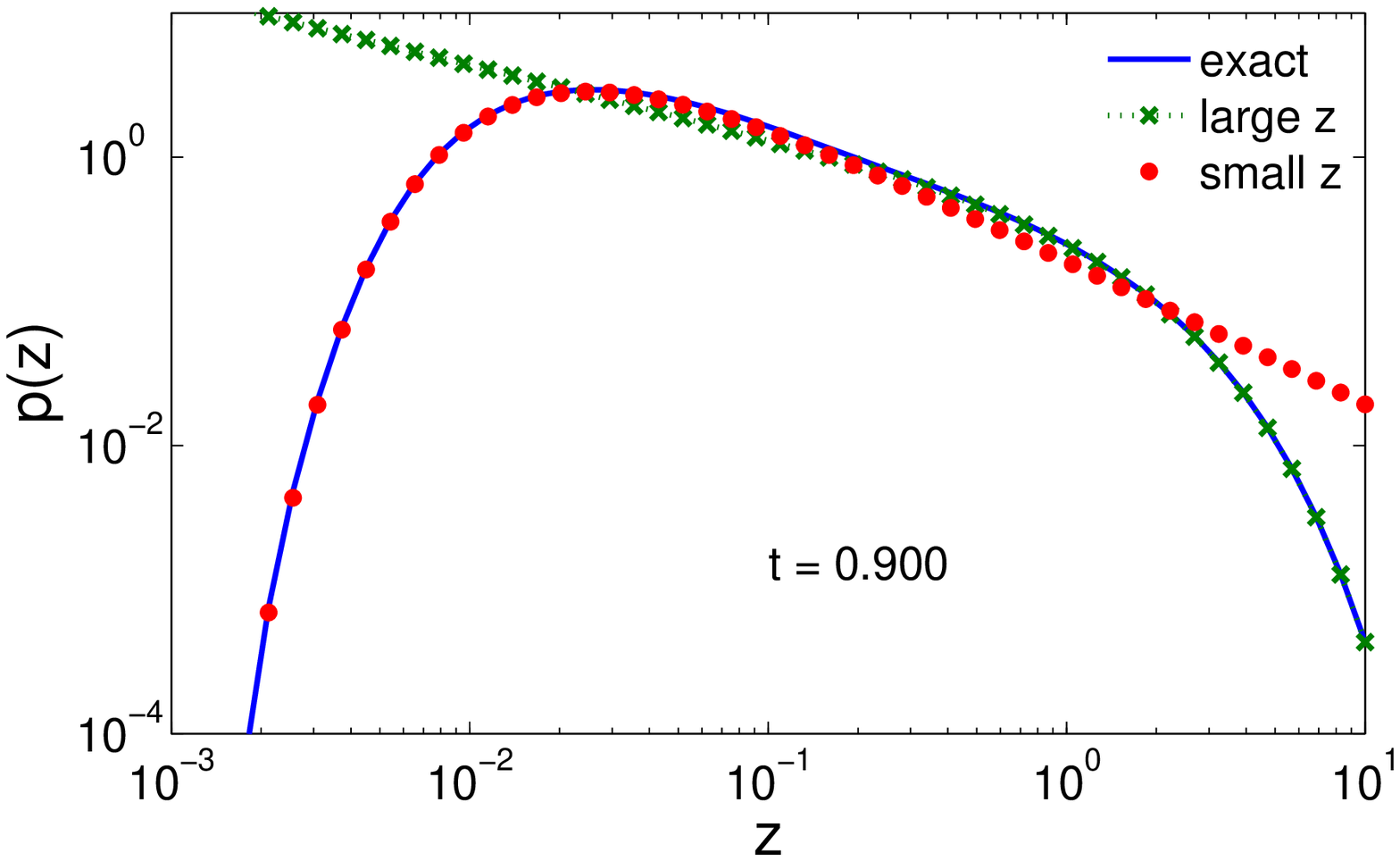}
\includegraphics[width=70mm]{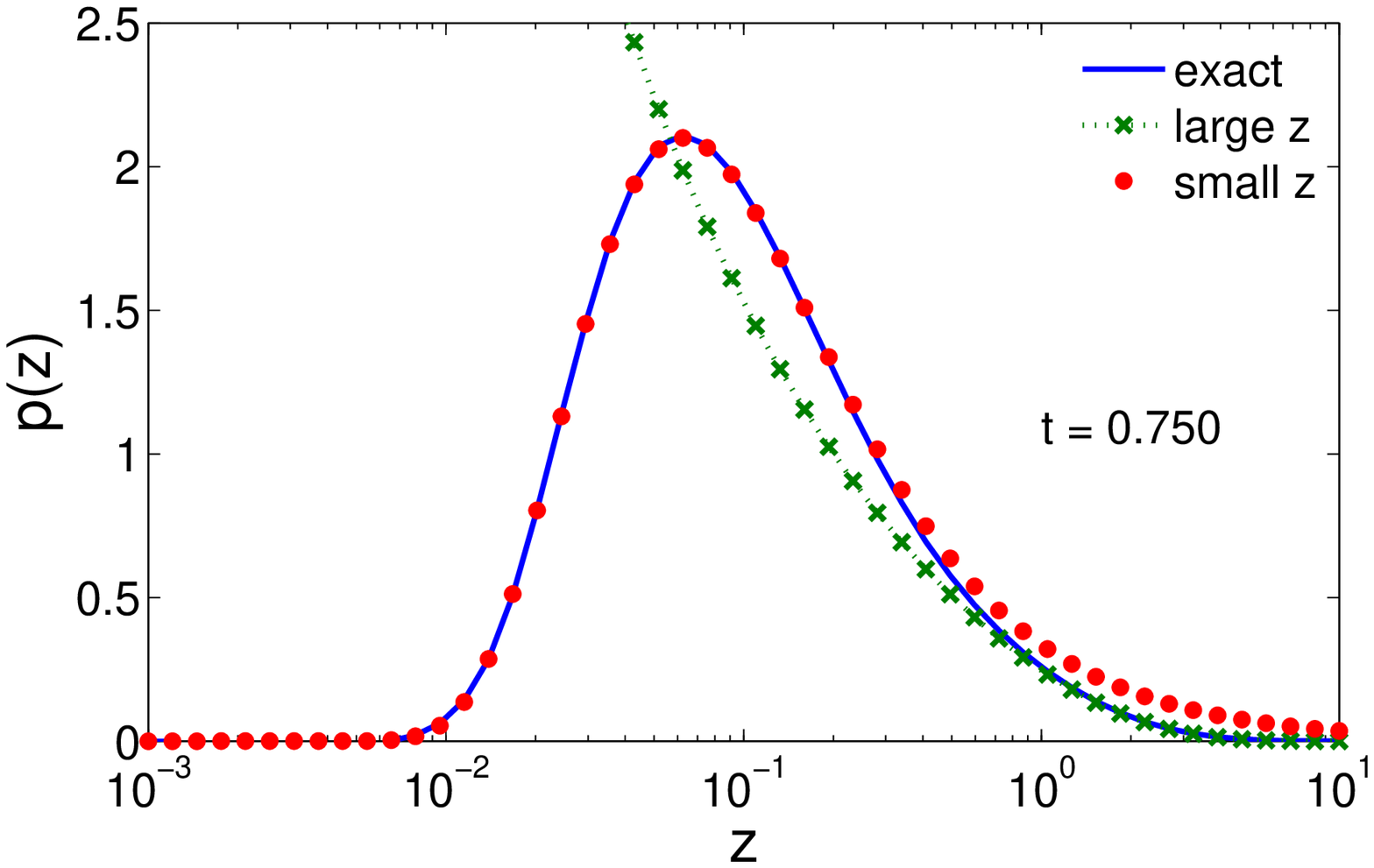}   \includegraphics[width=70mm]{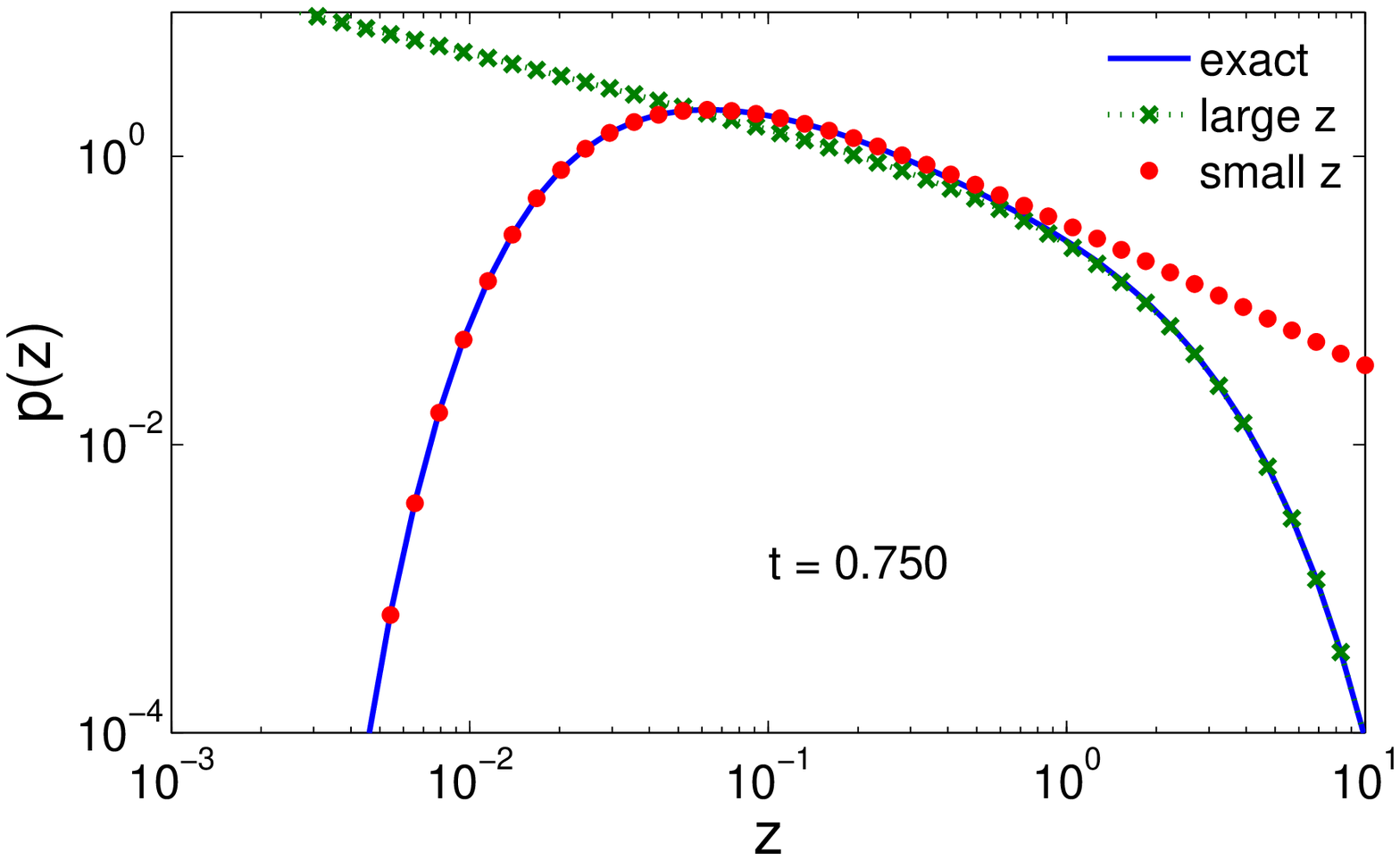}
\includegraphics[width=70mm]{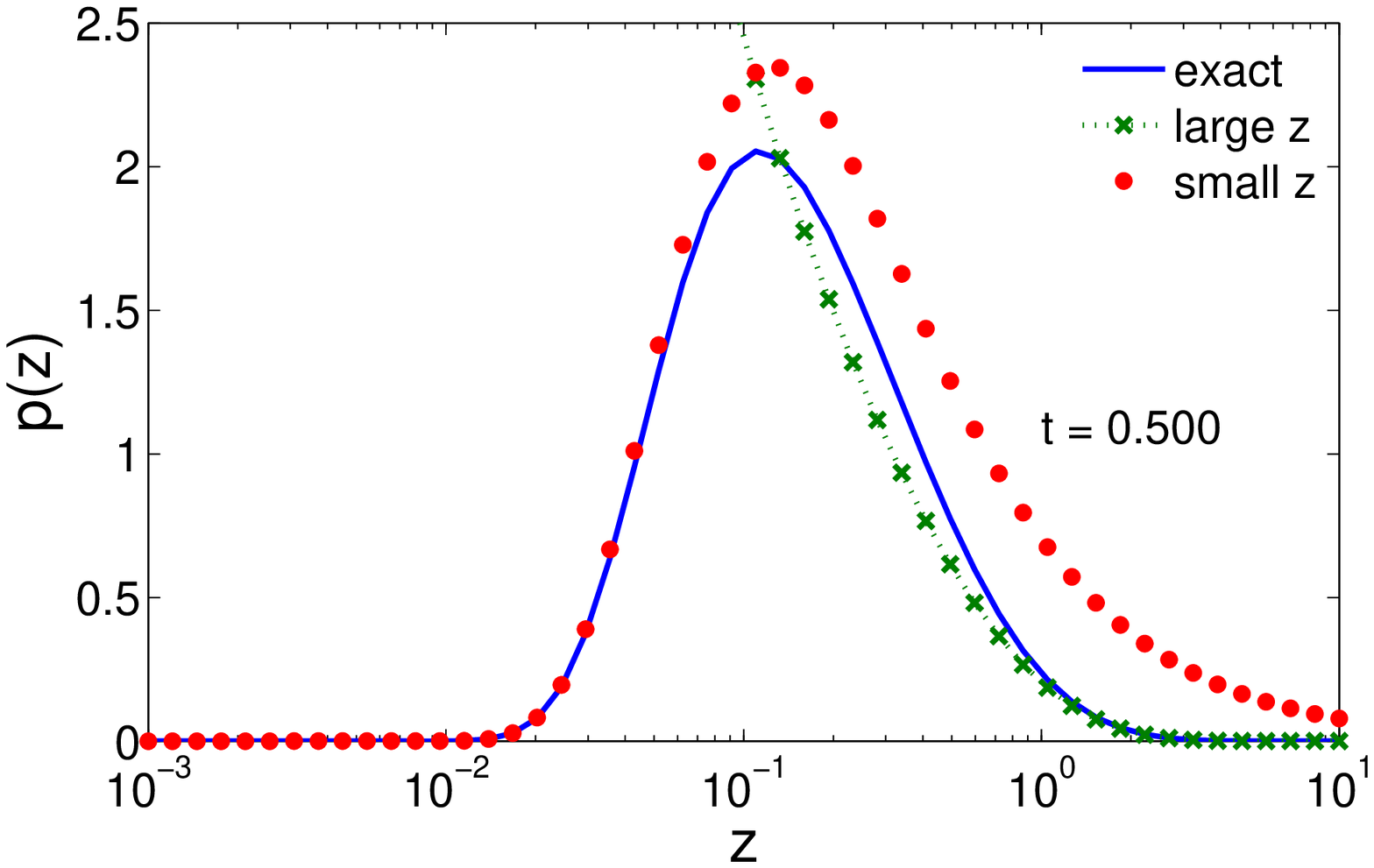}   \includegraphics[width=70mm]{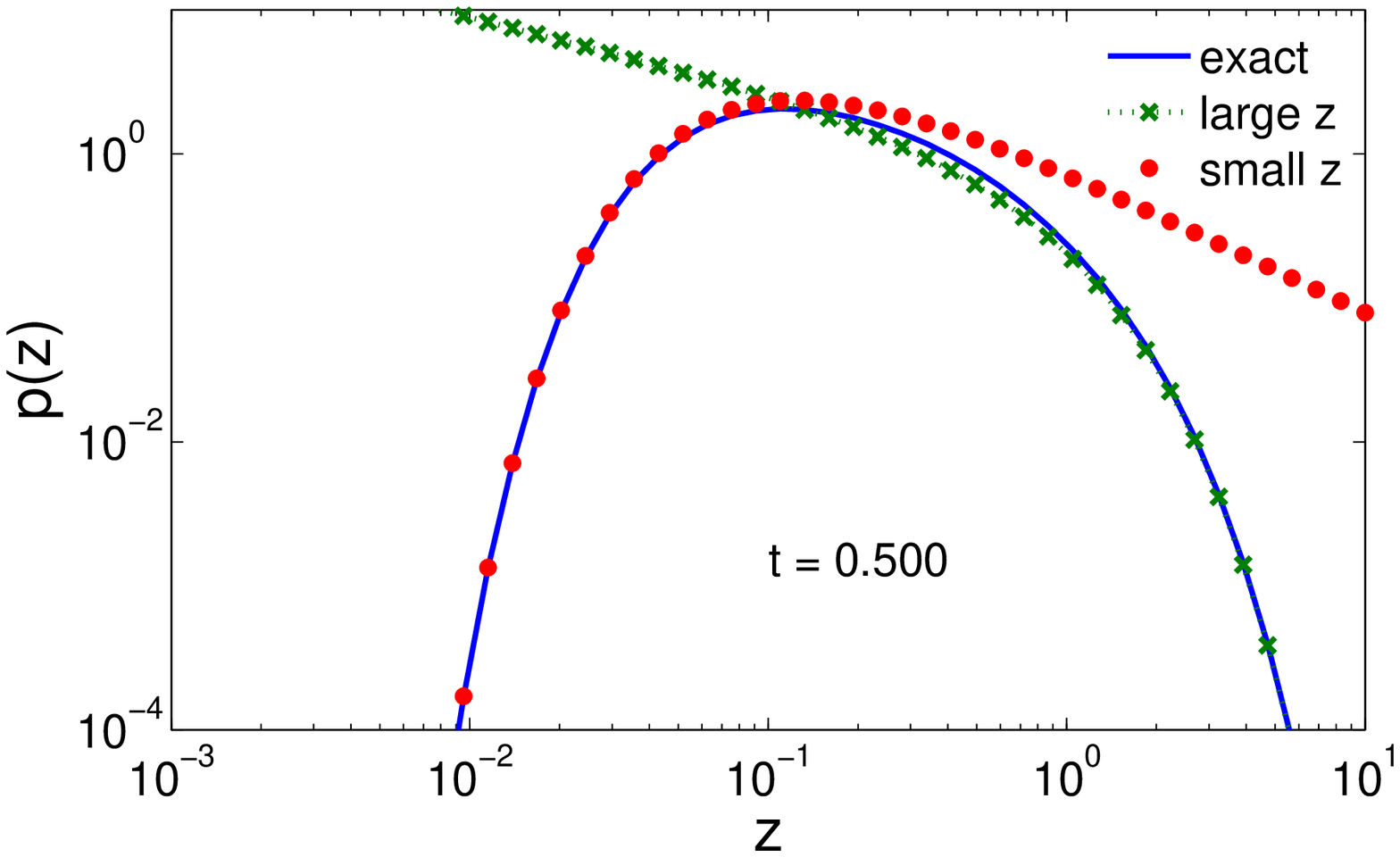}
\end{center}
\caption{
(Color online) The probability density $p_t(z)$ (solid line), its
small $z$ asymptotic relation \eref{eq:pz_approx1} (shown by circles),
and large $z$ asymptotic relation \eref{eq:pz_short} (shown by
crosses), at lag times $t = 0.9$, $t = 0.75$ and $t = 0.5$, in semilog
(left) and loglog (right) scales. }
\label{fig:pz_t}
\end{figure}

The accuracy of the asymptotic Eq. \eref{eq:pz_approx1} for small $z$
and Eq. \eref{eq:pz_short} for large $z$ is illustrated on
Fig. \ref{fig:pz_t}.  For all lag times,
Eqs. \eref{eq:pz_approx1},\eref{eq:pz_approx2} correctly describe the
asymptotic behavior of $p_t(z)$ at small $z$.  Moreover, for larger
$t$ (e.g., $t = 0.9$ and $t = 0.75$), Eq. \eref{eq:pz_approx1} turns
out to reproduce the whole shape of $p_t(z)$, except for large $z$.
The deviations at large $z$ are expected because the exponential decay
of $p_t(z)$ is not captured in the small $z$ asymptotic relation
\eref{eq:pz_approx1}.

%A_Andreanov_pz_fig.m;

\subsection{Limiting case $t = 1$}

In the limit $t\to 1$, one gets $\varphi_1(s) = 1/\sqrt{1+2s}$, from
which
\begin{equation}
\label{eq:p1}
p_1(z) = \frac{e^{-z/2}}{\sqrt{2\pi z}} .
\end{equation}
We retrieved therefore the limiting behavior of the discrete case for
the lag time $n=N-1$ \cite{Qian91,Grebenkov11b}.  Since the density
$p_1(z)$ describes the distribution of a squared Gaussian variable
$\chi_{1} = X^2(1)$, $\varphi_1(s)$ and then Eq. \eref{eq:p1} could be
directly obtained by integrating the Gaussian probability density
$\frac{1}{\sqrt{2\pi}} e^{-x^2/2}$ for $X(1)$ with $e^{-s x^2}$.

The function $p_1(z)$ diverges at $z = 0$ and monotonously decreases
on the positive semi-axis $z > 0$, in sharp contract to $p_t(z)$ (with
$t < 1$), which must have a maximum, as being equal to $0$ at $z = 0$
and $z\to \infty$.  In the limit $t\to 1$, the most probable value of
$\chi_t$ (i.e., the position of the maximum) approaches $0$, while the
mean value approaches $1$.  This also illustrates the fact that
$p_1(z)$ has the largest skewness among $p_t(z)$, as seen on
Fig. \ref{fig:kappa}.

\subsection{Limiting case $t\to 0$}

In the opposite limit of $t\to 0$ it is harder to get analytic
behavior from theoretical results, as the limit describes the
situation where the number of particles $m$ grows to infinity, which,
combined with complicated boundary conditions for normal modes, makes
a direct evaluation unfeasible.  We provide here a few qualitative
arguments, while a more rigorous analysis is reported in
\ref{sec:Asmall}.  Since the mean and variance of $\chi_t$ vanish as
$t\to 0$, it is convenient to rescale $\chi_t$ by $t$, in order to fix
the mean value to $1$.  For small $t$, the rescaled TAMSD can be
approximated as
\begin{equation*}
\tilde{\chi}_t \equiv \frac{\chi_t}{t} \simeq \frac{1}{m}\sum\limits_{k=1}^{m} \zeta_k^2 ,
\end{equation*}
where $m = [1/t]$ and $\zeta_k$ are independent Gaussian displacements
(standing for $X(t_0+t) - X(t_0)$, with $t_0 = k t$) with mean zero
and variance $1$.  The above sum is known to have a chi-squared (or
Gamma) distribution with $m$ degrees of freedom \cite{Hoel}, from
which
\begin{equation}
\label{eq:varphi_approx_gamma}
\tilde{\varphi}_t(s) \simeq (1 + 2s/m)^{-m/2} \simeq (1 + 2st)^{-1/(2t)} .
\end{equation}
In the limit $t\to 0$, we get $\tilde{\varphi}_0(s) = e^{-s}$ and
$\tilde{p}_0(z) = \delta(z-1)$ as expected since $\chi_t/t$
concentrates around $1$ in this limit.  Although the limiting results
for $\tilde{\varphi}_0(s)$ and $\tilde{p}_0(z)$ are correct, an
approximation \eref{eq:varphi_approx_gamma} of $\tilde{\varphi}_t(s)$
for small $t$ by a Gamma distribution is not accurate.

\subsection{Dimensional units}

All the above results were obtained for $D = 1/2$ and $T = 1$, with
$t$ being a dimensionless lag time between $0$ and $1$.  Dimensional
units can be easily incorporated by replacing $s$ by $2DTs$, $z$ by
$z/(2DT)$ and $t$ by $t/T$ in the formulas for $\varphi_t(s)$ and
$p_t(z)$.  For instance, Eq. \eref{eq:varphi_m1} reads for a given
diffusion coefficient $D$ and sample duration $T$ as
\begin{equation*}
\fl
\varphi_t(s) = \frac{2 e^{-\sqrt{2Ds(T-t)}}}{1 + e^{-2\sqrt{2Ds(T-t)}}} ~ \left(1 + \sqrt{2Ds(T-t)} ~\frac{3t-T}{T-t} ~
\frac{1- e^{-2\sqrt{2Ds(T-t)}}}{1+ e^{-2\sqrt{2Ds(T-t)}}} \right)^{-1/2} . 
\end{equation*}
In the formulas for the cumulant moments, one has to replace $t$ by
$t/T$ and multiply the right-hand side by $(2DT)^n$, with $n$ being
the order of the moment.

\subsection{Discrete versus continuous cases}
\label{sec:discrete}

As we discussed in Introduction, the TAMSD is commonly employed for
inferring the diffusion coefficient from an individual random
trajectory of a diffusing tracer.  For this purpose, one records
positions of a tracer at successive times, with a fixed time step
$\tau$.  An individual trajectory of the tracer is therefore
discretized, and the TAMSD becomes
\begin{equation}
\label{eq:chi_nN}
\chi_{n,N} = \frac{1}{N-n} \sum\limits_{k=1}^{N-n} (x_{k+n} - x_k)^2 ,
\end{equation}
where $x_k = X(k\tau)$ is the position of the tracer at time $k\tau$,
$n = t/\tau$ is the discrete lag time, and $N = T/\tau$ is the number
of points in the sample.  In the limit $\tau \to 0$, the discrete
TAMSD is expected to converge to the continuous one defined by
Eq. \eref{eq:MSD0}.  The probability distribution of the TAMSD for a
discrete Gaussian process was investigated in \cite{Grebenkov11b}.
For a given covariance matrix $\mathbf{C}$ (and zero mean), one can
use the classical matrix representation
\begin{equation}
\label{eq:varphi_discrete}
\varphi_{n,N}(s) \equiv \langle \exp(-s \chi_{n,N})\rangle = \frac{1}{\sqrt{\det(\mathbf{I} + s \mathbf{M}\mathbf{C})}} ,
\end{equation}
where the matrix $\mathbf{M}$ represents the discrete quadratic form
in Eq. \eref{eq:chi_nN}.  When the sample length $N$ is smaller than
few thousands, the Schur decomposition of the matrix
$\mathbf{M}\mathbf{C}$ allows one to rapidly compute the Laplace
transform $\varphi_{n,N}(s)$ or the characteristic function
$\phi_{n,N}(k) \equiv \varphi_{n,N}(-ik)$, from which the probability
density $p(z)$ can be reconstructed through the inverse Fourier
transform \cite{Grebenkov11b}.

\begin{figure}
\begin{center}
\includegraphics[width=75mm]{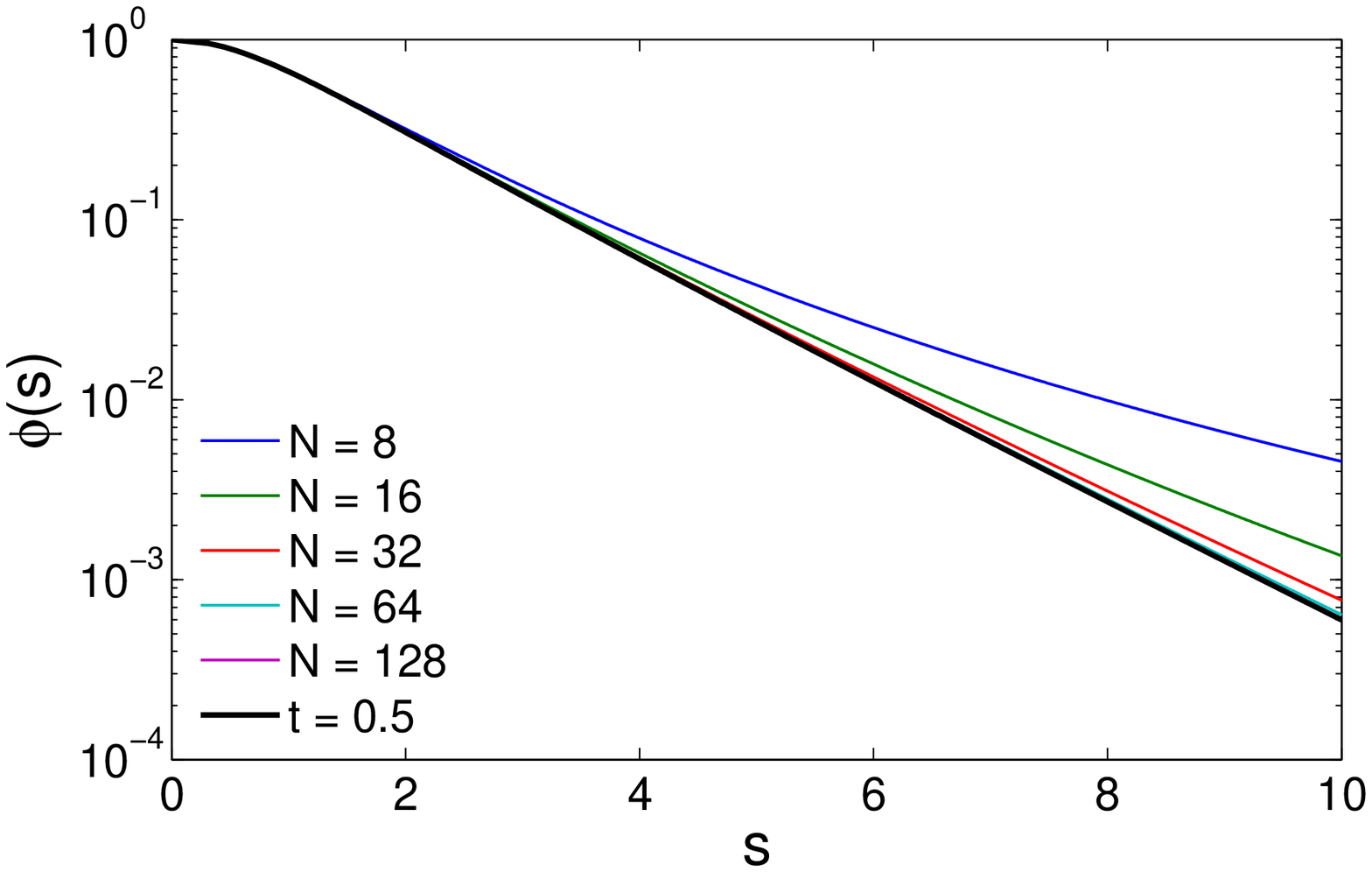}
\includegraphics[width=75mm]{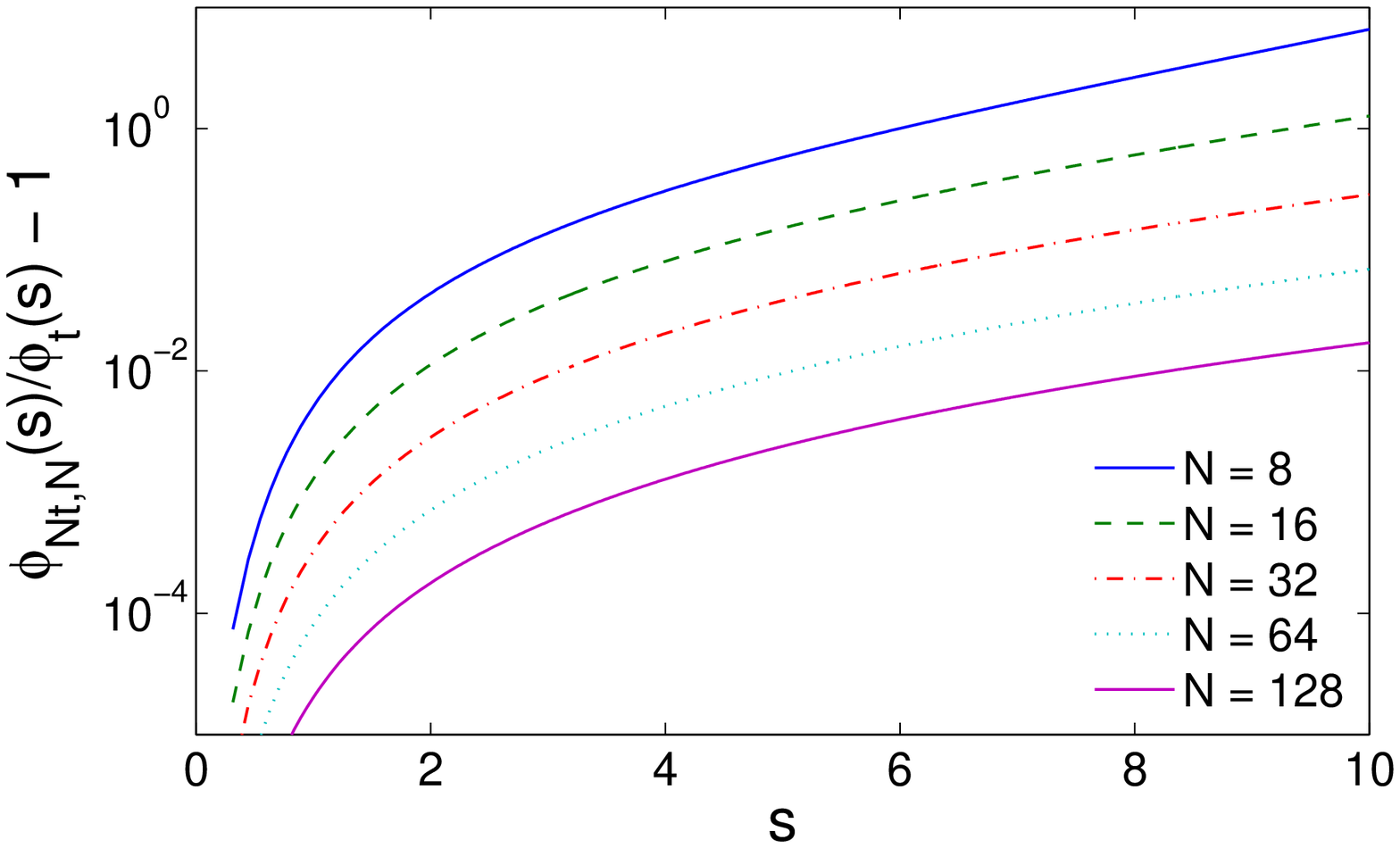}
\end{center}
\caption{
(Color online) {\bf (Left)} Comparison between $\varphi_t(s)$ for
Brownian motion (with $t = 0.5$) from Eq. \eref{eq:varphi_m1} (thick
black line) and $\varphi_{Nt,N}(s)$ for discrete random walks with
different sample length $N = 8, 16, 32, 64, 128$, with $n = t N$.  The
curves for $N = 64, 128$ are barely distinguishable from the
theoretical one.  {\bf (Right)} The relative error between
$\varphi_t(s)$ and $\varphi_{Nt,N}(s)$.  As expected, the relative
error increases with $s$ and it is larger for smaller $N$. }
\label{fig:varphi_05}
\end{figure}

Although Eq. \eref{eq:varphi_discrete} provides an exact solution for
$\varphi_{n,N}(s)$ and an efficient numerical way for computing $p(z)$
for moderate $N$, the dependence on the parameters (e.g., the lag
time) remains elusive.  In contrast, Eq. \eref{eq:varphi_m1} for
Brownian motion is explicit and allows one to derive the asymptotic
behavior and the dependence on the lag time.  Since $\chi_{n,N}$
approaches $\chi_{t,T}$ in the limit $\tau \to 0$,
Eq. \eref{eq:varphi_m1} is expected to be an accurate approximation to
$\varphi_{n,N}(s)$ for large enough $N$.  This point is illustrated on
Fig. \ref{fig:varphi_05}, in which $\varphi_t(s)$ for $t = 0.5$ is
compared to $\varphi_{N/2,N}(s)$, for different sample lengths $N$.
The curves for $N = 8, 16, 32$ exhibit deviations at large $s$, while
the curves for larger $N = 64, 128$ are barely distinguishable from
$\varphi_t(s)$ at the present range of $s$ values.  As a consequence,
one can use the explicit formula \eref{eq:varphi_m1} for discrete
samples with $N \gtrsim 50$.  At the same time, the functions
$\varphi_{n,N}(s)$ and $\varphi_t(s)$ will always be different for
large enough $s$.  In fact, for a fixed $N$, $\varphi_{n,N}(s)$
exhibits a power law asymptotic decay (as $\det(\mathbf{I} + s
\mathbf{M}\mathbf{C})$ is a polynomial in $s$), while $\varphi_t(s)$
decays exponentially fast.  However, the distinction between these
behaviors may appear for so large $s$, for which both functions become
already extremely small so that a distinction would be irrelevant for
practical applications.  In summary, Eq. \eref{eq:varphi_m1} and its
extensions for $t < 1/2$ can in practice be used for studying the
TAMSD of discretely sampled positions of a freely diffusing particle.

%A_Andreanov_fit_varphi.m;

\subsection{Relation to generalized Gamma distribution}

\begin{figure}
\begin{center}
\includegraphics[width=70mm]{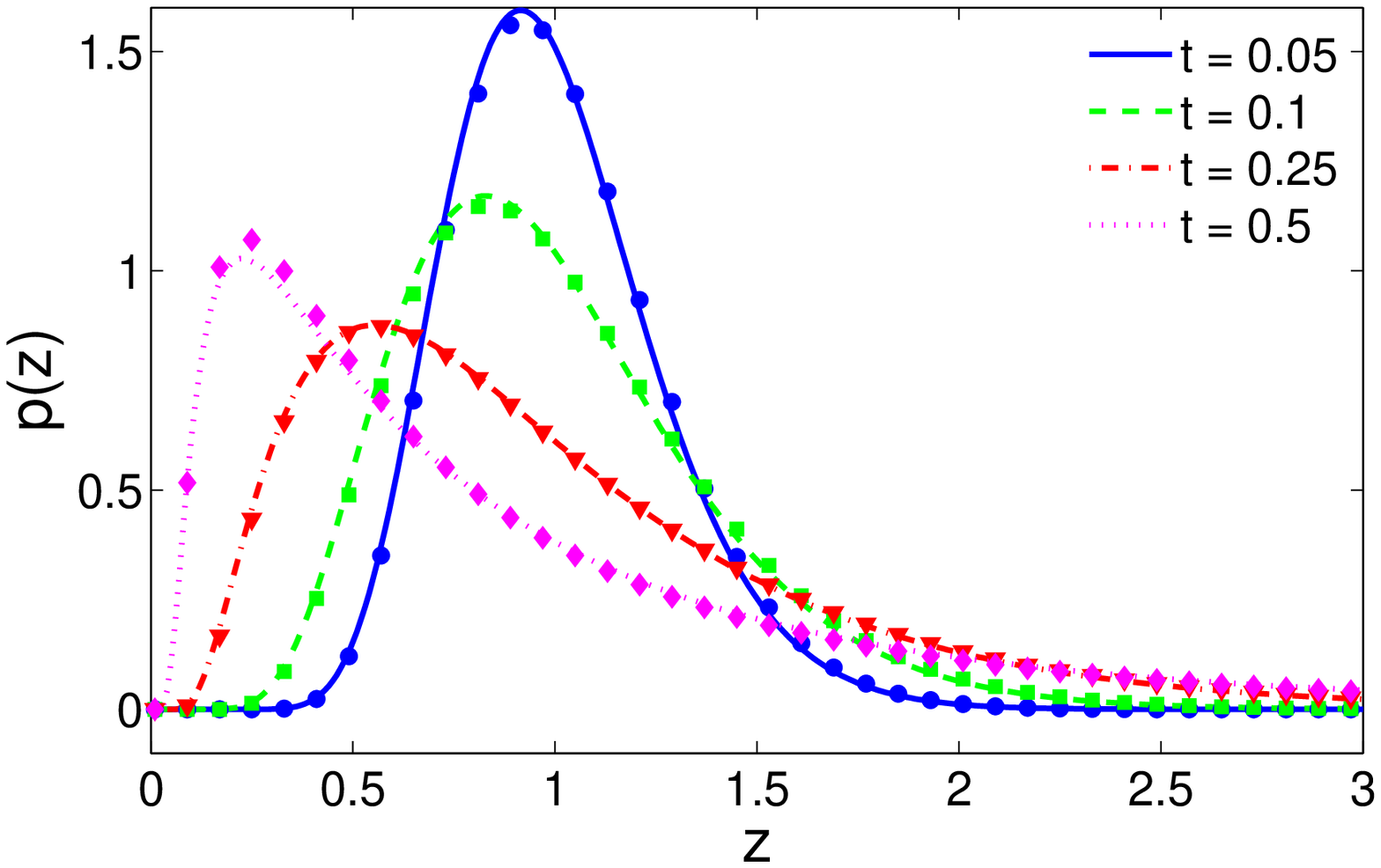}   \includegraphics[width=70mm]{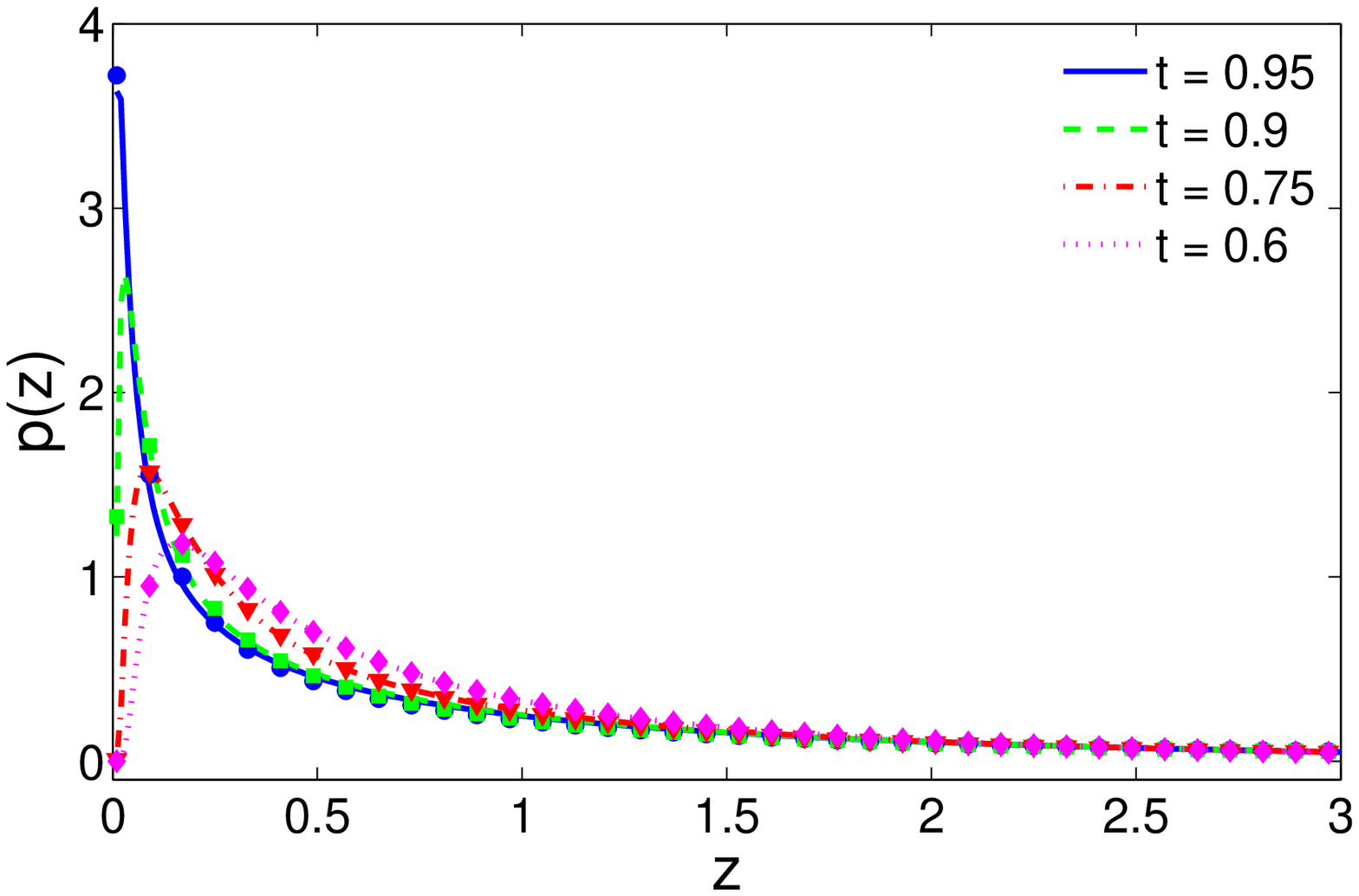}
\includegraphics[width=70mm]{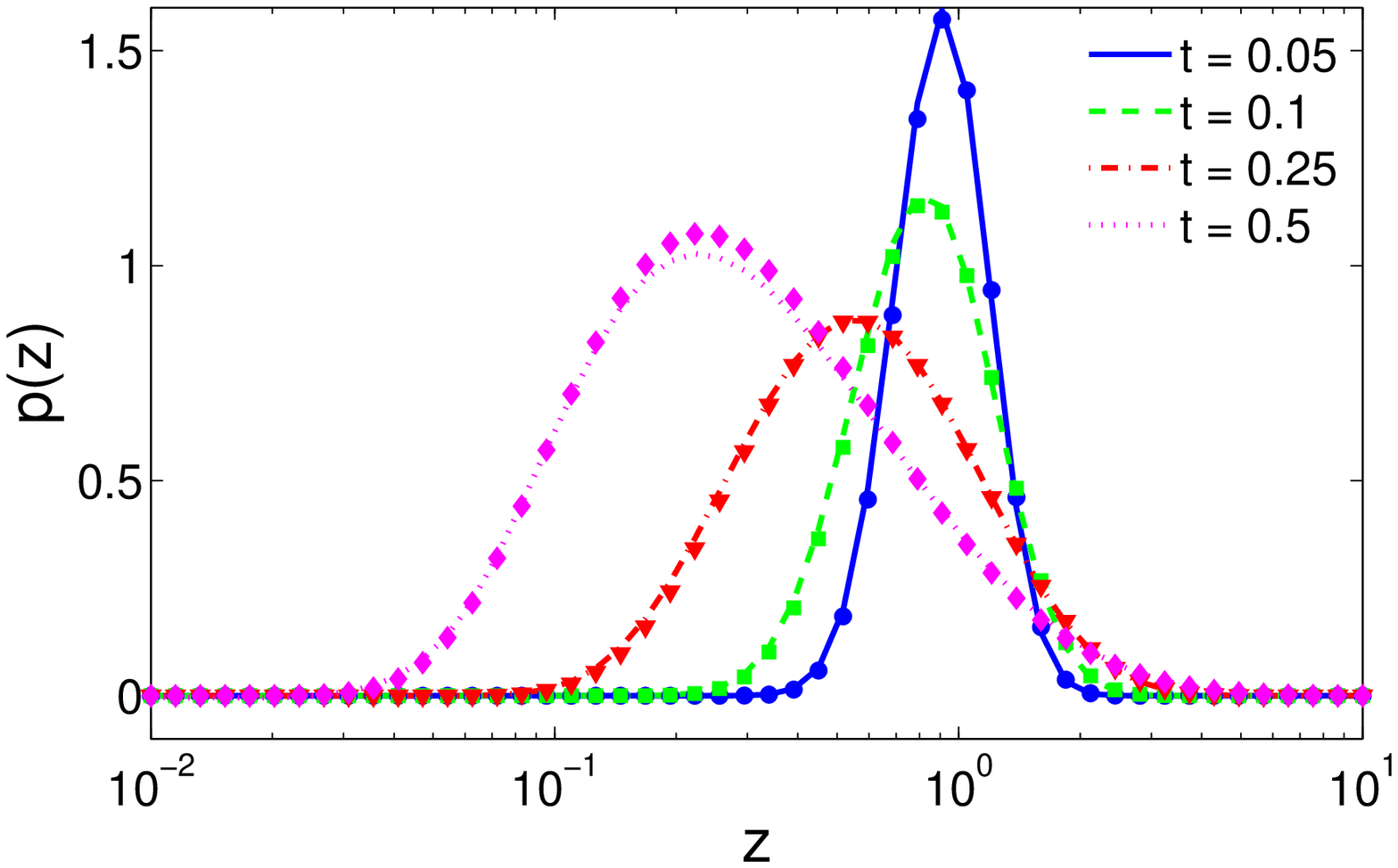}   \includegraphics[width=70mm]{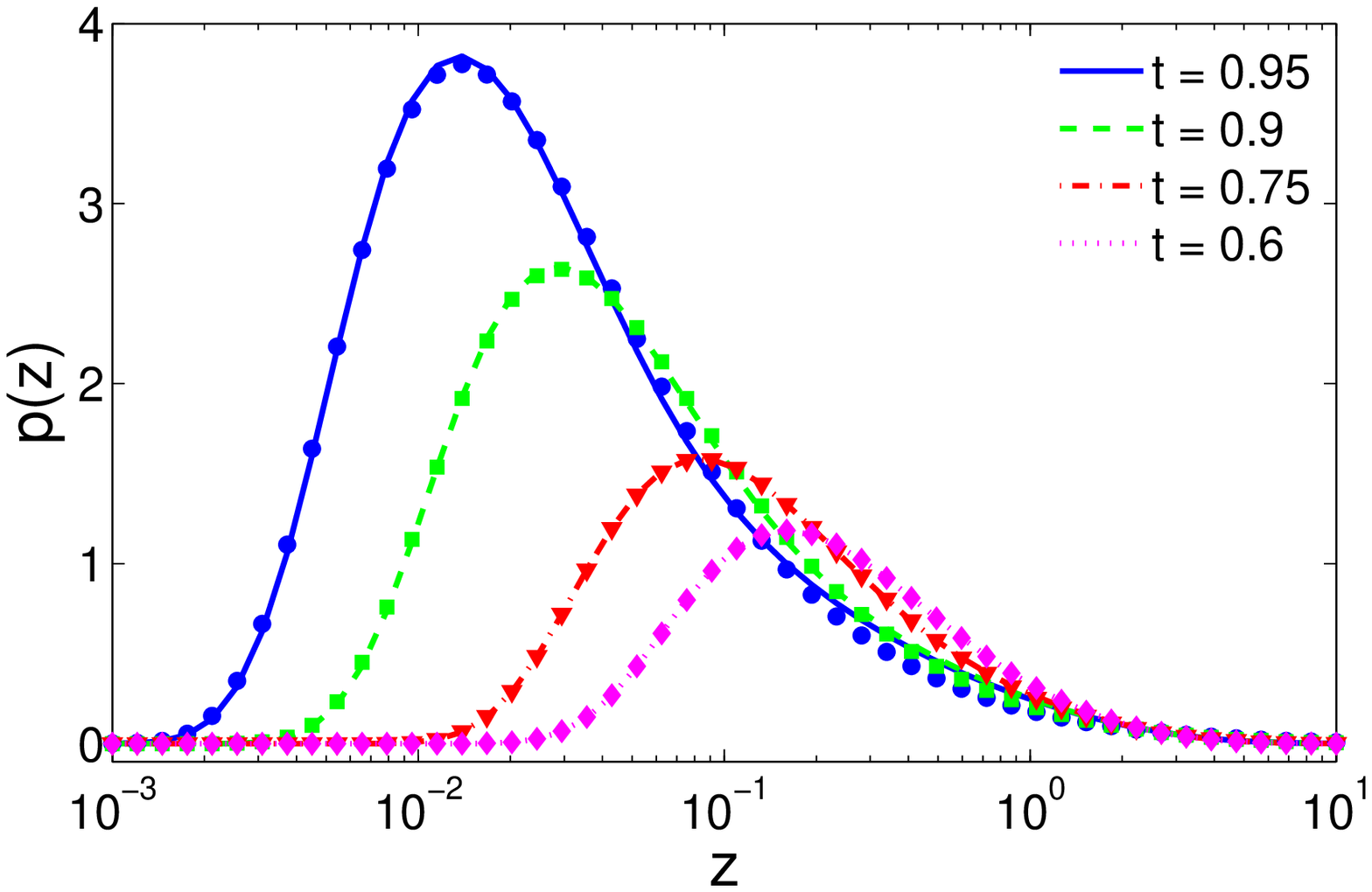} 
\includegraphics[width=70mm]{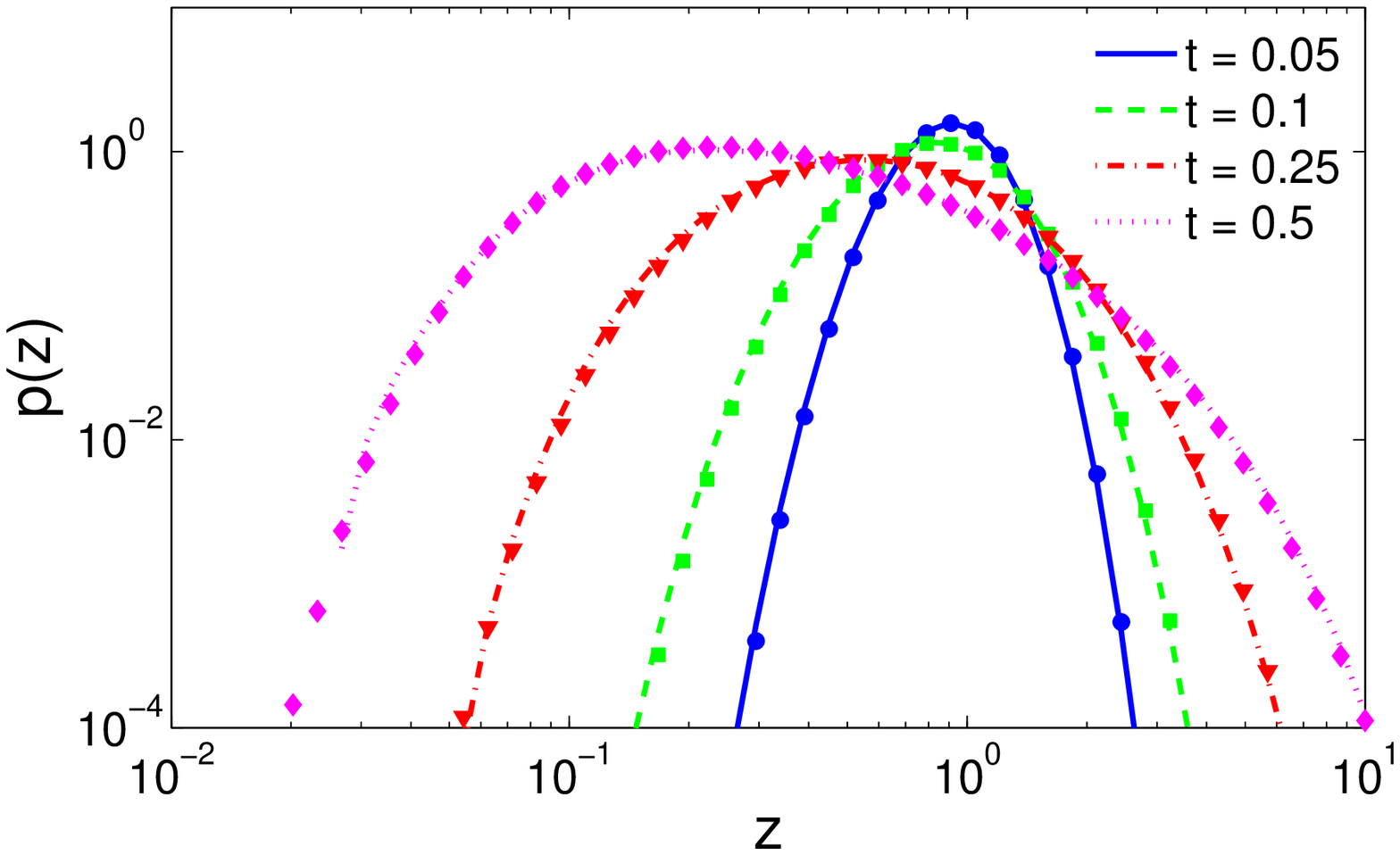}   \includegraphics[width=70mm]{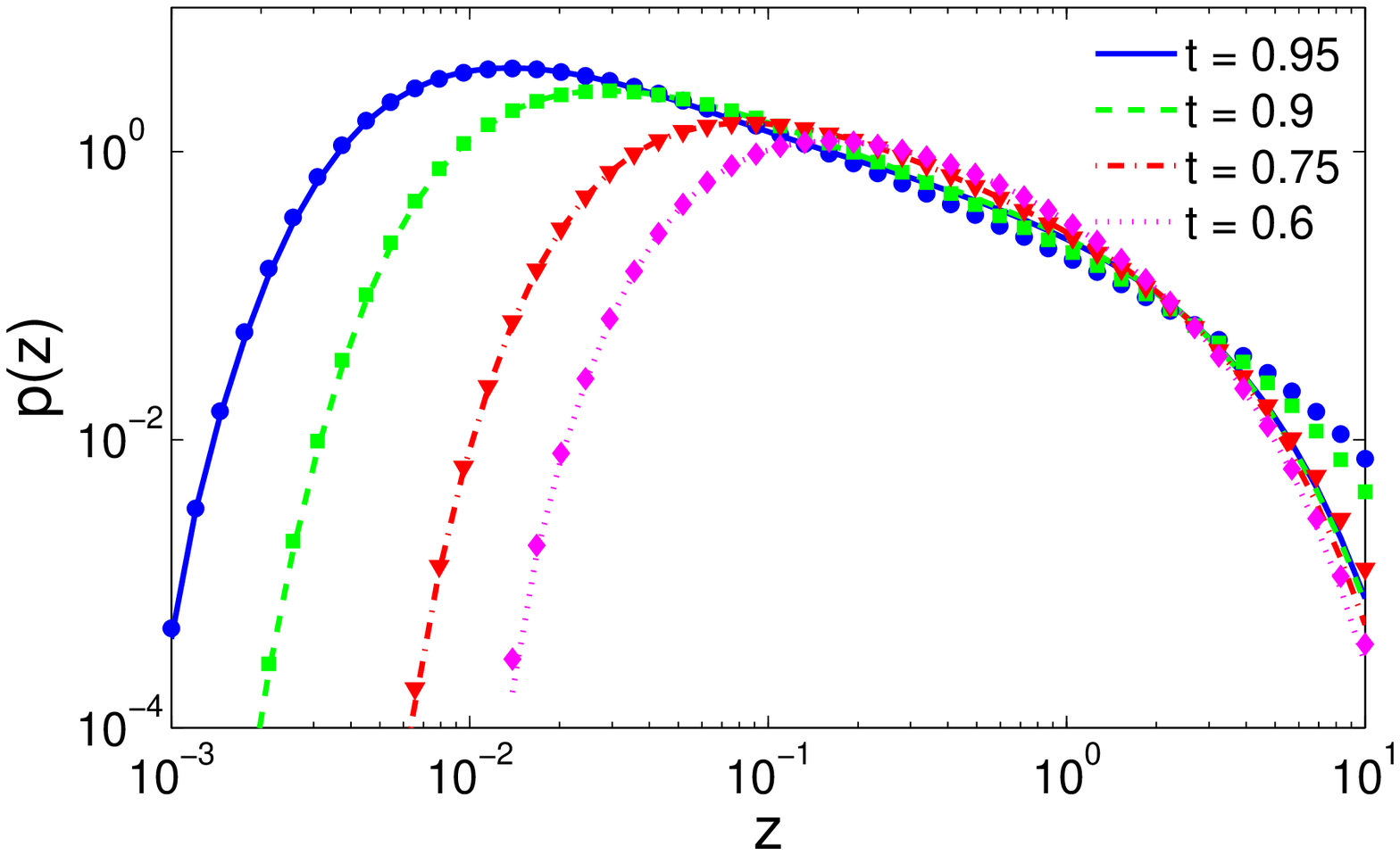}
\end{center}
\caption{
(Color online) The probability density $p_t(z)$ of the TAMSD rescaled
by $t$ for several lag times $t$, in normal, semilog and loglog
scales.  Lines show exact solutions for a discrete random walk, while
circles represent approximations by GGDs.  For $t \leq \frac12$ (left
panel), the parameters $a$, $b$ and $\nu$ are obtained from the
asymptotic formulas \eref{eq:nu_asympt}, \eref{eq:a_asympt},
\eref{eq:b_asympt}.  Even for the limiting value $t = \frac12$, the
approximation remains accurate.  For $t > \frac12$ (right panel), the
parameters $a$, $b$ and $\nu$ are extracted from the best fit of
$p_t(z)$ by a GGD.}
\label{fig:abnu_pz}
% A_Andreanov_article_fig.m
% A_Andreanov_article_fig1a.m
\end{figure}

As shown numerically in \cite{Grebenkov11b}, the probability density
$p_t(z)$ for a discrete random walk can be accurately approximated by
a generalized Gamma distribution (GGD) from Eq. \eref{eq:GGD}.  Its
Laplace transform reads as
\begin{equation}
\varphi(s) = (1 + sb)^{-\nu/2} \frac{K_\nu(2\sqrt{a(1+sb)/b})}{K_\nu(2\sqrt{a/b})} .
\end{equation}
On one hand, its comparison to Eqs. \eref{eq:varphi_m1},
\eref{eq:varphi_m}, \eref{eq:varphi_m2} immediately shows that a GGD is
not an exact distribution for $\chi_t$ (except for the trivial case $t
= 1$).  On the other hand, we have seen earlier that the GGD correctly
reproduces the asymptotic behavior at small and large $z$, and it
turns out to be a remarkably accurate approximation of $p_t(z)$ (see
Fig. \ref{fig:abnu_pz}).  Moreover, a simple form of Eq. \eref{eq:GGD}
is particularly attractive for practical purposes (e.g., inferring
statistics of diffusion coefficients by maximum likelihood methods,
estimating the confidence intervals, etc.).  For practical uses of
this approximation, it is important to relate the parameters $a$, $b$
and $\nu$ to the lag time $t$.

\begin{figure}
\begin{center}
\includegraphics[width=70mm]{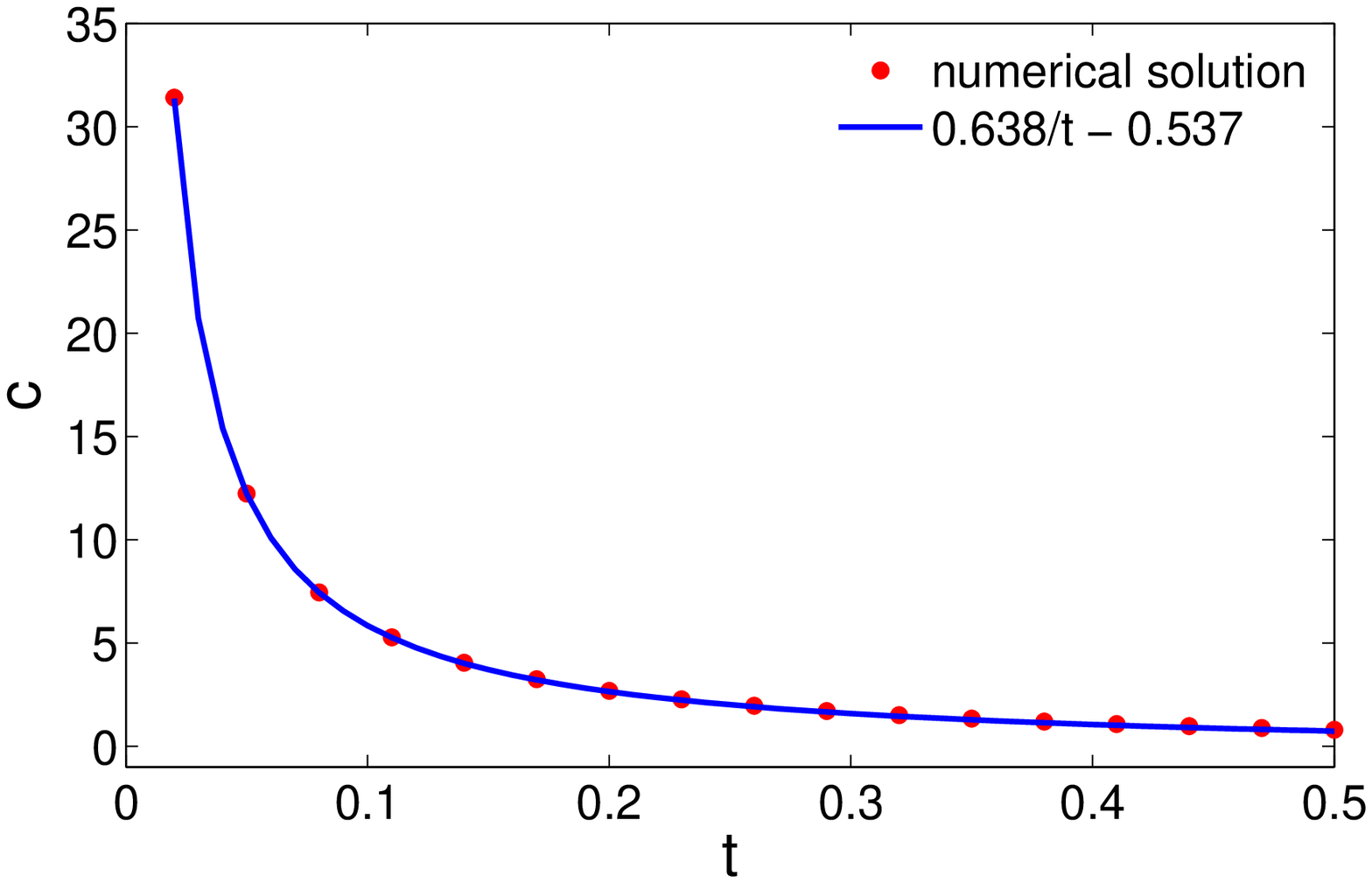}
\includegraphics[width=70mm]{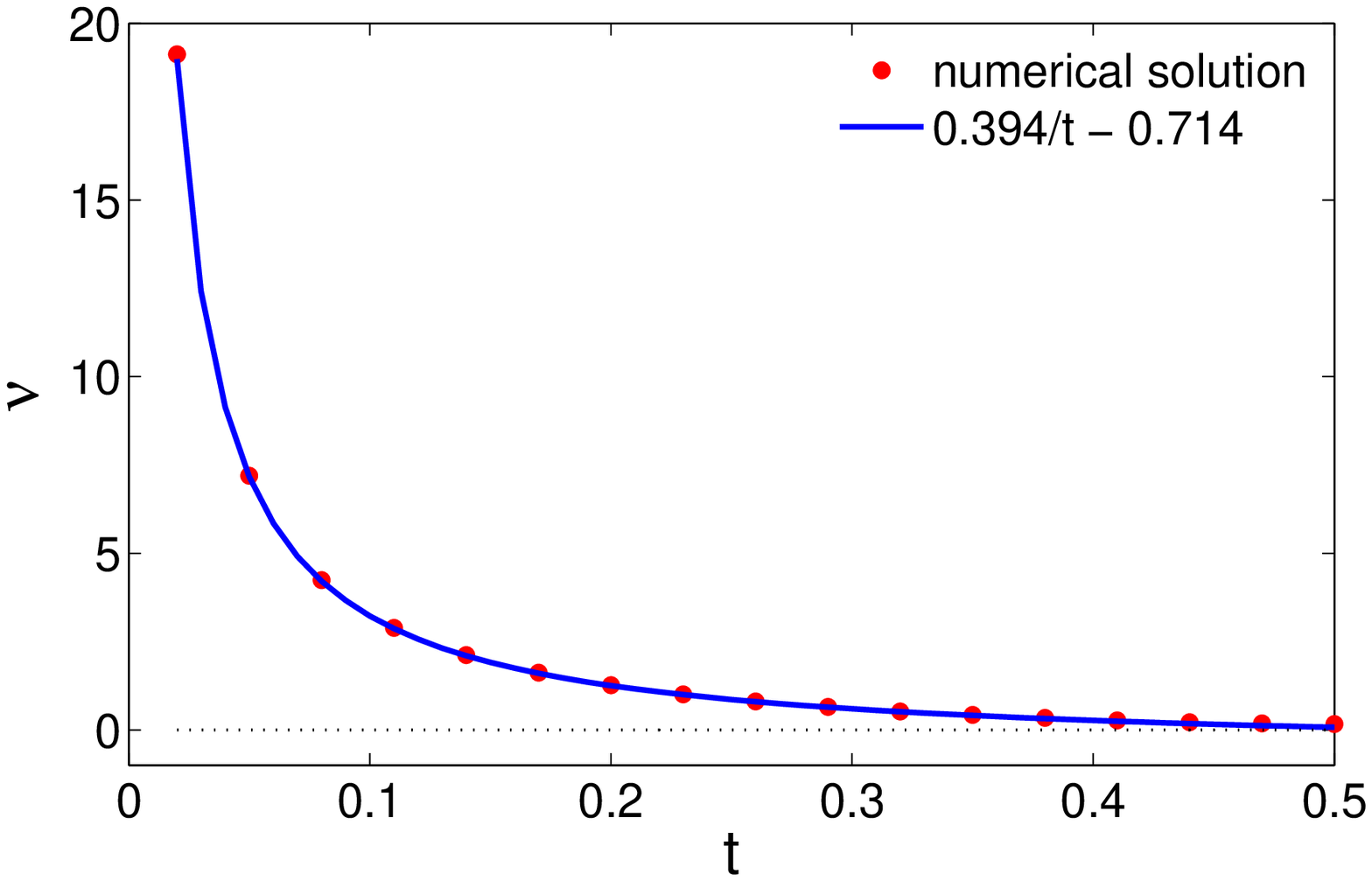}
\includegraphics[width=70mm]{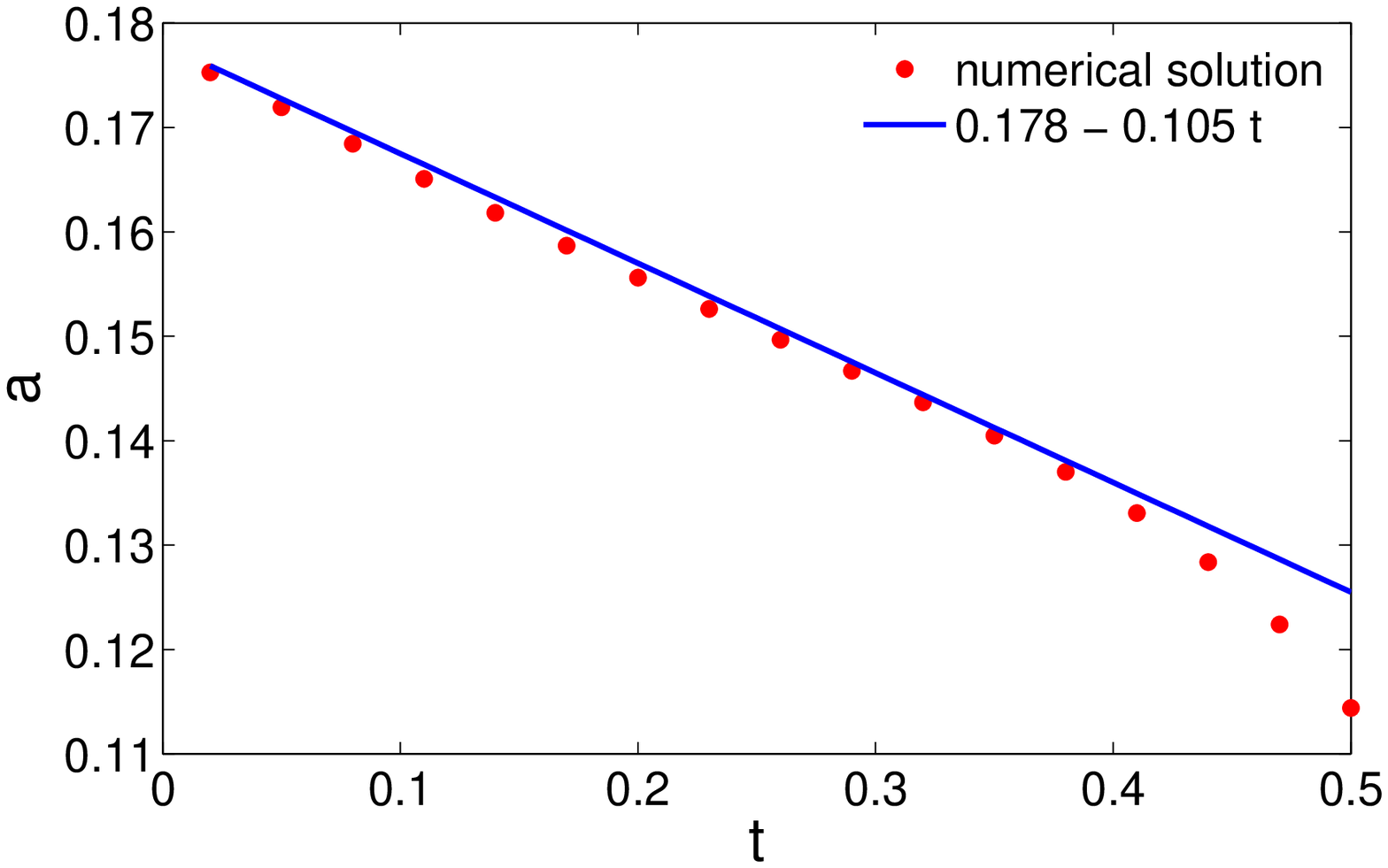}
\includegraphics[width=70mm]{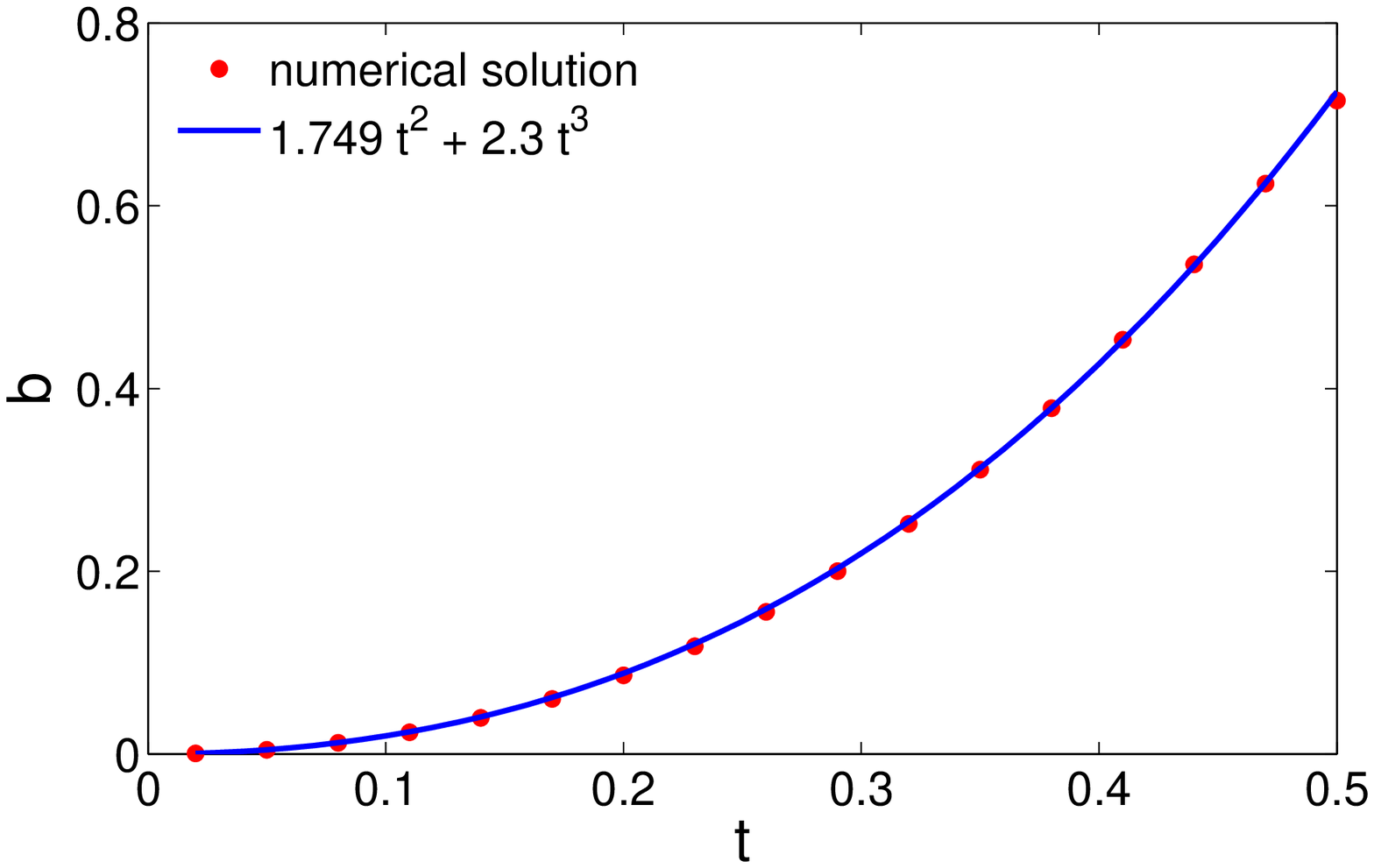}
\end{center}
\caption{
(Color online) The parameters $a$, $b$ and $\nu$ (and $c =
2\sqrt{a/b}$) of the GGD as functions of the lag time $t$.  Full
circles are obtained by a numerical resolution of
Eqs. \eref{eq:moments_system}, \eref{eq:ab_nuc} and solid lines
represent the short-time asymptotic formulas \eref{eq:c_asympt},
\eref{eq:nu_asympt}, \eref{eq:a_asympt}, \eref{eq:b_asympt} for $t <
\frac12$. }
\label{fig:abnu}
% [a,b,nu,tt] = A_Gauss_abnu_fig0(a,b,nu,tt);
\end{figure}

In \ref{sec:moments}, we derive the asymptotic relations for the
parameters $a$, $b$ and $\nu$ in the limit $t\to 0$ by matching the
first three moments:
\begin{eqnarray}
\label{eq:c_asympt}
c &=& x\nu \simeq \frac{3\sqrt{1159}}{160} ~t^{-1} + c_0 \approx 0.638 ~t^{-1} - 0.537 , \\
\label{eq:nu_asympt}
\nu &=& \frac{1}{\ve} \simeq \frac{63}{160} ~t^{-1} + \nu_0 \approx 0.394 ~t^{-1} - 0.714  ,  \\
\label{eq:a_asympt}
a &=& \frac{x}{1+\sqrt{1+x^2}} ~ \frac{tc}{2} \simeq \frac{57}{320} + a_0 t \approx 0.178 - 0.105 t , \\
\label{eq:b_asympt}
b &=& \frac{4a}{c^2} \simeq \frac{320}{183} ~t^2 + b_3 t^3 \approx 1.749 ~t^2 + 2.3 t^3 ,
\end{eqnarray}
where $c = 2\sqrt{a/b}$ and $x = \frac{\sqrt{1159}}{21} \approx
1.6211$.  Here, the leading terms were derived analytically, while the
values of the next-order corrections $\nu_0$, $c_0$, $a_0$ and $b_0$
were obtained by fitting the numerical curves of these quantities.
Although they could in principle be found from the higher-order terms,
such analysis goes beyond the scope of the paper.  The asymptotic
relations \eref{eq:c_asympt}, \eref{eq:nu_asympt}, \eref{eq:a_asympt},
\eref{eq:b_asympt} turn out to be accurate for the whole range of $t <
\frac12$, as shown on Fig. \ref{fig:abnu}.

We check the accuracy of the GGD with the parameters $a$, $b$ and
$\nu$ from the short-time asymptotic formulas \eref{eq:nu_asympt},
\eref{eq:a_asympt}, \eref{eq:b_asympt} by comparing this approximation
to the probability density obtained numerically for a discrete random
walk (Sect. \ref{sec:discrete}).  The latter is referred here as
``exact'' solution, although it was obtained by a numerical
computation of the inverse Fourier transform of the exact
characteristic function from Eq. \eref{eq:varphi_discrete}, as
explained in detail in \cite{Grebenkov11b}.  In order to plot several
curves onto the same figure, the TAMSD was rescaled by $t$ to set the
mean value to $1$ for all lag times $t$.  Figure
\ref{fig:abnu_pz}(left) shows an excellent agreement between exact
solutions and approximations by the GGD, for several lag times $t$
from $0$ to $\frac12$.

For larger lag times, the probability density $p_t(z)$ becomes more
and more skewed, with the maximum of $p_t(z)$ getting much smaller
than the mean value.  As a consequence, positive-order moments are
determined by the values of $p_t(z)$ far from the maximum.  Given the
objective to approximate the function $p_t(z)$ around the maximum, one
cannot therefore rely on matching positive-order moments, as done in
\ref{sec:moments}.  One needs therefore to identify other
characteristics of the true and approximate distributions to be
matched.  One can see that even with the exact solution at hand, the
analysis is intricate.

At the same time, GGD remains an accurate approximation of the
probability density $p_t(z)$ even for $t > \frac12$, as illustrated on
Fig. \ref{fig:abnu_pz} (right panel).  On the right panel, the
parameters $a$, $b$ and $\nu$ were obtained by fitting the exact
density $p_t(z)$ (obtained numerically) to a GGD.  As a perspective,
one may attempt to derive analytical asymptotic relations for these
parameters for $t > \frac12$.  Note also that, for $t > \frac12$, the
small $z$ asymptotic formula \eref{eq:pz_approx1} is also an accurate
approximation of $p_t(z)$, as discussed in Sect. \ref{sec:pz_small}.

\subsection{Two-dimensional and higher-dimensional case}

The above analysis was focused on a one-dimensional Brownian motion.
In turn, many SPT techniques allow one to record the displacements of
a tracer in two or three dimensions.  Under the assumption of
isotropic motion, one often considers the combined time-averaged MSD
\begin{equation}
\chi^{(d)}_{t,T} = \frac{1}{T-t}\int\limits_0^{T-t} dt_0 \sum\limits_{i=1}^d \bigl(X_i(t_0+t)-X_i(t_0)\bigr)^2 , 
\end{equation}
where $X_i(t)$ is the tracer's position at time $t$ along the $i$th
coordinate.  For $d$-dimensional Brownian motion, each process
$X_i(t)$ is a one-dimensional Brownian motion independent from the
others so that the Laplace transform $\varphi_t^{(d)}(s)$ for the
random variable $\chi^{(d)}_{t,T}$ is simply $[\varphi_t(s)]^d$.  One
can then easily extend the above 1D results to higher-dimensional
cases.  In particular, the probability density $p(z)$ would exhibit
similar asymptotic behavior at small and large $z$, with deviations
coming from power law corrections.

\section*{Conclusion}

We have studied the statistical properties of the time-averaged
mean-square displacements of Brownian motion.  Using the standard
tools of functional integrals, the original problem was mapped onto an
exactly solvable model of coupled harmonic oscillators.  When the lag
time $t$ was larger than a half of the trajectory length $T$ ($t >
T/2$), a simple closed formula \eref{eq:varphi_m1} for the Laplace
transform $\varphi_t(s)$ of the probability density $p_t(z)$ was
derived.  In the opposite case $t < T/2$, the analysis was quite
similar, although the resulting exact formulas were less explicit.
This reflects the many-body character of the original problem.  In the
special case $t = 1/m$ with $m = 2,3,4,...$, one dealt with a chain of
coupled harmonic oscillators and derived Eq. \eref{eq:varphi_m} for
$\varphi_t(s)$.  The general situation corresponded to two chains of
coupled harmonic oscillators, for which Eq. \eref{eq:varphi_m2} for
$\varphi_t(s)$ was deduced.  From these formulas, the probability
density $p_t(z)$ can be formally reconstructed either through the
inverse Laplace transform of $\varphi_t(s)$, or through the inverse
Fourier transform of the characteristic function $\phi_t(k) \equiv
\varphi_t(-ik)$ (the second option being more convenient for numerical
implementation).  To our knowledge, the probability distribution of
the TAMSD of Brownian motion was derived for the first time.

From the exact formulas for $\varphi_t(s)$, we deduced the first four
cumulant moments of the TAMSD.  While the first and second moments
(mean and variance) were reported earlier, the third and fourth
moments (related to skewness and kurtosis of the distribution) have
been obtained for the first time.  We also analyzed the asymptotic
behavior of the probability density $p_t(z)$ at small and large
arguments.  In particular, we have shown the sharp decay of $p_t(z)$
as $e^{-a/z}$ at small $z$.  This behavior justified the use of a
generalized Gamma distribution as an approximation for $p_t(z)$
suggested in \cite{Grebenkov11b}.  The fact that the GGD correctly
reproduces the asymptotic behavior of the probability density at both
small and large $z$ explained a remarkable accuracy of this
approximation, illustrated by Fig. \ref{fig:abnu_pz}.  Finally, we
expressed the parameters $a$, $b$ and $\nu$ of a GGD as functions of
the lag time $t$ by matching the first moments of the true and
approximate distributions.  The resulting asymptotic relations were
shown to be accurate for $t < T/2$ so that one could easily construct
an accurate approximation for the probability density $p_t(z)$ of the
TAMSD for a given lag time.  For larger $t$, the distribution was too
skewed so that the first moments did not really control the shape of
$p_t(z)$ around the maximum.  In that case, one had to resort to a
simple approximation of $p_t(z)$ provided by the asymptotic form
\eref{eq:pz_approx1} (i.e., by a GGD with $a = (1-t)/4$, $b = 0$ and
$\nu = 0.5$).  However, as illustrated on Fig. \ref{fig:abnu_pz},
better approximations by a GGD could be constructed.  Their explicit
derivation requires further analysis.

From a practical point of view, an accurate approximation of the
probability density $p_t(z)$ by a simple function (such as a
generalized Gamma distribution) opens several interesting perspectives
for a more reliable analysis of random trajectories of diffusing
tracers acquired in single-particle tracking experiments.  In
particular, one can develop efficient inferring schemes through
maximum likelihood estimation, or accurately quantify the confidence
intervals of measured diffusion coefficients.  While the present work
was focused on Brownian motion, future investigations may challenge
the statistical properties of the TAMSD for more sophisticated
processes (e.g., restricted diffusion, fractional Brownian motion or
generalized Langevin equation with memory) which are known to be
representative models for diffusive dynamics in viscoelastic media
(e.g., living cells) or for time series in finance.

%%%%%%%%%%%%%%%%%%%%%%%%%%%%%%%%%%%%%%%%%%%%%%%%%%%%%%%%%%%%%%%%%%%%%%%%%%%%%%%%%%%%%%%%%%%%%%%%%%%%%%%%%%
\appendix
\section{Computation for $t < \frac12$}
\label{sec:details}

The computation for the case $t<\frac{1}{2}$ is a straightforward
generalization of the calculus from Sect. \ref{sec:m1}.  For
convenience, we consider separately the special case $t = \frac{1}{m}$
in \ref{sec:details1} and the generic case $\frac{1}{m+1} < t <
\frac{1}{m}$ in \ref{sec:details2}.

\subsection{Special cases $t = \frac{1}{m}$}
\label{sec:details1}

For $t = \frac{1}{m}$ with $m = 2,3,4,...$, the whole time interval
$[0,1]$ is covered by $m$ equal subintervals so that the trajectory
can be mapped onto a chain of $m$ coupled harmonic oscillators as
illustrated on Fig.~\ref{fig:tsplit}b.  We define oscillators $Y_k$ as
\begin{equation*}
Y_k(u) \equiv X(kt + u)  \qquad u\in (0,t), ~~ (k = 0...m-1),
\end{equation*}
so that the non-local part of the action in Eq. \eref{eq:varphi_path}
becomes:
\begin{equation*}
\fl
\int\limits_0^{1-t}  \hspace*{-1mm} du (X(u+t) - X(u))^2 = \sum\limits_{k=0}^{m-2} \hspace*{-1mm} \int\limits_{kt}^{(k+1)t}  
\hspace*{-3mm} du (X(u+t) - X(u))^2 = \sum\limits_{k=0}^{m-2} \int\limits_0^t  \hspace*{-1mm} du (Y_{k+1}(u) - Y_k(u))^2 ,
\end{equation*}
from which the Laplace transform reads as
\begin{equation*}
\fl
\varphi_{t}(s)= \int\limits_{\R^m} dy_1...dy_m \prod_{k=0}^{m-1}
\int_{Y_{k}(0)=y_{k}}^{Y_{k}(t)=y_{k+1}}\mathcal{D}Y_{k}\,
\exp\left[-\frac{1}{2}\int_{0}^{t}(\partial_{u}Y_{k})^{2}- \frac{g^2}{2} \int_{0}^{t}(Y_{k+1}-Y_{k})^{2}\right] ,
\end{equation*}
with the starting point $y_0 = 0$ (and $y_m = x_1$).  The non-locality
is therefore eliminated at the expense of many-body character of the
resulting problem.  To advance further we need to diagonalize the
chain of coupled oscillators with open boundary conditions:
\begin{equation*}
\fl
\Sigma = \sum_{k=0}^{m-1} \int_{0}^{t}(\partial_{u}Y_{k})^{2} + g^2\sum_{k=0}^{m-2} \int_{0}^{t}(Y_{k+1}-Y_{k})^{2} 
= \sum_{k=0}^{m-1} \int_{0}^{t}\left[(\partial_{u}Y_{k})^{2} + g^2 \sum\limits_{l=0}^{m-1} Y_k L_{kl} Y_l\right],
\end{equation*}
where the matrix $\mathbf{L}$ is a discrete Laplacian of size $m\times
m$ with von Neumann boundary conditions:
\begin{equation*}
\mathbf{L} = \left(\begin{array}{c c c c c c c} 1 & -1 & 0 & 0 & \cdots & 0 & 0\\
-1 & 2 & -1 & 0 & \cdots & 0 & 0\\
0 & -1 & 2 & -1 & \cdots & 0 & 0\\
0 & 0 & -1 & 2 & \cdots & 0 & 0\\
0 & 0 & 0 & -1 & \cdots & 0 & 0\\
\cdots & \cdots & \cdots & \cdots & \cdots & \cdots & \cdots \\
0 & 0 & 0 &  0  & \cdots & -1 & 1 \\  \end{array} \right) .
\end{equation*}
The normal modes which diagonalize $\Sigma$ are defined as
\begin{equation*}
A_{k}=\sum_{l=0}^{m-1}\Psi_{kl}Y_{l} \qquad (k = 0,\dots,m-1), 
\end{equation*}
where $\mathbf{\Psi}$ is an orthogonal matrix with columns formed by
eigenvectors of $\mathbf{L}$: $\mathbf{L}\cdot\mathbf{\Psi} =
\mathbf{\Psi}\cdot\mathbf{\Lambda}$.  It is straightfoward to compute
the eigenvalues and eigenvectors of this matrix:
\begin{eqnarray}
\label{eq:lambda}
&& \Lambda_{kl} = \delta_{kl} \lambda_k, \qquad 
\lambda_{k} = 2\left(1-\cos\frac{\pi k}{m}\right)\qquad (k=0...m-1),  \\
\label{eq:psi}
&& \Psi_{kl} = \frac{\sin\frac{\pi k(l+1)}{m} - \sin\frac{\pi kl}{m}}{\sqrt{m(1-\cos\frac{\pi k}{m})}} 
\qquad (k=1...m-1, ~ l=0...m-1),\\ \nonumber
&& \Psi_{0l} = \frac{1}{\sqrt{m}} \qquad (l=0...m-1).
\end{eqnarray}
We obtain the diagonalized form as
\begin{equation}
\label{eq:diagonalized}
\Sigma = \sum_{k=0}^{m-1} \int_{0}^{t}\left[(\partial_{u}A_{k})^{2} + g^2\lambda_k A_k^{2}\right]
\end{equation}
(note that for $k = 0$ the second term vanishes as $\lambda_0 = 0$),
with the boundary conditions for $A_k$ translated from the boundary
conditions for $Y_k$ as
\begin{eqnarray}
\nonumber
a_{k0} &=& A_k(0) = \sum_{l=0}^{m-1}\Psi_{kl} Y_{l}(0) = \sum_{l=1}^{m-1}\Psi_{kl} y_{l}   \qquad (k = 0..m-1) , \\
\label{eq:BC_normal1}
a_{kt} &=& A_k(t) = \sum_{l=0}^{m-1}\Psi_{kl} Y_{l}(t) = \sum_{l=0}^{m-1}\Psi_{kl} y_{l+1}   \qquad (k = 0..m-1) .
\end{eqnarray}
These linear relations convert the original coordinates $y_1,\dots ,
y_m$ into new variables $a_{k0}$ and $a_{kt}$.  In the first relation,
the sum runs from $l=1$ as $y_0 = 0$ and the columns of the matrix
$\mathbf{\Psi}$ are shifted to the left, the last column being filled
with zeros (as there is no dependence on $y_m$).  This shift can be
represented by a shift matrix $\mathbf{S}$ (of size $m\times m$) as
$\mathbf{\Psi}\cdot\mathbf{S}$:
\begin{equation*}
\mathbf{S} = \left(\begin{array}{c c c c c} 
 0 & 0 &...& 0 & 0 \\
 1 & 0 &...& 0 & 0 \\
 0 & 1 &...& 0 & 0 \\
...&...&...&...&...\\
 0 & 0 &...& 1 & 0 \\
\end{array} \right).
\end{equation*}
Using the shift matrix, we can write
\begin{equation}
\label{eq:BC_normal2}
a_{k0} = \sum\limits_{l=0}^{m-1} [\mathbf{\Psi}\cdot\mathbf{S}]_{kl} y_{l+1}   \qquad (k = 0...m-1).
\end{equation}

Once we have the diagonalized form~\eref{eq:diagonalized}, the
Feynman-Kac formula is applicable and the Laplace transform
$\varphi_t(s)$ is again given as a product of $m-1$ propagators of
harmonic oscillators and one propagator of a free particle
(corresponding to $\lambda_0 = 0$):
\begin{equation*}
\varphi_t(s) =  \int\limits_{\R^m} dy_1\cdots dy_m\, G_0(a_{0t},t|a_{00},0)\,\prod\limits_{k=1}^{m-1} G_{g\sqrt{\lambda_k}}(a_{kt}, t | a_{k0}, 0).
\end{equation*}
We plug in the propagators as given by Eq.~\eref{eq:propagator} and
find the following expression:
\begin{eqnarray}
\label{eq:auxil2}
\fl
\varphi_t(s) &=& \int\limits_{\R^m} dy_1\cdots dy_m\,\frac{1}{\sqrt{2\pi t}}\,\exp\left(- \frac{(a_{0t} - a_{00})^2}{2t}\right) \\
\fl\nonumber
&\times& \prod\limits_{k=1}^{m-1} \frac{(g\sqrt{\lambda_k})^{1/2} e^{- t g\sqrt{\lambda_k}}}{\sqrt{\pi(1-q_k^2)}} 
\exp\left(- \frac{\alpha_k g\sqrt{\lambda_k}}{4} (a_{kt} + a_{k0})^2 - \frac{g\sqrt{\lambda_k}}{4\alpha_k} (a_{kt} - a_{k0})^2\right) ,
\end{eqnarray}
where
\begin{equation}
\label{eq:q_alpha}
q_k = e^{-t g \sqrt{\lambda_k}}, \qquad  \alpha_k = \frac{1-q_k}{1+q_k}. 
\end{equation}
The formula for $\varphi_t(s)$ can be rewritten as
\begin{equation}
\label{eq:varphi_ma}
\varphi_t(s) = \frac{\exp\left(-\frac{1}{2}g \eta(t)\right)}{\prod\limits_{k=1}^{m-1} (1+q_k)} ~\frac{1}{\sqrt{I_t(g)}},
\end{equation}
in which
\begin{equation}
\label{eq:eta_t}
\eta(t) = t\sum\limits_{k=1}^{m-1} \sqrt{\lambda_k} = t \biggl(\ctan\bigl(\frac{\pi}{4m}\bigr) - 1 \biggr),
\end{equation}
\begin{equation}
\label{eq:auxil4}
\frac{1}{\sqrt{I_t(g)}} = \frac{1}{\sqrt{2\pi t}}~ \frac{g^{(m-1)/2}}{\pi^{(m-1)/2}} 
\left(\prod\limits_{k=1}^{m-1} \frac{\sqrt{\lambda_k}}{\alpha_k}\right)^{1/2} 
\int\limits_{\R^m} d\y \exp\left(- \frac12 (\y^T\cdot\mathbf{A}\cdot\y)\right),
\end{equation}
where $\y = (y_1,...,y_m)^T$, and $\mathbf{A} =
\mathbf{V}^T\cdot\mathbf{U}\mathbf{V}$ is the matrix of size $m\times
m$ which combines both the structure of the quadratic form in
Eq. \eref{eq:auxil2} and the linear relations \eref{eq:BC_normal1},
\eref{eq:BC_normal2}, represented respectively by matrices
$\mathbf{U}$ and $\mathbf{V}$ of sizes $2m \times 2m$ and $2m \times
m$:
\begin{equation}
\label{eq:UV}
\mathbf{U} = \left(\begin{array}{c | c} 
\mathbf{U}^+ & \mathbf{U}^-  \\ \hline
\mathbf{U}^- & \mathbf{U}^+  \\  \end{array} \right),  \qquad
\mathbf{V} = \left(\begin{array}{c} 
\mathbf{\Psi}\cdot\mathbf{S} \\  \hline \mathbf{\Psi} \\ \end{array} \right),
\end{equation}
where $U^\pm$ are diagonal matrices of size $m\times m$ with the
elements
\begin{eqnarray*}
\fl
u_k^+ &=& \frac12 g\sqrt{\lambda_k}(\alpha_k + 1/\alpha_k) = \frac{g\sqrt{\lambda_k}}{\tanh(gt \sqrt{\lambda_k})}
\quad (k = 1...m-1), \qquad u_0^+ = \frac{1}{t},  \\
\fl
u_k^- &=& \frac12 g\sqrt{\lambda_k}(\alpha_k - 1/\alpha_k) = \frac{- g\sqrt{\lambda_k}}{\sinh(gt \sqrt{\lambda_k})}
\quad (k = 1...m-1), \qquad u_0^- = \frac{-1}{t}  .
\end{eqnarray*}
The elements $u_0^\pm$ can be seen as the limit of the generic
formulas with $\lambda_0\to 0$.  Performing the matrix multiplication
yields
\begin{equation}
\fl
\label{eq:A}
\mathbf{A} = \mathbf{V}^T \mathbf{U}\mathbf{V} = \mathbf{\Psi}^T\mathbf{U}^+\mathbf{\Psi}  +  
\mathbf{S}^T\mathbf{\Psi}^T\mathbf{U}^-\mathbf{\Psi} + \mathbf{\Psi}^T\mathbf{U}^-\mathbf{\Psi}\mathbf{S} + 
\mathbf{S}^T\mathbf{\Psi}^T\mathbf{U}^+\mathbf{\Psi}\mathbf{S}.
\end{equation}
The diagonal matrices $U^\pm$ contain the eigenvalues $\lambda_k$ of
the original matrix $\mathbf{L}$ and $\mathbf{\Psi}$ is formed by its
eigenvectors, implying a simple matrix representation for
$\mathbf{A}$:
\begin{equation}
\label{eq:A2}
\fl
\mathbf{A} = \frac{g\sqrt{\mathbf{L}}}{\tanh(gt\sqrt{\mathbf{L}})}  -  
\mathbf{S}^T\frac{g\sqrt{\mathbf{L}}}{\sinh(gt\sqrt{\mathbf{L}})} - 
\frac{g\sqrt{\mathbf{L}}}{\sinh(gt\sqrt{\mathbf{L}})}\mathbf{S} + 
\mathbf{S}^T\frac{g\sqrt{\mathbf{L}}}{\tanh(gt\sqrt{\mathbf{L}})}\mathbf{S}.
\end{equation}
This expression is convenient for a theoretical analysis, while the
original representation \eref{eq:A} is better suited for rapid
numerical construction of $\mathbf{A}$.

The computation of the Gaussian integral in Eq. \eref{eq:auxil4}
yields
\begin{equation}
\frac{1}{\sqrt{I_t(g)}} = \frac{(2g)^{(m-1)/2}}{\sqrt{t} ~ \sqrt{\det(\mathbf{A})}} 
\left(\prod\limits_{k=1}^{m-1} \frac{\sqrt{\lambda_k}}{\alpha_k}\right)^{1/2}.
\end{equation}
This relation, together with Eq. \eref{eq:varphi_ma}, completes the
computation of $\varphi_t(s)$ for special cases $t = 1/m$ (with $m =
2,3,4,...$) and provides a solution to the problem.  This solution can
alternatively be written in the form \eref{eq:varphi_m}.

\subsection{Generic situation}
\label{sec:details2}

For the generic case $\frac{1}{1+m} < t < \frac{1}{m}$ (for some $m =
2,3,4,...$), the interval $[0,1]$ is covered by $m$ subintervals of
duration $t$, plus a smaller interval $[mt,1]$ of duration $\delta = 1
- mt < t$.  In order to split the interval $[0,1]$ regularly, one
divides each subinterval in two parts, of durations $\delta$ and
$t-\delta$, respectively:
\begin{equation*}
[0,1]=[0,\delta]\cup[\delta,t]\cup\cdots\cup[t_{m-1},t_{m-1}+\delta]\cup[t_{m-1}+\delta,t_{m}]\cup[t_{m},1],
\end{equation*}
where $t_{k}=kt$.  This covering maps the whole trajectory onto two
interacting chains of $m+1$ and $m$ harmonic oscillators as
illustrated on Fig.~\ref{fig:tsplit}c:
\begin{eqnarray*}
Y_{k}(u) &=& X(t_{k}+u) \quad  u\in [0,\delta], \quad (k=0\ldots m),   \\  
Z_{k}(u) &=& X(t_{k}+u) \quad  u\in [\delta,t], \quad (k=0\ldots m-1).
\end{eqnarray*}
Here $m = [1/t]$, the integer part of $1/t$.  One gets
therefore
\begin{eqnarray*}
\fl
&& \varphi_{t}(s)= \hspace*{-3mm} \int\limits_{\R^{2m+1}}  \hspace*{-3mm} dy_1...dy_m dz_0...dz_m \prod_{k=0}^{m-1}
\int_{Y_{k}(0)=y_{k}}^{Y_{k}(\delta)=z_{k}} \hspace*{-3mm} \mathcal{D}Y_{k}\,
\exp\left[-\frac{1}{2}\int_{0}^{\delta}(\partial_{u}Y_{k})^{2}- \frac{g^2}{2} \int_{0}^{\delta}(Y_{k+1}-Y_{k})^{2}\right]  \\
\fl
&& \times \left\{\prod_{k=0}^{m-2}\int_{Z_{k}(0)=z_{k}}^{Z_{k}(t-\delta)=y_{k+1}}\mathcal{D}Z_{k}\,
\exp\left[-\frac{1}{2}\int_{\delta}^{t}(\partial_{u}Z_{k})^{2}- \frac{g^2}{2} \int_{\delta}^{t}(Z_{k+1}-Z_{k})^{2}\right]\right\}\\
\fl
&& \times \int_{Y_{m}(0)=y_{m}}^{Y_{m}(\delta)=z_m}\mathcal{D}Y_{m}\,\exp\left[-\frac{1}{2}\int_{0}^{\delta}(\partial_{u}Y_{m})^{2}\right]
\int_{Z_{m-1}(\delta)=z_{m-1}}^{Z_{m-1}(t)=y_{m}}\mathcal{D}Z_{m-1}\,
\exp\left[-\frac{1}{2}\int_{\delta}^{t}(\partial_{u}Z_{m-1})^{2}\right].
\end{eqnarray*}
(with $y_0 = 0$).  There is no direct interaction between the two
chains, but the boundary conditions for the chains are mixed since
$Y_k(\delta)=Z_k(0)$ and $Z_k(t-\delta)=Y_{k+1}(0)$ to ensure the
continuity of the original Brownian trajectory.  Introducing normal
modes:
\begin{equation*}
A_{k}=\sum_{l=0}^{m} \tilde{\Psi}_{kl} Y_{l} , \hskip 10mm  B_{k}=\sum_{l=0}^{m-1}\Psi_{kl} Z_{l} ,
\end{equation*}
this problem transforms into a set of uncoupled oscillators.  The
$\mathbf{\tilde{\Psi}}$ is an orthogonal matrix formed by the
eigenvectors of the discrete Laplacian $\mathbf{L}$ of size
$(m+1)\times (m+1)$ (tilde sign refers to quantities related to the
matrix $\mathbf{L}$ of size $(m+1)\times(m+1)$, in order to
distinguish them from similar quantities corresponding to $\mathbf{L}$
of size $m\times m$).  The boundary conditions read for each $k =
0\dots m$
\begin{eqnarray}
\label{eq:BC_normal_A1}
a_{k0}      &=& \sum_{k=1}^{m} \tilde{\Psi}_{kl} y_{l} ,     \qquad a_{k\delta} = \sum_{l=0}^{m} \tilde{\Psi}_{kl} z_{l} ,\\
\label{eq:BC_normal_A2}
b_{k\delta} &=& \sum_{l=0}^{m-1}\Psi_{kl} z_{l},      \qquad b_{kt}=\sum_{l=1}^{m}\Psi_{k,l-1} y_{l} .
\end{eqnarray}
We obtain then
\begin{equation*}
\fl 
\varphi_t(s) = \hspace*{-3mm} \int\limits_{\R^{2m+1}} \hspace*{-3mm}  dy_1...dy_m dz_0...dz_m
\left\{\prod\limits_{k=0}^m G_{g\sqrt{\tilde{\lambda}_k}}(a_{k\delta}, \delta | a_{k0}, 0)\right\} 
\left\{\prod\limits_{k=0}^{m-1} G_{g\sqrt{\lambda_k}}(b_{kt}, t | b_{k\delta}, \delta)\right\} .
\end{equation*}
Note that two factors with $k=0$ correspond to free diffusion
propagators (as $\lambda_0 = \tilde{\lambda}_0 = 0$).  Substituting
Eq. \eref{eq:propagator} for the propagators, one gets Gaussian
integrals:
\begin{eqnarray*}
\fl
\varphi_t(s) &=& \int\limits_{\R^{2m+1}} dy_1...dy_m dz_0...dz_m \frac{1}{2\pi\sqrt{\delta(t-\delta)}}
\exp\left(- \frac{(a_{0\delta} - a_{00})^2}{2\delta} - \frac{(b_{0t}-b_{0\delta})^2}{2(t-\delta)}\right) \\
\fl  
&\times& \prod\limits_{k=1}^m \frac{(g\sqrt{\tilde{\lambda}_k})^{1/2} e^{-\delta g\sqrt{\tilde{\lambda}_k}/2}}{\sqrt{\pi(1-\tilde{q}_k^2)}} 
\exp\left(- \frac{\tilde{\alpha}_k g\sqrt{\tilde{\lambda}_k}}{4} (a_{k\delta} + a_{k0})^2 - 
\frac{g\sqrt{\tilde{\lambda}_k}}{4\tilde{\alpha}_k} (a_{k\delta} - a_{k0})^2\right) \\
\fl  
&\times& \prod\limits_{k=1}^{m-1} \frac{(g\sqrt{\lambda_k})^{1/2}  e^{-(t-\delta) g\sqrt{\lambda_k}/2}}{\sqrt{\pi(1-q_k^2)}} 
\exp\left(- \frac{\alpha_k g\sqrt{\lambda_k}}{4} (b_{kt} + b_{k\delta})^2 - 
\frac{g\sqrt{\lambda_k}}{4\alpha_k} (b_{kt} - b_{k\delta})^2\right) . 
\end{eqnarray*}
The argument of the exponential function can be represented as a
quadratic form $\frac{1}{2}({\bf a}^T\mathbf{U}{\bf a})$:
\begin{equation*}
\fl
\left(\begin{array}{c} 
a_{00}         \\
 ...           \\
a_{m0}         \\
b_{0t}         \\
 ...           \\
b_{m-1,t}      \\
a_{0\delta}    \\
 ...           \\
a_{m\delta}    \\ 
b_{0\delta}    \\ 
 ...           \\
b_{m-1,\delta} \\ 
\end{array} \right)^T
\left(\begin{array}{c c c c c c c c c c c c} 
\u_0^+& ... &   0  &  0  & ... &  0      &\u_0^-& ... &   0  &  0  & ... &  0       \\
  ... & ... &  ... & ... & ... & ...     &  ... & ... &  ... & ... & ... & ...      \\
   0  & ... &\u_m^+&  0  & ... &  0      &   0  & ... &\u_m^-&  0  & ... &  0       \\
   0  & ... &   0  &u_0^+& ... &  0      &   0  & ... &   0  &u_0^-& ... &  0       \\
  ... & ... &  ... & ... & ... & ...     &  ... & ... &  ... & ... & ... & ...      \\
   0  & ... &   0  &  0  & ... &u_{m-1}^+&   0  & ... &   0  &  0  & ... &u_{m-1}^- \\
\u_0^-& ... &   0  &  0  & ... &  0      &\u_0^+& ... &   0  &  0  & ... &  0       \\
  ... & ... &  ... & ... & ... & ...     &  ... & ... &  ... & ... & ... & ...      \\
   0  & ... &\u_m^-&  0  & ... &  0      &   0  & ... &\u_m^+&  0  & ... &  0       \\ 
   0  & ... &   0  &u_0^-& ... &  0      &   0  & ... &   0  &u_0^+& ... &  0       \\ 
  ... & ... &  ... & ... & ... & ...     &  ... & ... &  ... & ... & ... & ...      \\
   0  & ... &   0  &  0  & ... &u_{m-1}^-&   0  & ... &  ... &  0  & ... &u_{m-1}^+ \\ 
\end{array} \right)
\left(\begin{array}{c} 
a_{00}         \\
 ...           \\
a_{m0}         \\
b_{0t}         \\
 ...           \\
b_{m-1,t}      \\
a_{0\delta}    \\
 ...           \\
a_{m\delta}    \\ 
b_{0\delta}    \\ 
 ...           \\
b_{m-1,\delta} \\ 
\end{array} \right)
\end{equation*}
where ${\bf a}$ is the $(4m+2)$-dimensional vector, $\mathbf{U}$ is a
$(4m+2)\times(4m+2)$ matrix with the elements:
\begin{eqnarray*}
\tilde{u}_k^\pm &=& \frac12 g\sqrt{\tilde{\lambda}_k}(\tilde{\alpha}_k \pm 1/\tilde{\alpha}_k) \quad (k = 1...m), \quad 
\tilde{u}_0^\pm = \frac{\pm 1}{\delta},  \\
u_k^\pm &=& \frac12 g\sqrt{\lambda_k}(\alpha_k \pm 1/\alpha_k) \quad (k = 1...m-1), \quad u_0^\pm = \frac{\pm 1}{t-\delta}. 
\end{eqnarray*}
The intermediate variables $a_{00}$, $a_{0\delta}$,$\cdots$ forming
the vector ${\bf a}$, are expressed through
Eqs. \eref{eq:BC_normal_A1}, \eref{eq:BC_normal_A2} which can be
written in a matrix form ${\bf a} = \mathbf{V} {\bf x}$, where ${\bf
x}$ is $(2m+1)$-dimensional vector ${\bf x} = (y_1, y_2, ..., y_m, ~
z_0, z_1, ..., z_m)^T$, and matrix $\mathbf{V}$ has a block structure
\begin{equation*}
\fl
\left(\begin{array}{c} 
a_{00}         \\
a_{10}         \\
 ...           \\
a_{m0}         \\ \hline
b_{0t}         \\
 ...           \\
b_{m-1,t}      \\ \hline
a_{0\delta}    \\
a_{1\delta}    \\
 ...           \\
a_{m\delta}    \\ \hline
b_{0\delta}    \\ 
 ...           \\
b_{m-1,\delta} \\ 
\end{array} \right) = \left(\begin{array}{c c c | c c c | c} 
\tilde{\Psi}_{01}& ... &\tilde{\Psi}_{0m}  &      0           & ... &          0          &        0          \\
\tilde{\Psi}_{11}& ... &\tilde{\Psi}_{1m}  &      0           & ... &          0          &        0          \\
       ...       & ... &       ...         &     ...          & ... &         ...         &       ...         \\
\tilde{\Psi}_{m1}& ... &\tilde{\Psi}_{mm}  &      0           & ... &          0          &        0          \\  \hline
\Psi_{00}        & ... & \Psi_{0,m-1}      &      0           & ... &          0          &        0          \\
     ...         & ... &       ...         &     ...          & ... &         ...         &       ...         \\
\Psi_{m-1,0}     & ... & \Psi_{m-1,m-1}    &      0           & ... &          0          &        0          \\  \hline
        0        & ... &        0          &\tilde{\Psi}_{00} & ... &\tilde{\Psi}_{0,m-1} &\tilde{\Psi}_{0m}  \\
        0        & ... &        0          &\tilde{\Psi}_{10} & ... &\tilde{\Psi}_{1,m-1} &\tilde{\Psi}_{1m}  \\
       ...       & ... &       ...         &     ...          & ... &         ...         &       ...         \\
        0        & ... &        0          &\tilde{\Psi}_{m0} & ... &\tilde{\Psi}_{m,m-1} &\tilde{\Psi}_{mm}  \\  \hline
        0        & ... &        0          & \Psi_{00}        & ... & \Psi_{0,m-1}        &        0          \\
       ...       & ... &       ...         &     ...          & ... &         ...         &       ...         \\
        0        & ... &        0          & \Psi_{m-1,0}     & ... & \Psi_{m-1,m-1}      &        0          \\
\end{array} \right)
\left(\begin{array}{c} 
y_1  \\
 ... \\
y_m  \\  \hline
z_0  \\
...  \\  
z_m  \\ 
\end{array} \right).
\end{equation*}
We obtain therefore the quadratic form $\frac{1}{2}({\bf
x}^T\mathbf{V}^T\mathbf{U}\mathbf{V}{\bf x}) = \frac{1}{2}({\bf
x}^T\mathbf{A}{\bf x})$, where 
\begin{equation}
\label{eq:A_gen}
\mathbf{A} = \mathbf{V}^T\mathbf{U}\mathbf{V}
\end{equation}
is the $(2m+1)\times(2m+1)$ matrix.

The computation of Gaussian integrals yields an exact representation:
\begin{equation}
\label{eq:varphi_m2a}
\varphi_t(s) = \left(\prod\limits_{k=1}^m \frac{1}{1+\tilde{q}_k}\right) \left(\prod\limits_{k=1}^{m-1} 
\frac{1}{1+q_k}\right)\frac{\exp\left(-\frac{g}{2}\,\eta(t)\right)}{\sqrt{I_t(g)}} , 
\end{equation}
where 
\begin{eqnarray}
\nonumber
\tilde{q}_k &=& e^{-\delta g \sqrt{\tilde{\lambda}_k}}, \qquad  \tilde{\alpha}_k = \frac{1-\tilde{q}_k}{1+\tilde{q}_k}, 
\qquad \tilde{\lambda}_k = 2\biggl(1 - \cos\biggl(\frac{\pi k}{m+1} \biggr) \biggr) , \\
\nonumber
q_k &=& e^{-(t-\delta) g \sqrt{\lambda_k}}, \qquad  \alpha_k = \frac{1-q_k}{1+q_k}, 
\qquad \lambda_k = 2\biggl(1 - \cos\biggl(\frac{\pi k}{m} \biggr) \biggr) , \\
\label{eq:ft}
\eta(t) &=& \delta\sum\limits_{k=1}^m \sqrt{\tilde{\lambda}_k} + (t-\delta)\sum\limits_{k=1}^{m-1} \sqrt{\lambda_k} , \\
\frac{1}{\sqrt{I_t(g)}} &=& \frac{(2g)^{m-\frac12}}{\sqrt{\delta(t-\delta)} ~ \sqrt{\det(\mathbf{A})}} 
\left(\prod\limits_{k=1}^m \frac{\sqrt{\tilde{\lambda}_k}}{\tilde{\alpha}_k}\right)^{1/2}
\left(\prod\limits_{k=1}^{m-1} \frac{\sqrt{\lambda_k}}{\alpha_k}\right)^{1/2} .
\end{eqnarray}
One can alternatively rewrite \eref{eq:varphi_m2a} in the form
\eref{eq:varphi_m2}.

\begin{figure}
\begin{center}
\includegraphics[width=85mm]{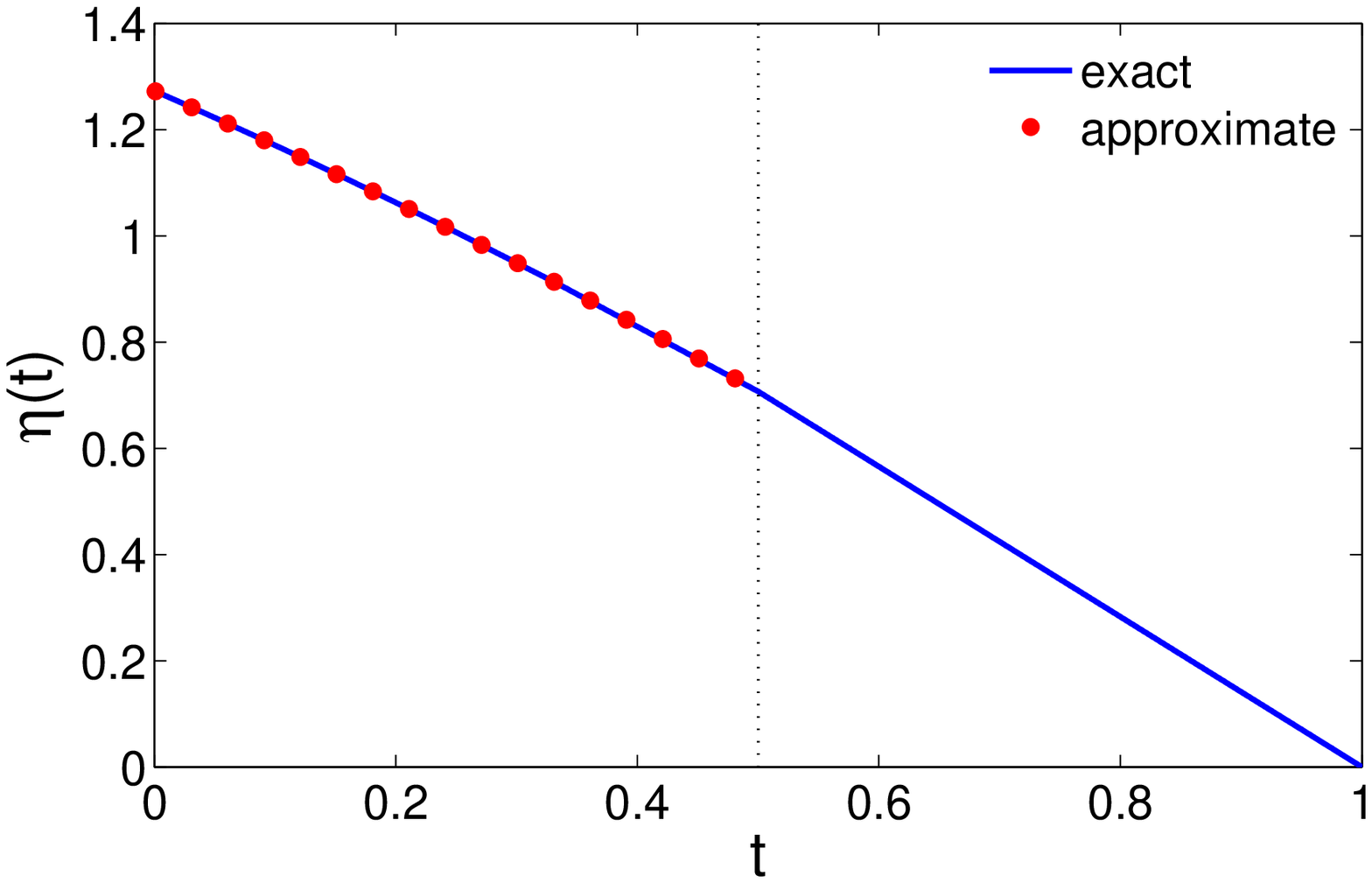}
\end{center}
\caption{
(Color online) The function $\eta(t)$ and its approximation
\eref{eq:ft_approx} for $t < 0.5$.  For $t > 0.5$, one has the exact
relation $\eta(t) = \sqrt{2}(1-t)$.}
\label{fig:ft}
\end{figure}

Using explicit expressions for $\lambda_k$, $\tilde{\lambda}_k$ and
formula \eref{eq:eta_t}, we get
\begin{eqnarray}
\label{eq:ft1}
\eta(t) &=& \left(\ctan\left(\frac{\pi}{4(m+1)}\right) - \ctan\left(\frac{\pi}{4m}\right)\right) \\ \nonumber
&-& t \left( 1 + m~ \ctan\left(\frac{\pi}{4(m+1)}\right) - (m+1)\ctan\left(\frac{\pi}{4m}\right)\right) .
\end{eqnarray}
If $t$ is small, then we can approximate $m\approx 1/t$, from which
the series expansion of $\ctan(x)$ leads to
\begin{equation}
\label{eq:ft_approx}
\eta(t) \approx \frac{4}{\pi} - t - \frac{\pi}{12}~ t^2 + O(t^3)  \qquad (t < 0.5).
\end{equation}
This expansion accurately approximates the function $\eta(t)$ for $t <
0.5$, as illustrated on Fig. \ref{fig:ft}.  For $t > 0.5$, one has $m
= 1$, and Eq. \eref{eq:ft1} simplifies to $\eta(t) = \sqrt{2}(1-t)$,
while Eq. \eref{eq:varphi_m2a} is reduced to Eq. \eref{eq:varphi_m1}.

% see 'A_Andreanov_ft.m'

%%%%%%%%%%%%%%%%%%%%%%%%%%%%%%%%%%%%%%%%%%%%%%%%%%%%%%%%%%%%%%%%%%%%%%%%%%%%%%%%%%%%%%%%%%%

\subsection{The case $1/3<t<1/2$}
\label{sec:case_m2}

For the case $1/3<t<1/2$ ($m = 2$), one performs the computation
analytically and gets the following result
\begin{equation}
\varphi_t(s) = \frac{\exp\left(-\frac{g}{2}\left[(1-2t)(1+\sqrt{3}) + (3t-1)\sqrt{2}\right]\right)}
{(1+\tilde{q}_1)(1+\tilde{q}_2)(1+q_1) \sqrt{c_0 + g c_1}} ,
\end{equation}
where
\begin{eqnarray*}
\fl
c_0(g) &=& \frac{1}{576} \biggl(9 + 4\sqrt{3}\tilde{\alpha}_1\tilde{\alpha}_2 + 10\sqrt{6}\tilde{\alpha}_2\alpha_1 + \tilde{\alpha}_1^2\tilde{\alpha}_2^2
+16\tilde{\alpha}_2^2 \alpha_1^2 + 6\sqrt{2} \tilde{\alpha}_1\alpha_1 \tilde{\alpha}_2^2\biggr), \\
\fl
c_1(g) &=& \frac{5t-1}{576} \biggl(6\tilde{\alpha}_1 + \frac{3}{\sqrt{2}} \tilde{\alpha}_1^2\tilde{\alpha}_2^2\alpha_1 
+ 6\tilde{\alpha}_1\tilde{\alpha}_2^2\alpha_1^2 + 2\sqrt{3}\tilde{\alpha}_2\alpha_1^2
+ \frac{3}{\sqrt{2}} \alpha_1 + 6\sqrt{6} \tilde{\alpha}_1\tilde{\alpha}_2 \alpha_1 + 2\sqrt{3}\tilde{\alpha}_1^2\tilde{\alpha}_2\biggr) .
\end{eqnarray*}
It is worth recalling that the ``coefficients'' $c_0$ and $c_1$, as
well as $\tilde{q}_1$, $\tilde{q}_2$, $q_1$, $\tilde{\alpha}_1$,
$\tilde{\alpha}_2$, $\alpha_1$ all depend on $g$ which enters through
an exponential function.  As $t$ gets smaller, expressions are getting
even more cumbersome.

\subsection{ Exact formulas for $t = 1/3$ and $t = 1/4$ }

For the case $t = 1/m$ with small $m = 2,3,4,...$, the matrix
$\mathbf{A}$ from Eq. \eref{eq:A2} has a small size so that one can
compute $\det(\mathbf{A})$ explicitly.  For instance, one has for $m =
3$,
\begin{equation*}
\det(\mathbf{A}) = \frac{g^2}{12 \alpha_1\alpha_2} \biggl[4g\sqrt{3}\alpha_1 + 4g\alpha_2\alpha_1^2 + 
\alpha_1^2\alpha_2^2\sqrt{3} + 9\sqrt{3} + 12\alpha_1\alpha_2\biggr] ,
\end{equation*}
and for $m = 4$
\begin{eqnarray*}
\fl
\det(\mathbf{A}) &=& \frac{g^3}{64\alpha_1\alpha_2\alpha_3}\biggl[ -4g\alpha_2^2\alpha_3^2\alpha_1\sqrt{2}\nu_1 
+ 12g\alpha_2^2\alpha_3\alpha_1^2\nu_1\sqrt{2} -32\alpha_2\alpha_3\nu_1\sqrt{2} \\
\fl
&& + 128\alpha_1\alpha_2\sqrt{2}\nu_1 + 12\sqrt{2}g\nu_1\alpha_3 + 28g\alpha_1\sqrt{2}\nu_1 + 8g\alpha_2^2\alpha_3^2\alpha_1\nu_1 \\
\fl
&& +16g\alpha_2^2\alpha_3\alpha_1^2\nu_1 + 160\alpha_1\alpha_2\nu_1 + 96\nu_1\alpha_2\alpha_3 + 40g\alpha_1\nu_1 - 16g\nu_1\alpha_3 \\
\fl
&& - 16\alpha_2^2\sqrt{2}\alpha_3^2 + 16\alpha_2^2\alpha_1^2\sqrt{2} + 16\alpha_1\alpha_2^2\sqrt{2}\alpha_3 + 17g\alpha_2\alpha_3^2\sqrt{2} \\
\fl
&& + 17g\alpha_1^2\alpha_2\sqrt{2} + 64 
+24\alpha_2^2\alpha_3^2 + 24\alpha_2^2\alpha_1^2 + 24g\alpha_1^2\alpha_2 - 24g\alpha_2\alpha_3^2 + 12g\alpha_2\alpha_3\alpha_1\biggr] ,
\end{eqnarray*}
where $\nu_1 = \sqrt{\mu_1}/2 = \sin(\pi/8) = \sqrt{2-\sqrt{2}}/2$.
For larger $m$, explicit formulas are much more cumbersome.

\section{Cumulant moments}
\label{sec:cumulant}

The simple explicit formula \eref{eq:varphi_m1} for $\varphi_t(s)$
allows one to easily compute high-order cumulant moments of $\chi_t$
for $t > \frac12$.  The analysis of the opposite case $t < \frac12$ is
more difficult, especially for small $t$.  For this reason, we present
an alternative approach to evaluation of these moments.

We start by establishing a formal representation of $\varphi_t(s)$ as
a determinant of some integral operator.  Using the standard Gaussian
identity
\begin{equation*}
e^{-x^2/2} = \int\limits_\R \frac{dy}{\sqrt{2\pi}} ~ e^{-y^2/2 + ixy } ,
\end{equation*}
one can rewrite Eq. \eref{eq:varphi_path} by introducing an auxiliary
Gaussian field $Y$:
\begin{eqnarray*}
\fl
\varphi_t(s) &=& \int \mathcal{D}Y(u) \exp\left(-\frac12\int\limits_0^{1-t} Y^2(u)\right)
\int\limits_\R dx_1 \hspace*{-3mm} \int\limits_{X(0)=0}^{X(1)=x_1}  \hspace*{-3mm} \mathcal{D}X(u)  \\
\fl
&& \times 
\exp\left(-\frac{1}{2}\int\limits_0^1 (\partial_u X(u))^2 + i \left(\frac{2s}{1-t}\right)^{1/2} \int\limits_0^{1-t} Y(u) (X(u+t)-X(u))\right).
\end{eqnarray*}
Splitting the trajectory $X(u)$ into a linear part $t x_1$ and
Brownian bridge $\tilde{X}(u)$ started from $0$ and arrived at $0$ at
time $t=1$, one integrates over the whole path $X(u)$ to get
\begin{equation*}
\fl
\varphi_t(s) = \int \mathcal{D}Y(u) \exp\left(-\frac12\int\limits_0^{1-t} du Y^2(u) - \frac{s}{1-t} \int\limits_0^{1-t} du \int\limits_0^{1-t}
dv ~ Y(u) G(u,v) Y(v)\right),
\end{equation*}
where $G(u,v)$ is a combination of the two-point correlators $C(u,v)$
of Brownian bridge $\tilde{X}(u)$
\begin{eqnarray*}
C(u,v) &=& (1-v)u - (u-v)\Theta(u-v),   \\
G(u,v) &=& t^2 + C(u+t,v+t) - C(u+t,v) - C(u,v+t) + C(u,v) , 
\end{eqnarray*}
where $\Theta(x)$ is the Heaviside function: $\Theta(x) = 1$ for $x>0$
and $0$ otherwise.  One gets then
\begin{equation}
\label{eq:Guv}
\fl
G(u,v) = (u-v + t) \Theta(u-v + t) + (u-v - t) \Theta(u-v - t) - 2(u-v) \Theta(u-v)  ,
\end{equation}
which also reduces to $t - |u-v|$ for $|u-v| < t$, and $0$ otherwise.

Evaluating Gaussian integrals over the auxiliary field $Y(u)$ yields
\begin{equation*}
\varphi_t(s) = \frac{1}{\sqrt{\det\left(1 + \frac{2s}{1-t} G\right)}} ,
\end{equation*}
where $G$ is an integral operator with the kernel $G(u,v)$ acting from
$L_2([0,1-t])$ to $L_2([0,1-t])$.  This is a direct extension of the
classical representation (as Eq. \eref{eq:varphi_discrete}) for
discrete Gaussian processes to the continuous case (see
\cite{Grebenkov11b}).  Rewriting the above formula as
\begin{equation*}
- \ln \varphi_t(s) = \frac12 \Tr \ln\left(1 + \frac{2sG}{1-t} \right) = \frac12 
\sum\limits_{n=1}^\infty \frac{1}{n} \left(\frac{-2s}{1-t}\right)^n \Tr (G^n) ,
\end{equation*}
one gets a representation of cumulant moments:
\begin{equation*}
\kappa_n = (-1)^{n-1} \lim\limits_{s\to 0} \left(\frac{\partial^n \ln \varphi_t(s)}{\partial s^n}\right) = 
\frac{2^{n-1} (n-1)!}{(1-t)^n} \Tr(G^n) .
\end{equation*}
More explicitly, one has
\begin{equation}
\label{eq:kappan_G}
\kappa_n = \frac{2^{n-1} (n-1)!}{(1-t)^n} \int\limits_0^{1-t} du_1 ... \int\limits_0^{1-t} du_n~ G_{u_1,u_2} ~ G_{u_2,u_3} ~ ... ~ G_{u_n,u_1} ,
\end{equation}
with an explicit function $G_{uv}$ from \eref{eq:Guv}.

\section{Large $g$ asymptotic behavior for $t < \frac{1}{2}$}
\label{sec:Alarge}

\subsection{Case $t = 1/m$: representation of the determinant} 

In this section, we only consider the case $t = 1/m$, with $m =
2,3,4,...$, although the general situation could be analyzed in a
similar way.  The matrix $\mathbf{U}$ from Eq. \eref{eq:UV} can be
split into two parts: $\mathbf{U} = \mathbf{U}_0 + g\,\mathbf{U}_1$,
where the matrix $\mathbf{U}_0$ has only 4 nonzero elements which do
not depend on $g$:
\begin{equation*}
\mathbf{U}_0 = \left(\begin{array}{c c c c c c} 
u_0^+& ... &  0      &u_0^-& ... &  0       \\
 ... & ... & ...     & ... & ... & ...      \\
  0  & ... &  0      &  0  & ... &  0       \\
u_0^-& ... &  0      &u_0^+& ... &  0       \\
 ... & ... & ...     & ... & ... & ...      \\
  0  & ... &  0      &  0  & ... &  0       \\
\end{array} \right),  \qquad
\mathbf{U}_1 = g^{-1}(\mathbf{U} - \mathbf{U}_0).
\end{equation*}
Given that $\Psi_{0k} = 1/\sqrt{m}$ and $u_0^+ = -u_0^- =
1/t = m$, one finds
\begin{equation*}
\mathbf{A}_0 = \mathbf{V}^T\mathbf{U}_0\mathbf{V} = \left(\begin{array}{c c c c} 
  0  & ... &  0  &  0   \\
 ... & ... & ... & ...  \\
  0  & ... &  0  &  0   \\
  0  & ... &  0  &  1   \\
\end{array} \right).
\end{equation*}
One has therefore
\begin{equation*}
\fl
\det(\mathbf{A}) = \det(\mathbf{A}_0 + g\,\mathbf{A}_1) = \det(\mathbf{A}_1) 
\det(g\mathbf{I} + \mathbf{A}_1^{-1}\mathbf{A}_0) = \det(\mathbf{A}_1) \prod\limits_{k=1}^{m} (g + \alpha_k) ,
\end{equation*}
where $\mathbf{A}_1 = \mathbf{V}^T\mathbf{U}_1\mathbf{V}$ and
$\alpha_k$ are the eigenvalues of the matrix
$\mathbf{A}_1^{-1}\mathbf{A}_0$.  Since the matrix $\mathbf{A}_0$ has
only one nonzero element, the product $\mathbf{A}_1^{-1}\mathbf{A}_0$
is a matrix in which only the last column is nonzero.  As a
consequence, all the eigenvalues of this product are zero, except one,
which is equal to $(\mathbf{A}_1^{-1})_{m-1,m-1}$.  We obtain
therefore
\begin{equation}
\det(\mathbf{A}) = \det(\mathbf{A}_1) g^{m-1} (g + (\mathbf{A}_1^{-1})_{m-1,m-1}) = g^{m-1} (a_0 + a_1 g) ,
\end{equation}
where the coefficients $a_0 =
\det(\mathbf{A}_1)(\mathbf{A}_1^{-1})_{m-1,m-1}$ and
$a_1 = \det(\mathbf{A}_1)$ depend on $g$ only through the coefficients
$\alpha_k$ in the matrix $\mathbf{A}_1$.  Denoting
\begin{equation*}
c_i =  \frac{a_i}{m 2^{m-1}} \prod\limits_{k=1}^{m-1} \frac{\alpha_k}{\sqrt{\lambda_k}}  \qquad (i=0,1),
\end{equation*}
one gets $I_t(g) = c_0 + g c_1$ so that
\begin{equation*}
\varphi_t(s) = \frac{\exp\left(-\frac{g}{2} t \sum\limits_{k=1}^{m-1} \sqrt{\lambda_k}\right)}{(1+q_1)...(1+q_{m-1}) \sqrt{c_0 + g c_1}}  .
\end{equation*}
This representation is exact but quite formal as the dependence on $g$
and $t$ is still ``hidden'' in coefficients $c_0$ and $c_1$.

\subsection{Case $t = 1/m$: large $g$ asymptotic behavior}

For large $g$, all $\alpha_k$ are close to $1$, while $q_k$ are
negligible so that
\begin{equation}
\label{eq:c1}
c_i \simeq  \frac{a_i}{m^{3/2} 2^{m-1}}  \qquad (g\gg 1, ~i=0,1),
\end{equation}
where we used the identity
\begin{equation}
\prod\limits_{k=1}^{m-1} \sqrt{\lambda_k} = \prod\limits_{k=1}^{m-1} 2\sin(\frac{\pi k}{2m}) = \sqrt{m} .
\end{equation}

\begin{figure}
\begin{center}
\includegraphics[width=85mm]{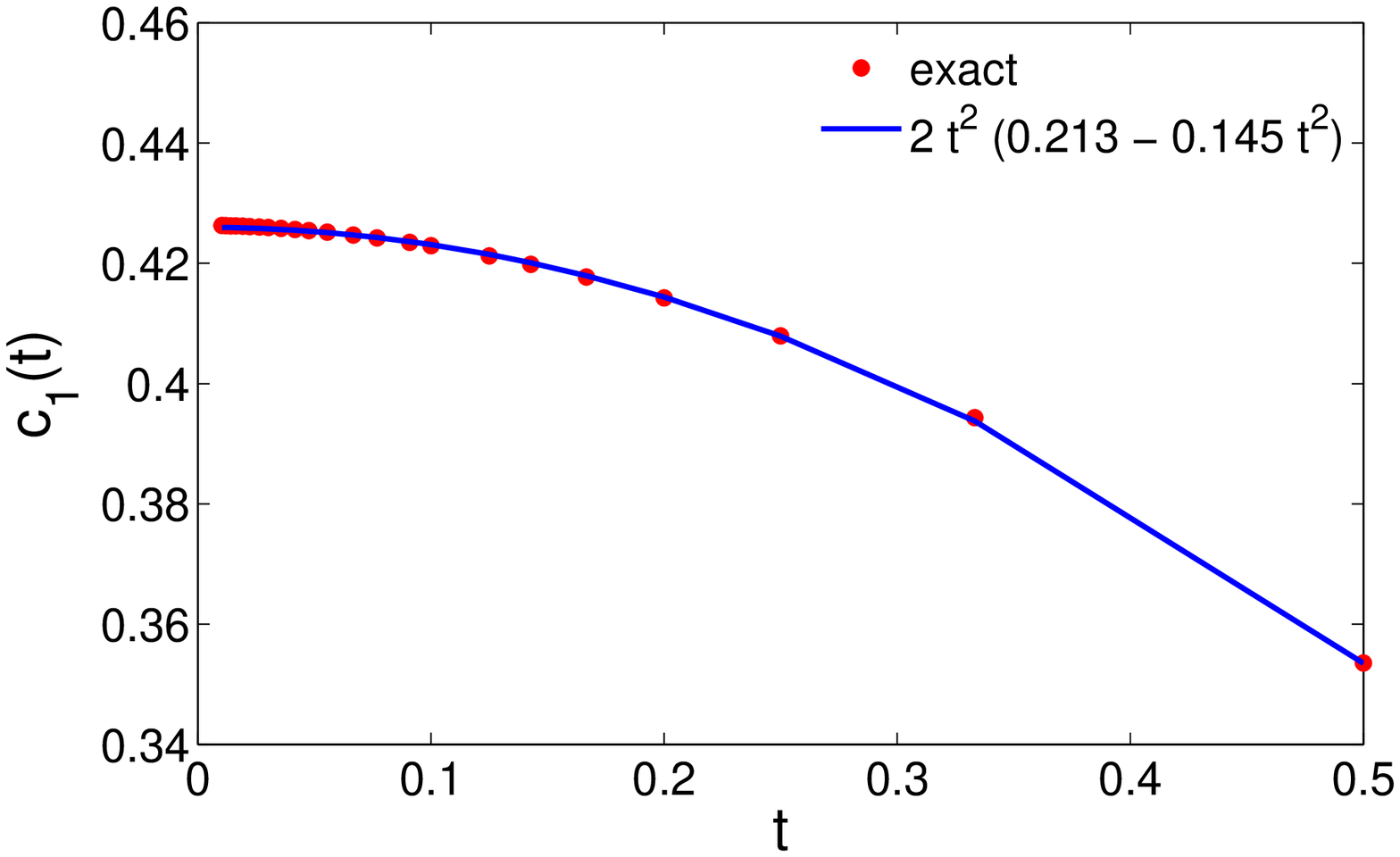}
\end{center}
\caption{
(Color online) The exact coefficient $c_1(t)$ for $m = 2,...,101$
computed through Eq.~\eref{eq:c1} and its fit as $2t^2(0.213 -
0.145t^2)$. }
\label{fig:c1t}
\end{figure}

Since the coefficients $a_i$ depend on $g$ only through $q_k$, they
rapidly converge to the limiting values.  In the limit $g\to \infty$,
the matrix $\mathbf{U}_1$ gets a diagonal form
\begin{equation*}
\mathbf{U}_1 \simeq \left(\begin{array}{c | c} 
\sqrt{\mathbf{\Lambda}} & 0 \\ \hline   0  & \sqrt{\mathbf{\Lambda}} \\ 
\end{array} \right),
\end{equation*}
from which
\begin{equation*}
\mathbf{A}_1 = \mathbf{V^T}\mathbf{U}_1\mathbf{V}\simeq\mathbf{V}^T\sqrt{\mathbf{\Lambda}}\mathbf{V} + 
\mathbf{S}^T\mathbf{V}^T\sqrt{\mathbf{\Lambda}}\mathbf{V}\mathbf{S} = \sqrt{\mathbf{L}} + \mathbf{S}^T\sqrt{\mathbf{L}}\mathbf{S} \qquad (g\gg 1).
\end{equation*}
The limiting matrix on the right-hand side, which is explicit and
independent of $g$, determines the limiting values $a_0$ and $a_1$.
Moreover, the explicit formulas for $\lambda_k$ and $\Psi_{kl}$ allow
one to get the elements of the matrix $\sqrt{\mathbf{L}}$ ($k,l =
0,...,m-1$):
\begin{equation*}
(\sqrt{\mathbf{L}})_{kl} = - \frac{\sin(\frac{\pi}{2m})}{m} \left(\frac{1}{\cos(\frac{\pi}{2m}) - \cos(\frac{\pi(k-l)}{m})}
+ \frac{1}{\cos(\frac{\pi}{2m}) - \cos(\frac{\pi(k+l+1)}{m})} \right).
\end{equation*}
Using this representation, we obtain numerically an accurate
approximation
\begin{equation*}
a_1 = \det(\mathbf{A}_1) \simeq \sqrt{t} ~2^m ~ (0.213 - 0.145 t^2 + ... ) ,
\end{equation*}
from which
\begin{equation}
\label{eq:c1t}
c_1(t) \simeq  \frac{a_1}{m^{3/2} 2^{m-1}} = 2t^2 (0.213 - 0.145 t^2 + ... ) .
\end{equation}
Its accuracy is illustrated on Fig. \ref{fig:c1t}.

% A_Andreanov_c1t_fig(C1,C0,t);

\subsection{General case}

In the general case $\frac{1}{m+1} < t < \frac{1}{m}$ (for some $m =
2,3,4,...$), the analysis is quite similar but formulas are more
cumbersome.  The contribution $I_t(g)$ from the Gaussian integral
reads as
\begin{equation*}
I_t(g) = \frac{\delta(t-\delta)}{2^{2m-3} g^{2m-1} \sqrt{m(m+1)}} \det(\mathbf{A}) .
\end{equation*}
By construction, the matrix $\mathbf{A}$ can be represented as
$\mathbf{A} = \mathbf{A}_0 + g \mathbf{A}_1$, where the matrix
$\mathbf{A}_0$ is independent of $g$, while the matrix $\mathbf{A}_1$
depends on $g$ only through $\alpha_k$ and $\tilde{\alpha}_k$.  As a
consequence, $\mathbf{A}_1$ becomes independent of $g$ for large $g$
(or $s$).  The determinant of the matrix $\mathbf{A}_0 + g
\mathbf{A}_1$ is in general a polynomial of $g$ of degree $2m+1$.  We
conclude that the asymptotic behavior of $\varphi_t(s)$ in
Eq.~\eref{eq:varphi_m2} is essentially determined by the exponential
function $\exp(-\sqrt{s} \frac{\eta(t)}{\sqrt{2(1-t)}})$, with an
algebraic prefactor from $\det(\mathbf{A})$.

For the cases $m = 1$ and $m = 2$, we obtain that $\det(\mathbf{A}) =
g^{2m-1}(a_0 + a_1 g)$, with the coefficients $a_0$ and $a_1$
depending on $t$.  As a consequence, one gets $I_t(g) = c_0 + c_1 g$,
with the new coefficients $c_0$ and $c_1$.  We expect that this
representation is valid for any $m$, but the related analysis is
beyond the scope of the paper.

\section{Small $g$ asymptotic behavior for $t < \frac12$}
\label{sec:Asmall}

We only consider the specific cases $t = 1/m$, with $m = 2,3,4,...$.
One can split the explicit representation~\eref{eq:A2} into two
parts: $\mathbf{A} = \mathbf{H}_0 + \mathbf{H}_1$, with
\begin{equation}
\label{eq:H0H1}
\fl
\mathbf{H}_0 = \frac{g\sqrt{\mathbf{L}}}{\tanh(gt\sqrt{\mathbf{L}})} ,  \quad \mathbf{H}_1 =  
-  \mathbf{S}^T \frac{g\sqrt{\mathbf{L}}}{\sinh(gt\sqrt{\mathbf{L}})} - \frac{g\sqrt{\mathbf{L}}}{\sinh(gt\sqrt{\mathbf{L}})}\mathbf{S}
+ \mathbf{S}^T\frac{g\sqrt{\mathbf{L}}}{\tanh(gt\sqrt{\mathbf{L}})}\mathbf{S} .
\end{equation}
One gets then $\det(\mathbf{A}) = \det(\mathbf{H}_0) \det(\mathbf{I} +
\mathbf{H}_0^{-1}\mathbf{H}_1)$, from which the explicit computation
of the determinant $\det(\mathbf{H}_0)$ yields
\begin{equation}
\label{eq:varphi_aux}
\fl
\varphi_t(s) = \exp\left(-\frac{gt}{2} \sum\limits_{k=1}^{m-1} \sqrt{\lambda_k}\right) 
\left(\prod\limits_{k=1}^{m-1} \frac{1+q_k^2}{2}\right)^{-1/2} ~ \frac{1}{\sqrt{\det(\mathbf{I} + \mathbf{H}_0^{-1}\mathbf{H}_1)}}.
\end{equation}
Up to this point, this is just another representation of the previous
result.

In the limit $g\to 0$, one can expand $q_k$ from Eq.~\eref{eq:q_alpha}
into Taylor series to get
\begin{equation*}
\fl
\ln\left(\prod\limits_{k=1}^{m-1} \frac{1+q_k^2}{2}\right) = \sum\limits_{k=1}^{m-1} \ln\left(\frac{1+q_k^2}{2}\right)
\simeq -tg \sum\limits_{k=1}^{m-1} \sqrt{\lambda_k} + \frac{t^2g^2}{2}
\sum\limits_{k=1}^{m-1} \lambda_k - \frac{t^4 g^4}{12} \sum\limits_{k=1}^{m-1} \lambda_k^2 + \cdots 
\end{equation*}
Given that
\begin{equation*}
\sum\limits_{k=1}^{m-1} \lambda_k = 2(m-1) , \qquad  \sum\limits_{k=1}^{m-1} \lambda_k^2 = 6m-8 ,
\end{equation*}
one finds
\begin{equation*}
\left(\prod\limits_{k=1}^{m-1} \frac{1+q_k^2}{2}\right)^{-1/2} \simeq \exp\left(\frac{gt}{2} \left(\sum\limits_{k=1}^{m-1} \sqrt{\lambda_k} \right)
- st + \frac{s^2 t^3(1 - 4t/3)}{(1-t)^2} + \cdots\right).
\end{equation*}
Note that the first term here exactly compensates the first factor in
Eq.~\eref{eq:varphi_aux}.

Now we turn to the analysis of the determinant $\det(\mathbf{I} +
\mathbf{H}_0^{-1} \mathbf{H}_1)$.  Expanding expressions for
$\mathbf{H}_0^{-1}$ and $\mathbf{H}_1$ from Eqs.~\eref{eq:H0H1} into
Taylor series, one gets
\begin{eqnarray*}
\fl
\mathbf{H}_0^{-1}\mathbf{H}_1 &=& (\mathbf{S}^T\mathbf{S} - \mathbf{S}^T - \mathbf{S}) + \frac{z}{6} 
\biggl(2\mathbf{S}^T\mathbf{L}\mathbf{S} + \mathbf{S}^T\mathbf{L} + \mathbf{S}\mathbf{L} - 2\mathbf{LS}^T\mathbf{S} 
+ 2\mathbf{LS}^T + 2\mathbf{LS}\biggr) \\
\fl
&& - \frac{z^2}{90} \biggl(2\mathbf{S}^T\mathbf{L}^2\mathbf{S} + \frac{7}{4}(\mathbf{S}^T\mathbf{L}^2 
+ \mathbf{L}^2 \mathbf{S}) + 10 \mathbf{LS}^T\mathbf{LS} + 5\mathbf{LS}^T\mathbf{L} + 5\mathbf{LSL}\biggr) + O(z^3) ,
\end{eqnarray*}
where $z = g^2 t^2$.  Denoting $\mathbf{J} = \mathbf{I} +
\mathbf{S}^T\mathbf{S} - \mathbf{S}^T - \mathbf{S}$, one has
\begin{equation*}
\det(\mathbf{I} + \mathbf{H}_0^{-1}\mathbf{H}_1) = \det(\mathbf{J}) \det\left(\mathbf{I} + z\mathbf{J}_1 + \frac{z^2}{2}\mathbf{J}_2 + O(z^3)\right)  ,
\end{equation*}
where
\begin{eqnarray*}
\mathbf{J}_1 &=& \frac16 \mathbf{J}^{-1} \biggl(2\mathbf{S}^T\mathbf{LS} + \mathbf{S}^T\mathbf{L} + 
\mathbf{SL} - 2\mathbf{LS}^T\mathbf{S} + 2\mathbf{LS}^T + 2\mathbf{LS}\biggr),  \\
\mathbf{J}_2 &=& \frac{1}{45} \mathbf{J}^{-1} \biggl(2\mathbf{S}^T \mathbf{L}^2\mathbf{S} + 
\frac{7}{4}(\mathbf{S}^T \mathbf{L}^2 + \mathbf{L}^2\mathbf{S}) + 10\mathbf{LS}^T\mathbf{LS} + 5\mathbf{LS}^T\mathbf{L} + 5\mathbf{LSL}\biggr) .
\end{eqnarray*}
Checking that $\det(\mathbf{J}) = 1$ and using a perturbative
expansion for the second determinant,
\begin{equation*}
\det(\mathbf{I} + \mathbf{X}) = 1 + \tr(\mathbf{X}) + \frac{1}{2} \bigl[\tr(\mathbf{X})^2 - \tr(\mathbf{X}^2)\bigr] + \cdots ,
\end{equation*}
one finds, up to $z^2$:
\begin{equation*}
\det(\mathbf{I} + \mathbf{H}_0^{-1}\mathbf{H}_1) = 1 + z \tr(\mathbf{J}_1) + \frac{z^2}{2} \biggl( \tr(\mathbf{J}_2) + 
(\tr(\mathbf{J}_1))^2 - \tr(\mathbf{J}_1^2) \biggr) + O(z^3) .
\end{equation*}
Using the explicit form of the matrices $\mathbf{J}_1$ and
$\mathbf{J}_2$, one can check that $\tr(\mathbf{J}_1) = 0$ and compute
exactly two other traces, from which
\begin{equation}
\det(\mathbf{I}+ \mathbf{H}_0^{-1}\mathbf{H}_1) \simeq 1 + s^2 \frac{t^3(2/3 - t)}{(1-t)^2} + O(s^3) .
\end{equation}

Bringing all these results together, we obtain
\begin{equation}
\varphi_t(s) \simeq e^{-ts} ~ \frac{\exp(s^2 t^3\frac{(1 - 4t/3)}{(1-t)^2})}{\sqrt{1 + s^2 t^3 \frac{2/3 - t}{(1-t)^2}}} + O(s^3).
\end{equation}
One can see that the factor in front of $e^{-ts}$ gives the
second-order correction in the form $s^2 t^3$.  If one rescales the
TAMSD $\chi_t$ by $t$ (to set the mean value to $1$), the Laplace
transform $\tilde{\varphi}_t(s')$ of the probability density for
$\chi_t/t$ is obtained by replacing $st$ by $s'$.  In the limit $t\to
0$, one can show that $\tilde{\varphi}_t(s')$ converges to $e^{-s'}$,
as expected.

\section{Comparison of moments}
\label{sec:moments}

Given that a generalized Gamma distribution turns out to be an
accurate approximation for the probability density $p_t(z)$, one can
attempt to retrieve the parameters $a$, $b$ and $\nu$ of this
distribution by matching the moments of the true and approximate
distributions.  For a GGD, one has
\begin{equation}
\mu_\alpha \equiv \langle \chi_t^\alpha \rangle = (ab)^\alpha \frac{K_{\nu+\alpha}(c)}{K_\nu(c)} , \qquad  c = 2\sqrt{a/b} .
\end{equation}
Following Ref. \cite{Grebenkov11b}, one can solve the system of two
nonlinear equations involving the first three moments:
\begin{equation}
\label{eq:moments_system}
\frac{\mu_2}{\mu_1^2} = \frac{K_{\nu+2}(c) K_\nu(c)}{K_{\nu+1}^2(c)}  , \qquad  
\frac{\mu_3}{\mu_1^3} = \frac{K_{\nu+3}(c) K_\nu^2(c)}{K_{\nu+1}^3(c)}  .
\end{equation}
Once the pair $\{c, \nu\}$ is found, one retrieves the other two
parameters as
\begin{equation}
\label{eq:ab_nuc}
b = 2\mu_1 \frac{K_\nu(c)}{c K_{\nu+1}(c)}, \qquad  a = bc^2/4
\end{equation}
(note that the first relation was misprinted in \cite{Grebenkov11b}).
Using the explicit formulas~\eref{eq:kappa1_1}, \eref{eq:kappa2_1},
\eref{eq:kappa3_1} for the cumulant moments for $t < \frac12$, one
finds
\begin{eqnarray}
\label{eq:mu2}
\fl
\frac{\mu_2}{\mu_1^2} &=& 1 + \frac{\kappa_2}{\kappa_1^2} = 1 + \frac{t(4-5t)}{3(1-t)^2} \simeq 1 + \frac43 t + t^2 + O(t^3) , \\
\label{eq:mu3}
\fl
\frac{\mu_3}{\mu_1^3} &=& 1 + 3\frac{\kappa_2}{\kappa_1^2} + \frac{\kappa_3}{\kappa_1^3} 
= 1 + \frac{t(60-69t-19t^2)}{15(1-t)^3} \simeq 1 + 4 t + \frac{37}{5} t^2 + O(t^3) . 
\end{eqnarray}
We have solved numerically Eqs. \eref{eq:moments_system} and found the
dependence of the parameters $a$, $b$ and $\nu$ on the lag time $t$,
as shown on Fig. \ref{fig:abnu}.

One can see that when $t\to 0$, both $c$ and $\nu$ increase inversely
proportional to $t$.  In this limit, one can approximate the modified
Bessel function $K_\nu(\nu x)$ as
\begin{equation}
K_\nu(\nu x) \simeq \sqrt{\pi/\nu} \left(\frac{1+\sqrt{1+x^2}}{x}\right)^\nu (1 + x^2)^{-1/4} e^{-\nu\sqrt{1+x^2}} .
\end{equation}
Considering $\ve = 1/\nu$ as a small parameter and assuming $c/\nu$ to
be a constant in the limit $t\to 0$, one takes $c/\nu = x - \alpha \ve
+ \beta \ve^2$ and gets
\begin{eqnarray}
\label{eq:K2_eps}
\fl
\frac{K_{\nu+2}(c) K_\nu(c)}{K_{\nu+1}^2(c)} &\simeq& 1 + \frac{\ve}{\sqrt{1+x^2}} 
+ \Biggl(\frac{\alpha x}{(1+x^2)^{3/2}} - \frac{\sqrt{1+x^2} - 1}{(1+x^2)^2}\Biggr) \ve^2 + O(\ve^3) , \\
\label{eq:K3_eps}
\fl
\frac{K_{\nu+3}(c) K_\nu^2(c)}{K_{\nu+1}^3(c)} &\simeq& 1 + \frac{3\ve}{\sqrt{1+x^2}} 
+ \Biggl(\frac{3\alpha x (2+\sqrt{1+x^2})}{(1+x^2) (1 + \sqrt{1+x^2})^2} \\
\nonumber
&-& \frac{2(2+x^2)\sqrt{1+x^2} + 3x^4 + 4x^2 + 4}{(1+x^2)^2 (1 + \sqrt{1+x^2})^2}\Biggr) \ve^2 + O(\ve^3).
\end{eqnarray}
It is worth noting that these terms are independent of $\beta$ (which
appears in higher-order terms).

Comparing these expansions to Eqs.~\eref{eq:mu2}, \eref{eq:mu3}, one
sees that $\ve$ is of the order of $t$.  Taking $\ve = \gamma t +
\delta t^2 + O(t^3)$, one immediately gets $\gamma = \frac43
\sqrt{1+x^2}$ from the first-order terms.  Substituting $\ve = \frac43
\sqrt{1+x^2} ~t + \delta ~t^2$ into Eqs.~\eref{eq:K2_eps},
\eref{eq:K3_eps}, one gets a system of two equations with three
unknowns $\alpha$, $\delta$ and $x$:
\begin{eqnarray*}
\fl
&& \frac{\delta}{\sqrt{1+x^2}} + \Biggl(\frac{\alpha x}{\sqrt{1+x^2}} - \frac{\sqrt{1+x^2} - 1}{(1+x^2)}\Biggr) \frac{16}{9} = 1 , \\
\fl
&& \frac{3\delta}{\sqrt{1+x^2}} + \Biggl(\frac{3\alpha x (2+\sqrt{1+x^2})}{(1 + \sqrt{1+x^2})^2} 
 - \frac{2(2+x^2)\sqrt{1+x^2} + 3x^4 + 4x^2 + 4}{(1+x^2)(1 + \sqrt{1+x^2})^2}\Biggr) \frac{16}{9} = \frac{37}{5} .
\end{eqnarray*}
These equations are linear in $\alpha$ and $\delta$.  Expressing
$\alpha$ from the first equation and substituting into the second one
yields, after algebraic simplifications, a simple equation on $x$
which turns out to be independent of $\delta$: $2x^2 - 38 +
\sqrt{1+x^2} (21x^2 - 38) = 0$.  Its solution is
\begin{equation*}
x = \frac{\sqrt{1159}}{21} \approx 1.6211 .
\end{equation*}
As a consequence, one finds the leading terms for $\nu$ and $c$ and
then for $a$ and $b$ as $t\to 0$ which are summarized in
Eqs.~\eref{eq:nu_asympt}, \eref{eq:a_asympt}, \eref{eq:b_asympt}.


\vskip 5mm
\begin{thebibliography}{10}

\bibitem{Weiss}            Weiss G H, 1994
                           {\it Aspects and Applications of the Random Walk}
                           (North-Holland, Amsterdam)

\bibitem{Ben-Avraham}      Ben-Avraham D and Havlin S, 2000
                           {\it Diffusion and reaction in disordered systems} 
                           (Cambridge University Press)

\bibitem{Saxton97b}        Saxton M J and Jacobson K, 1997
%                           ``Single-particle tracking: Applications to Membrane Dynamics'',
                           {\it Annu. Rev. Biophys. Biomol. Struct.} {\bf 26} 373

\bibitem{Majumdar05}       Majumdar S N, 2005
%                           ``Brownian functionals in physics and computer science'',
                           {\it Curr. Sci.} {\bf 89} 2076

\bibitem{Grebenkov07}      Grebenkov D S, 2007
%                           ``NMR Survey of Reflected Brownian Motion'',
                           {\it Rev. Mod. Phys.} {\bf 79} 1077



\bibitem{Bouchaud}         Bouchaud J-P and Potters M, 2000
                           {\it Theory of Finantial Risks: From Statistical Physics to Risk Management}
                           (Cambridge University Press)



\bibitem{Qian91}           Qian H, Sheetz M P, and Elson E L, 1991
%                           ``Single particle tracking. Analysis of diffusion and flow in two-dimensional systems'',
                           {\it Biophys. J.} {\bf 60} 910

\bibitem{Saxton93}         Saxton M J, 1993
%                           ``Lateral diffusion in an archipelago. Single-particle diffusion'',
                           {\it Biophys. J.} {\bf 64} 1766

\bibitem{Saxton97}         Saxton M J, 1997
%                           ``Single-particle tracking: the distribution of diffusion coefficients'',
                           {\it Biophys. J.} {\bf 72} 1744

\bibitem{Goulian00}        Goulian M and Simon S M, 2000
%                           ``Tracking Single Proteins within Cells'',
                           {\it Biophys. J.} {\bf 79} 2188

\bibitem{Tolic04}          Toli\'c-Norrelykke I M, Munteanu E-L, Thon G, Oddershede L, and Berg-Sorensen K, 2004
%                           ``Anomalous Diffusion in Living Yeast Cells'',
                           {\it Phys. Rev. Lett.} {\bf 93} 078102

\bibitem{Golding06}        Golding I and Cox E C, 2006
%                           ``Physical Nature of Bacterial Cytoplasm'',
                           {\it Phys. Rev. Lett.} {\bf 96} 098102

\bibitem{Arcizet08}        Arcizet D, Meier B, Sackmann E, R\"adler J O, and Heinrich D, 2008
%                           ``Temporal Analysis of Active and Passive Transport in Living Cells'',
                           {\it Phys. Rev. Lett.} {\bf 101} 248103

\bibitem{Wilhelm08}        Wilhelm C, 2008
%                           ``Out-of-Equilibrium Microrheology inside Living Cells'',
                           {\it Phys. Rev. Lett.} {\bf 101} 028101

\bibitem{Wirtz09}          Wirtz D, 2009
%                           ``Particle-Tracking Microrheology of Living Cells: Principles and Applications'', 
                           {\it Ann. Rev. Biophys.} {\bf 38} 301

\bibitem{Metzler09}        Metzler R, Tejedor V, Jeon J-H, He Y, Deng W H, Burov S, and Barkai E, 2009
%                           ``Analysis of Single Particle Trajectories: From Normal to Anomalous Diffusion'',
                           {\it Acta Phys. Pol. B} {\bf 40} 1315






\bibitem{Feynman}          Feynman R P and Hibbs A R, 1965
                           {\it Quantum Mechanics and Path Integrals}
                           (New York, McGraw-Hill)

\bibitem{Freidlin}         Freidlin M, 1985
                           {\it Functional Integration and Partial Differential Equations}, 
                           Annals of Mathematics Studies (Princeton University, Princeton, New Jersey) 


\bibitem{Bray80}           Bray A J and Moore M A, 1980
%                           ``Replica theory of quantum spin glasses'',
                           {\it J. Phys. C: Solid State Phys.} {\bf 13} L655

\bibitem{Kirkpatrick87}    Kirkpatrick T R and Thirumalai D, 1987
%                           ``p-spin-interaction spin-glass models: Connections with the structural glass problem'',
                           {\it Phys. Rev. B} {\bf 36} 5388



%\bibitem{Elsgolts}         L. E. El'sgol'ts and S. B. Norkin,
%                           {\it Introduction to thet Theory and Application of Differential Equations with Deviating Arguments}
%                           Mathematics in Science and Engineering, Vol. 105 (Academic Press, New York, London 1973).

\bibitem{Ruben62}         Ruben H, 1962
%                          ``Probability Content of Regions Under Spherical Normal Distributions, IV: The Distribution 
%                          of Homogeneous and Non-Homogeneous Quadratic Functions of Normal Variables'',
                          {\it Ann. Math. Stat.} {\bf 33} 542

\bibitem{Ruben63}         Ruben H, 1963
%                          ``A New Result on the Distribution of Quadratic Forms'',
                          {\it Ann. Math. Stat.} {\bf 34} 1582

\bibitem{Robbins48}       Robbins H, 1948
%                          ``The Distribution of a Definite Quadratic Form'',
                          {\it Ann. Math. Stat.} {\bf 19} 266

\bibitem{Robbins49}       Robbins H and Pitman E J G, 1949
%                          ``Application of the Method of Mixtures to Quadratic Forms in Normal Variates'',
                          {\it Ann. Math. Stat.} {\bf 20} 552

\bibitem{Pachares55}      Pachares J, 1955
%                          ``Note on the Distribution of a Definite Quadratic Form'',
                          {\it Ann. Math. Stat.} {\bf 26} 128

\bibitem{Shah61}          Shah B K and Khatri C G, 1961
%                          ``Distribution of a Definite Quadratic Form for Non-Central Normal Variates'',
                          {\it Ann. Math. Stat.} {\bf 32} 883



\bibitem{Gurland53}        Gurland J, 1953
%                           ``Distribution of Quadratic Forms and Ratios of Quadratic Forms'',
                           {\it Ann. Math. Stat.} {\bf 24} 416

\bibitem{Gurland55}        Gurland J, 1955
%                           ``Distribution of Definite and of Indefinite Quadratic Forms'',
                           {\it Ann. Math. Stat.} {\bf 26} 122

\bibitem{Gurland56}        Gurland J, 1956
%                           ``Quadratic Forms in Normally Distributed Random Variables'',
                           {\it Ind. J. Stat.} {\bf 17} 37

\bibitem{Kotz67a}          Kotz S, Johnson N L and Boyd D W, 1967
%                           ``Series Representations of Distributions of Quadratic Forms in Normal Variables. I. Central Case'',
                           {\it Ann. Math. Stat.} {\bf 38} 823

\bibitem{Kotz67b}          Kotz S, Johnson N L and Boyd D W, 1967
%                           ``Series Representations of Distributions of Quadratic Forms in Normal Variables II. Non-Central Case'',
                           {\it Ann. Math. Stat.} {\bf 38} 838.


\bibitem{Duits09}          Duits M H G, Li Y, Vanapalli S A, and Mugele F, 2009
%                           ``Mapping of spatiotemporal heterogeneous particle dynamics in living cells'',
                           {\it Phys. Rev. E} {\bf 79} 051910

\bibitem{Boyer11}          Boyer D and Dean D S, 2011
%                           ``On the distribution of estimators of diffusion constants for Brownian motion'',
                           {\it J. Phys. A: Math. Theor.} {\bf 44} 335003

\bibitem{Boyer12}          Boyer D, Dean D S, Mejia-Monasterio C, and Oshanin G, 2012
%                           ``Optimal estimates of the diffusion coefficient of a single Brownian trajectory'',
                           {\it Phys. Rev. E} {\bf 85} 031136

\bibitem{Voisinne10}       Voisinne G, Alexandrou A, and Masson J-B, 2010
%                           ``Quantifying biomolecule diffusivity using an optimal Bayesian method'',
                           {\it Biophys. J.} {\bf 98} 596

\bibitem{Grebenkov11a}     Grebenkov D S, 2011
%                           ``Time-Averaged Quadratic Functionals of a Gaussian Process'', 
                           {\it Phys. Rev. E} {\bf 83} 061117

\bibitem{Grebenkov11b}     Grebenkov D S, 2011
%			   ``Probability Distribution of the Time-Averaged Mean-Square Displacement of a Gaussian Process'', 
			   {\it Phys. Rev. E} {\bf 84} 031124 

\bibitem{Nakagawa07}       Nakagawa K, 2007
%                           ``Application of Tauberian Theorem to the Exponential Decay of the Tail Probability of a Random Variable'',
                           {\it IEEE Trans. Inf. Theory} {\bf 53} 3239

\bibitem{Hoel}             Hoel P G, 1962
                           {\it Introduction to Mathematical Statistics}
                           (John Wiley and Sons, New York)


\end{thebibliography}
\end{document}